\documentclass[11pt,showpacs,preprintnumbers,superscriptaddress,amsmath,amssymb,nofootinbib,aps]{revtex4}
\usepackage{graphicx}
\usepackage{dcolumn}
\usepackage{bm}
\usepackage{amssymb}
\usepackage{amsmath}
\usepackage{epsfig}
\usepackage{color}
\usepackage{slashed}
\usepackage{hhline}
\usepackage{ulem}
\usepackage{bbm}
\usepackage{tikz-feynman}
\tikzfeynmanset{compat=1.1.0}

\def\be{\begin{equation}}
\def\ee{\end{equation}}
\newcommand{\bea}{\begin{eqnarray}}
\newcommand{\eea}{\end{eqnarray}}
\newcommand{\nn}{\nonumber}

\newcommand{\ourNIS}{$\mathcal{FR} (6)$}

\begin{document}


\title{Two-loop rainbow neutrino masses in a non-invertible symmetry}


\author{Hiroshi Okada}
\email{hiroshi3okada@htu.edu.cn}
\affiliation{Department of Physics, Henan Normal University, Xinxiang 453007, China}

\author{Yoshihiro Shigekami}
\email{shigekami@htu.edu.cn}
\affiliation{Department of Physics, Henan Normal University, Xinxiang 453007, China}

\date{\today}

\begin{abstract}
{
We propose two-loop rainbow type of the neutrino mass model via $\mathbbm{Z}_2$ gauging of $\mathbbm{Z}_6$ non-invertible symmetry in which we introduce three families of isospin doublet vector-like fermions, heavy right-handed neutrinos and isospin doublet and singlet bosons. 
All new fields, which have nonzero charges under the non-invertible symmetry, can be dark matter candidates, since the non-invertible symmetry possesses a remnant $\mathbbm{Z}_2$ symmetry that plays a role in assuring the stability of our dark matter candidate. 
Even though the non-invertible symmetry is dynamically broken at one-loop level, its violation does not affect our scenario. 
In this paper, we especially consider the lightest mode of the neutral components in the doublet vector-like fermions as our main dark matter candidate. 
The dark matter is potentially degenerated to the other two families of neutral fermions, since the mass difference is induced at one-loop level. 
Thus, we consider our dark matter candidate in rather simpler co-annihilation system among their particles. 
Considering all the constraints of neutrino oscillation data, lepton flavor violations, muon $g-2$ and the relic density of dark matter, we perform the numerical analysis and show some allowed regions for these phenomenology. 
Due to our dark matter nature, the sum of neutrino masses in case of normal hierarchy is larger than that in case of inverted hierarchy, which is opposite situation compared with typical active neutrino models. 
}
\end{abstract}
\maketitle
\newpage

\section{Introduction}
\label{sec:intro}

It is one of our important tasks to discover nature of the neutrino sector; masses and their mixing beyond the Standard Model (SM). 
Even though there exists a vast literature of the theory and experimental results, we have not found a unique solution yet. 
Hence, along one direction of the methodology, it becomes more popular that neutrino models include and correlate with several verifiable phenomenology such as lepton flavor violations (LFVs), muon anomalous magnetic dipole moment (muon $g-2$), dark matter candidate, and etc. 

Radiative seesaw scenario is one of the promising ideas in order to achieve such explanations among the phenomenology.\footnote{See some representative radiative models in refs.~\cite{Ma:2006km,Zee:1980ai} for one-loop, refs.~\cite{Zee:1985id,Babu:1988ki,Kajiyama:2013rla} for two-loop and refs.~\cite{Krauss:2002px,Aoki:2008av,Gustafsson:2012vj,Nishiwaki:2015iqa} for three-loop.} 
To construct these models, we need to add new fields and {\it often} impose additional symmetry to forbid tree-level contributions of neutrino masses and/or consider dark matter (DM) candidate~\cite{Bertone:2004pz,Cirelli:2024ssz}. 
In particular, it would be an intriguing idea that neutrino masses are so tiny, because the neutrino cannot directly interact with the SM Higgs but interact with DM candidate. 
In this case, it is inevitable to impose an additional symmetry to stabilize the DM candidate and various types of symmetries, such as Abelian discrete/continuous and non-Abelian discrete/continuous symmetries,\footnote{For specific symmetries, there are enormous studies with, e.g., discrete $\mathbbm{Z}_2$~\cite{Ma:2006km,LopezHonorez:2006gr,Pierce:2007ut,Chen:2009ab,Okada:2010wd,LopezHonorez:2010eeh,Eby:2011qa,Chao:2012sz,Farzan:2012hh,Kashiwase:2013uy,Biswas:2013nn,Aoki:2013gzs,Klasen:2013jpa,Goudelis:2013uca,Basak:2013cga,Ibarra:2014qma,Alanne:2014bra,Keus:2014jha,Okada:2014qsa,Vicente:2014wga,Bonilla:2014xba,Jin:2014glp,Chakrabarty:2015yia,Ibarra:2015fqa,Blinov:2015qva,Okada:2016gsh,Kanemura:2016sos,Singirala:2016kam,Arcadi:2016kmk,Okada:2016tci,DuttaBanik:2016jzv,Oda:2017kwl,Chao:2017rwv,Cox:2017rgn,Okada:2018ktp,Escudero:2018fwn,Bandyopadhyay:2018qcv,Bhattacharya:2018cgx,Borah:2018smz,Das:2019pua,Dutta:2020xwn,Barman:2020ifq,Nam:2020byw,Chen:2020ark,Arcadi:2021mag,Sarazin:2021nwo,Datta:2021gyi,Bandyopadhyay:2022xlp,Khaw:2022qxh,VanLoi:2023pkt,Zhang:2024sox,Lu:2025vif,Liu:2025swd}, global/gauged $U(1)$~\cite{Khalil:2008kp,Mambrini:2010dq,Kanemura:2011vm,Lindner:2011it,Okada:2012sg,Lindner:2013awa,Kanemura:2014rpa,Ko:2014uka,Rodejohann:2015lca,Biswas:2016ewm,Escudero:2016tzx,Patra:2016ofq,Singirala:2017see,Nomura:2017vzp,Bandyopadhyay:2017bgh,DeRomeri:2017oxa,Nomura:2017jxb,Nomura:2017wxf,Nanda:2017bmi,Singirala:2017cch,Okada:2018tgy,Das:2018tbd,Biswas:2019ygr,Mohapatra:2019ysk,Okada:2020evk,Seto:2020udg,Nagao:2020azf,Asai:2020qlp,Ghosh:2021khk,Nath:2021uqb,Barman:2021yaz,Okada:2022cby,Okada:2023mdv,deBoer:2023phz,Figueroa:2024tmn,Babu:2024zoe,He:2025uvm,Abdelrahim:2025fiz}, discrete $A_4$~\cite{Hirsch:2010ru,Meloni:2010sk,Boucenna:2011tj,Hamada:2014xha,deMedeirosVarzielas:2015ybd,deMedeirosVarzielas:2015lmh,Lamprea:2016egz,DeLaVega:2018bkp,Nomura:2019jxj,Nomura:2019yft,Nomura:2019lnr,Okada:2020dmb,Behera:2020lpd,Hutauruk:2020xtk,Nagao:2021rio,Kobayashi:2021ajl,Dasgupta:2021ggp,Otsuka:2022rak,Kang:2022psa,Bonilla:2023pna,Kim:2023jto,Nomura:2023kwz,Kumar:2024zfb,Borah:2024gql,Kumar:2025zvv} and $SU(2)$~\cite{Hambye:2008bq,Walker:2009en,Chen:2009ab,Diaz-Cruz:2010czr,Chiang:2013kqa,Baek:2013dwa,Khoze:2014woa,Chen:2015nea,Chen:2015cqa,Gross:2015cwa,Chen:2015dea,Karam:2015jta,Ko:2016yfb,Choi:2017zww,Barman:2018esi,Choi:2019zeb,Barman:2019lvm,Abe:2020mph,Ko:2020qlt,Ghosh:2020ipy,Nomura:2020zlm,Baouche:2021wwa,Chowdhury:2021tnm,Borah:2022phw,Borah:2022dbw,Otsuka:2022zdy,Frigerio:2022kyu,Coleppa:2023vfh,Chen:2025ihr} symmetries. 
For more details as well as other possible symmetries, see also references therein.} have been widely applied. 
Recently, interesting but (a bit) weird symmetries are arisen, and they are applied to several directions of phenomenology such as quark and lepton mass matrix textures, radiatively induced tiny masses via dynamical breaking, axion, stability of DM, and so on~\cite{Choi:2022jqy,Cordova:2022ieu,Cordova:2022fhg,Cordova:2024ypu,Kobayashi:2024yqq,Kobayashi:2024cvp,Kobayashi:2025znw,Suzuki:2025oov,Liang:2025dkm,Kobayashi:2025ldi,Kobayashi:2025cwx,Kobayashi:2025lar,Nomura:2025sod,Dong:2025jra,Nomura:2025yoa,Chen:2025awz,Okada:2025kfm,Kobayashi:2025thd,Suzuki:2025bxg,Jangid:2025krp,Kobayashi:2025rpx,Jiang:2025psz,Nomura:2025tvz,Okada:2025adm,Nakai:2025thw}. 

In this paper, we apply $\mathbbm{Z}_2$ gauging of $\mathbbm{Z}_6$ non-invertible symmetry to the neutrino sector, in which the dominant neutrino mass matrix is induced at the two-loop level. 
We introduce three families of isospin doublet vector fermions, heavy right-handed neutrinos and isospin doublet and singlet bosons. 
These new fields are assumed to be charged under our non-invertible symmetry, and electrically neutral components of these can be DM candidates, due to a remnant $\mathbbm{Z}_2$ symmetry that plays a role in assuring their stability. 
Even though the non-invertible symmetry is dynamically broken at one-loop level, its violation does not affect our scenario. 
In the current work, we especially consider the lightest mode of the neutral components in the doublet vector-like fermions as our main DM candidate. 
It would be emphasized that the DM is potentially degenerated to the other two families of neutral fermions, since the mass difference is induced at one-loop level under the assumption of degenerated bare masses. 
This affects our DM phenomenology, and the correct relic density will be explained via simple co-annihilation system among these particles. 
In addition to the relic density, we can discuss allowed regions for the other phenomenological constraints, e.g., neutrino oscillation data, LFVs and the muon $g-2$, with numerical analysis. 
Due to our DM nature, the sum of neutrino masses in case of the normal hierarchy (NH) is larger than that in case of the inverted hierarchy (IH). 
This result is opposite situation compared with typical active neutrino models, and therefore, we may be able to see and/or test some signature of our model. 

This paper is organized as follows. 
We first review our model setup formulating isospin doublet neutral fermion mass matrix, active neutrino mass matrix at two-loop level and relevant phenomenology for our discussion in Sec.~\ref{sec:II}. 
Then, in Sec.~\ref{sec:III}, we discuss our DM candidate and demonstrate how to explain the correct relic density of the DM in our co-annihilation system. 
In Sec.~\ref{sec:IV}, we perform our numerical analysis satisfying neutrino observables, LFVs, muon $g-2$ and relic density of DM. 
Finally, we summarize and conclude in Sec.~\ref{sec:summary}.

\section{Model setup}
\label{sec:II}

In this section, we review our model setup. 
As for the fermion sector, we introduce three families of right-handed neutrinos $N_R$ and vector fermions $L' \equiv [n', e']^T$ in addition to the SM fermions. 
$N_R$ is totally singlet under the SM gauge groups with \ourNIS \ charge of $\rho$, while $L'_{L, R}$ have same SM charges as $L_L$ with \ourNIS \ charge of $\sigma$. 
Here, $n'$ and $e'$ indicate the neutral and charged component of $L'_{L, R}$, respectively. 
As for the boson sector, we add an inert doublet boson $\eta \equiv [\eta^+, \eta^0]^T$ and an inert singlet boson $S_0$. 
$\eta$ and $S_0$ respectively have $\epsilon$ and $\sigma$ charge under \ourNIS. 
The relevant particle contents and its charge assignments are summarized in Table~\ref{tab:chaged-assignments}. 
For our discussion, SM charged leptons and Higgs doublet as well as all quarks are singlet under \ourNIS. 
\begin{table}[!t]
\begin{center}
\begin{tabular}{|c||c|c|c|c|c||c|c|c|}\hline\hline
& \multicolumn{5}{c||}{Fermions} & \multicolumn{3}{c|}{Scalars} \\ \hline
Fields & ~$L_L$~ & ~$\ell_R$~ & ~$N_R$~ & ~$L'_L$~ & ~$L'_R$~ & ~~$H$~~ & ~~$\eta$~~ & ~$S_0$~ \\ \hline
~$SU(2)_L$~ & $\mathbbm{2}$& \multicolumn{2}{c|}{$\mathbbm{1}$} & \multicolumn{2}{c||}{$\mathbbm{2}$} & \multicolumn{2}{c|}{$\mathbbm{2}$} & $\mathbbm{1}$ \\ \hline
$U(1)_Y$ & $- \frac{1}{2}$ & $-1$ & $0$ & \multicolumn{2}{c||}{$- \frac{1}{2}$} & \multicolumn{2}{c|}{$\frac{1}{2}$} & $0$ \\ \hline
\ourNIS & \multicolumn{2}{c|}{$\mathbbm{1}$} & $\rho$ & \multicolumn{2}{c||}{$\sigma$} & $\mathbbm{1}$ & $\epsilon$ & $\sigma$ \\ \hline
\end{tabular}
\caption{The charge assignments of relevant particles. 
Other particles not listed in this table have same SM charges with singlet under \ourNIS. }
\label{tab:chaged-assignments}
\end{center}
\end{table}

With these charge assignments and fusion rules of \ourNIS \ in Appendix~\ref{app:FusionRule}, we find the following renormalizable Lagrangian terms:
\begin{align}
- \mathcal{L} &\supset y^{\ell}_{i} \overline{L_{L_i}} H \ell_{R_i} + f_{ib} \overline{L_{L_i}} L'_{R_b} S_0 + f'_{ab} \overline{L'_{L_a}} L'_{R_b} S_0 + g_{a\alpha} \overline{L'_{L_a}} N_{R_{\alpha}} \widetilde{\eta} + h_{ab} \overline{L^{\prime c}_{R_a}} N_{R_b} \eta \nn \\
&\hspace{1.2em} + M_{R_{\alpha}} \overline{N^c_{R_{\alpha}}} N_{R_{\alpha}} + M'_a \overline{L'_{L_a}} L'_{R_a} + {\rm h.c.} \, , \label{eq:ma_1}
\end{align}
where $y^\ell$, $M_R$, and $M'$ are supposed to be diagonal without loss of generality, and $\tilde{\eta} \equiv (i\sigma_2) \eta^{\dag}$ with $\sigma_2$ being the second Pauli matrix. 
The scalar potential is found as
\begin{align}
V &= \sum_{\phi = H, \eta, S_0} \Bigl[ - \mu_{\phi}^2 |\phi|^2 + \lambda_{\phi} |\phi|^4 \Bigr] + \kappa_S S_0^3 + \mu |\eta|^2 S_0 + \lambda''_{H \eta}(H^{\dag} \eta)^2 + {\rm c.c.} \nonumber \\[0.3ex]
&\hspace{1.2em} + \lambda_{H \eta} |H|^2 |\eta|^2 + \lambda'_{H \eta} |H^{\dagger} \eta|^2 + \lambda_{H S} |H|^2 |S_0|^2 + \lambda_{\eta S} |\eta|^2 |S_0|^2 \, .
\label{eq:ma_2}
\end{align}
Note that we assume that additional scalar bosons $\eta, S_0$ do not acquire non-zero vacuum expectation values (VEVs), and therefore, potential analysis for minimization conditions and each mass eigenvalue can be straightforwardly done. 
Therefore, we omit it in this paper, and we move to focus on the specific mass generation as well as its spectrum for additional fermions below. 

\begin{figure}
\centering
\begin{tikzpicture}
\begin{feynman}
\vertex[label=left:\(L' (\sigma)\)] (a) at (0,0);
\vertex[label=right:\(L' (\sigma)\)] (b) at (6,0);
\vertex (c) at (1,0);
\vertex[label=above:\(N_R (\rho)\)] (m) at (3,0);
\vertex (d) at (5,0);
\vertex (e) at (3,2);
\vertex[label=above:\(\langle H \rangle\)] (f) at (1.5,3.2);
\vertex[label=above:\(\langle H \rangle\)] (g) at (4.5,3.2);
\diagram* {
(a) -- (c) -- [insertion={[size=2.5pt]0.5}] (d) -- (b),
(c) -- [scalar, quarter left, edge label=\(\eta (\epsilon)\)] (e) -- [scalar, quarter left, edge label=\(\eta (\epsilon)\)] (d),
(f) -- [scalar] (e) -- [scalar] (g),
};
\end{feynman}
\end{tikzpicture}
\caption{Majorana mass terms $\delta m_L$, $\delta m_R$ at one-loop level. 
$\epsilon, \sigma, \rho$ in the parentheses are \ourNIS \ charge for corresponding fields, and $\langle H \rangle$ denotes the VEV of the SM Higgs doublet. }
\label{fig:MajoranaMass}
\end{figure}
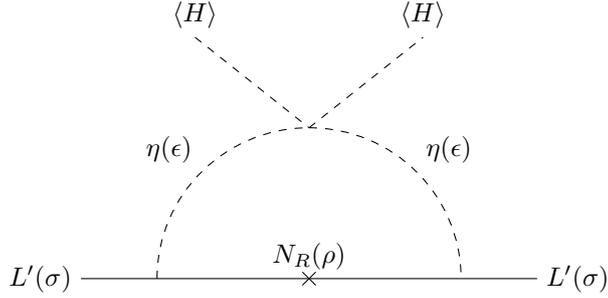

\subsection{Mass matrix for isospin doublet neutral fermions}

The mass matrix for isospin doublet neutral fermion consists of the bare Dirac mass term $M'$ and smaller Majorana mass terms $\delta m_R$, $\delta m_L$ at one-loop level, as shown in Fig.~\ref{fig:MajoranaMass}. 
The $6 \times 6$ mass matrix in basis of $[n'_R,n'_L]^T$ is then given by
\begin{align}
\left( \begin{array}{cc} 
\overline{n'^C_R} & \overline{n'_L} 
\end{array} \right) \left( \begin{array}{cc} 
\delta m_R & M' \\
M' & \delta m_L
\end{array} \right) \left( \begin{array}{c} 
{n'_R} \\ {n'^C_L}
\end{array} \right) \equiv \left( \begin{array}{cc} 
\overline{n'^C_R} & \overline{n'_L} 
\end{array} \right) M_N \left(\begin{array}{c} 
{n'_R} \\ {n'^C_L}
\end{array} \right) \, ,
\label{eq:cgd-mtrx}
\end{align}
where $M'$ is diagonal and can be real and positive, we abbreviate generation indices, and each block is understood as $3 \times 3$ matrix. 
As we will see below, $\delta m_R$ and $\delta m_L$ are $3 \times 3$ symmetric matrices, and hence, $M_N$ is understood as a $6 \times 6$ symmetric matrix. 
$\delta m_{R, L}$ are estimated as follows:
\begin{align}
(\delta m_R^{\dag})_{a b} &\approx \frac{h^*_{a \alpha} M_{R_{\alpha}} h^{\dag}_{\alpha b}}{(4 \pi)^2} \left[ \frac{m_{\eta_R}^2}{m_{\eta_R}^2 - M_{R_{\alpha}}^2} \ln \left( \frac{m_{\eta_R}^2}{M_{R_{\alpha}}^2} \right) - \frac{m_{\eta_I}^2}{m_{\eta_I}^2 - M_{R_{\alpha}}^2} \ln \left( \frac{m_{\eta_I}^2}{M_{R_{\alpha}}^2} \right) \right] \, , \label{eq:delmR} \\
(\delta m_L)_{a b} &\approx \frac{g_{a \alpha} M_{R_{\alpha}}^{\dag} g^T_{\alpha b}}{(4 \pi)^2} \left[ \frac{m_{\eta_R}^2}{m_{\eta_R}^2 - M_{R_{\alpha}}^2} \ln \left( \frac{m_{\eta_R}^2}{M_{R_{\alpha}}^2} \right) - \frac{m_{\eta_I}^2}{m_{\eta_I}^2 - M_{R_{\alpha}}^2} \ln \left( \frac{m_{\eta_I}^2}{M_{R_{\alpha}}^2} \right) \right] \, , \label{eq:delmL}
\end{align}
where $m_{\eta_R}$ and $m_{\eta_I}$ are masses for $\eta_R$ and $\eta_I$, respectively, with $\eta^0 = (\eta_R + i \eta_I) / \sqrt{2}$. 

Now we consider how to diagonalize $M_N$. 
First of all, it is convenient to transform $M_N$ as
\begin{align}
M'_N \equiv V^{(1)T}_N M_N V^{(1)}_N &\equiv \left( \begin{array}{cc} 
\frac{1}{\sqrt2} \cdot {\bf 1} & \frac{1}{\sqrt2} \cdot {\bf 1} \\
- \frac{1}{\sqrt2} \cdot {\bf 1} & \frac{1}{\sqrt2} \cdot {\bf 1} \end{array} \right) \left( \begin{array}{cc} 
\delta m_R & M' \\
M' & \delta m_L
\end{array} \right) \left( \begin{array}{cc} 
\frac{1}{\sqrt2} \cdot {\bf 1} & - \frac{1}{\sqrt2} \cdot {\bf 1} \\
\frac{1}{\sqrt2} \cdot {\bf 1} & \frac{1}{\sqrt2} \cdot {\bf 1} \end{array} \right) \nn \\
&= \left( \begin{array}{cc} 
M' + \frac{\delta m_L + \delta m_R}{2} & \frac{\delta m_L - \delta m_R}{2} \\
\frac{\delta m_L - \delta m_R}{2} & - M' + \frac{\delta m_L + \delta m_R}{2}
\end{array} \right) \equiv \left( \begin{array}{cc} 
M' + m_+ & m_- \\
m_- & - M' + m_+
\end{array} \right) \, , \label{eq:MNpDef}
\end{align}
where bold number ${\bf 1}$ represents $3 \times 3$ unit matrix, and we define $m_{\pm} \equiv \left( \delta m_L \pm \delta m_R \right) / 2$. 
Since $\delta m_{R, L}$ are generated at the one-loop level, it is natural to assume that $\delta m_R, \delta m_L \ll M'$, and hence, $m_- \lesssim m_+ \ll M'$. 
This helps to diagonalize $M'_N$ as
\begin{align}
(\Omega P)^T M'_N (\Omega P) &\approx \left( \begin{array}{cc} 
M' - m_+ & {\bf 0} \\
{\bf 0} & M' + m_+
\end{array} \right) \approx \left( \begin{array}{cc} 
D_R & {\bf 0} \\
{\bf 0} & D_L^C
\end{array} \right) \, , \label{eq:MNpOmegaP} \\[0.5ex]
\Omega P &\approx \left( \begin{array}{cc} 
{\bf 1} & - \frac{1}{2} M'^{-1} m_- \\
\frac{1}{2} m_-^{\dag} M'^{-1} & {\bf 1}
\end{array} \right) \left( \begin{array}{cc} 
{\bf 0} & {\bf 1} \\
i \cdot {\bf 1} & {\bf 0}
\end{array} \right) \, . \label{eq:OmegaP}
\end{align}
Here, bold number ${\bf 0}$ represents $3 \times 3$ zero matrix. 
In general, $m_+$ is not diagonal, and we further need to diagonalize $M' \pm m_+$. 
However, we regard diagonal components of Eq.~\eqref{eq:MNpOmegaP} as mass eigenvalues in this paper, with in good approximation.\footnote{In order to diagonalize $M_N$, we need to assume that $M'$ is proportional to the unit matrix, $M' = M'_0 \times {\mathbbm 1}_{3 \times 3}$, and we take this assumption hereafter. 
For more details about further diagonalization of $M'_N$, see, e.g., refs.~\cite{Catano:2012kw,Kajiyama:2012xg,Nomura:2025tvz}.} 
As a result, the diagonalizing matrix for $M_N$ is approximately obtained from $V_N^{(1)}$ in Eq.~\eqref{eq:MNpDef} and $\Omega P$ in Eq.~\eqref{eq:OmegaP} as $V_N \approx V_N^{(1)} \Omega P$. 
With this diagonalization matrix, we explicitly write down $n'_{R, L}$ as
\begin{align}
&n'_R \approx \frac{1}{\sqrt2} 
\left(
i(-1-\frac{1}{2} M'^{-1} m_-) \psi_R + (1-\frac{1}{2} m_-^{\dag} M'^{-1})\psi_L^C 
\right) \, , \label{eq:masseigenNR} \\
&n'^C_L \approx \frac{1}{\sqrt2} 
\left(
i(1-\frac{1}{2} M'^{-1} m_-) \psi_R + (1+\frac{1}{2} m_-^{\dag} M'^{-1})\psi_L^C 
\right) \, , \label{eq:masseigenNL}
\end{align}
where $\psi_{R}$ and $\psi^C_L$ are the mass eigenvectors for $n'_R$ and $n'^C_L$ and have mass eigenvalues of $D_R$ and $D_L^C$, respectively.

\subsection{Active neutrino mass matrix}
\label{subsec:neut}

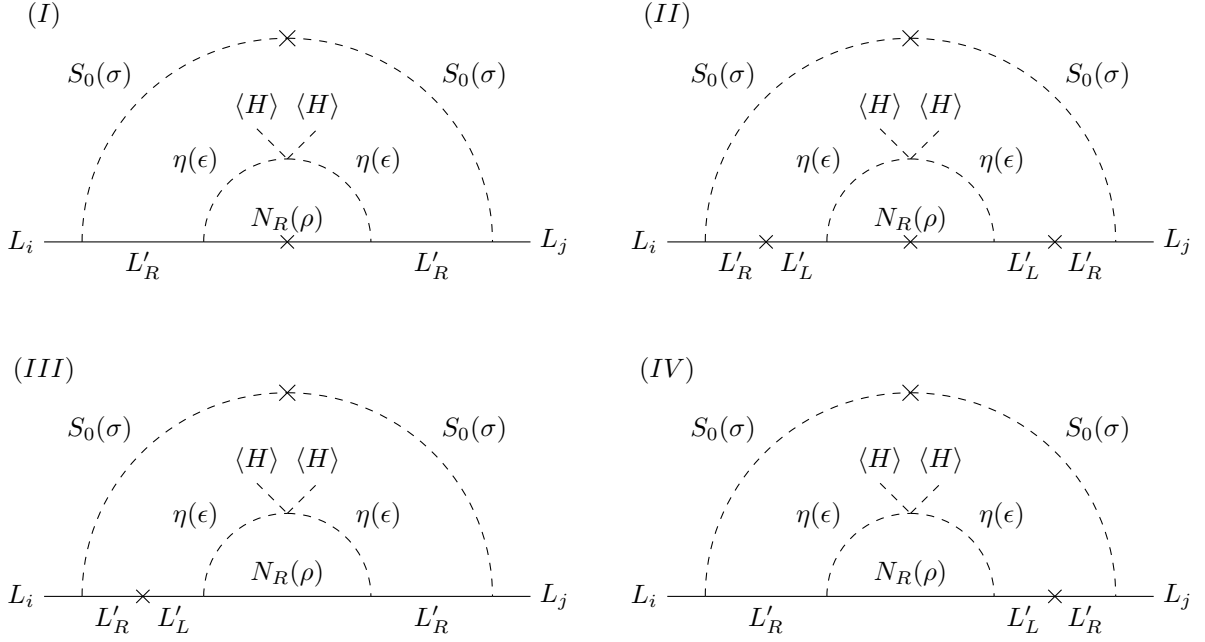
\begin{figure}
\centering
\begin{tikzpicture}
\begin{feynman}
\vertex[label=above:\((I)\)] (l1) at (0,2.7);
\vertex[label=left:\(L_i\)] (a1) at (0,0);
\vertex[label=right:\(L_j\)] (b1) at (6.4,0);
\vertex (c1) at (2.1,0);
\vertex[label=above:\(N_R (\rho)\)] (m1) at (3.2,0);
\vertex (d1) at (4.3,0);
\vertex (e1) at (3.2,1.1);
\vertex[label=above:\(\langle H \rangle\)] (f1) at (2.8,1.5);
\vertex[label=above:\(\langle H \rangle\)] (g1) at (3.6,1.5);
\vertex (h1) at (0.5,0);
\vertex (i1) at (5.9,0);
\vertex (j1) at (3.2,2.7);
\diagram* {
(a1) -- (h1) -- [edge label'=\(L'_R\)] (c1) -- [insertion={[size=2.5pt]0.5}] (d1) -- [edge label'=\(L'_R\)] (i1) -- (b1),
(c1) -- [scalar, quarter left, edge label=\(\eta (\epsilon)\)] (e1) -- [scalar, quarter left, edge label=\(\eta (\epsilon)\)] (d1),
(f1) -- [scalar] (e1),
(e1) -- [scalar] (g1),
(h1) -- [scalar, quarter left, edge label=\(S_0 (\sigma)\),insertion={[size=3pt]0.9999}] (j1) -- [scalar, quarter left, edge label=\(S_0 (\sigma)\)] (i1),
};
\vertex[label=above:\((II)\)] (l2) at (8.2,2.7);
\vertex[label=left:\(L_i\)] (a2) at (8.2,0);
\vertex[label=right:\(L_j\)] (b2) at (14.6,0);
\vertex (c2) at (10.3,0);
\vertex[label=above:\(N_R (\rho)\)] (m2) at (11.4,0);
\vertex (d2) at (12.5,0);
\vertex (e2) at (11.4,1.1);
\vertex[label=above:\(\langle H \rangle\)] (f2) at (11,1.5);
\vertex[label=above:\(\langle H \rangle\)] (g2) at (11.8,1.5);
\vertex (h2) at (8.7,0);
\vertex (i2) at (14.1,0);
\vertex (j2) at (11.4,2.7);
\diagram* {
(a2) -- (h2) -- [insertion={[size=2.5pt]0.5}, edge label'=\(L'_R ~~~ L'_L\)] (c2) -- [insertion={[size=2.5pt]0.5}] (d2) -- [insertion={[size=2.5pt]0.5}, edge label'=\(L'_L ~~~ L'_R\)] (i2) -- (b2),
(c2) -- [scalar, quarter left, edge label=\(\eta (\epsilon)\)] (e2) -- [scalar, quarter left, edge label=\(\eta (\epsilon)\)] (d2),
(f2) -- [scalar] (e2),
(e2) -- [scalar] (g2),
(h2) -- [scalar, quarter left, edge label=\(S_0 (\sigma)\),insertion={[size=3pt]0.9999}] (j2) -- [scalar, quarter left, edge label=\(S_0 (\sigma)\)] (i2),
};
\vertex[label=above:\((III)\)] (l3) at (0,-2);
\vertex[label=left:\(L_i\)] (a3) at (0,-4.7);
\vertex[label=right:\(L_j\)] (b3) at (6.4,-4.7);
\vertex (c3) at (2.1,-4.7);
\vertex[label=above:\(N_R (\rho)\)] (m3) at (3.2,-4.7);
\vertex (d3) at (4.3,-4.7);
\vertex (e3) at (3.2,-3.6);
\vertex[label=above:\(\langle H \rangle\)] (f3) at (2.8,-3.2);
\vertex[label=above:\(\langle H \rangle\)] (g3) at (3.6,-3.2);
\vertex (h3) at (0.5,-4.7);
\vertex (i3) at (5.9,-4.7);
\vertex (j3) at (3.2,-2);
\diagram* {
(a3) -- (h3) -- [insertion={[size=2.5pt]0.5}, edge label'=\(L'_R ~~~ L'_L\)] (c3) -- (d3) -- [edge label'=\(L'_R\)] (i3) -- (b3),
(c3) -- [scalar, quarter left, edge label=\(\eta (\epsilon)\)] (e3) -- [scalar, quarter left, edge label=\(\eta (\epsilon)\)] (d3),
(f3) -- [scalar] (e3),
(e3) -- [scalar] (g3),
(h3) -- [scalar, quarter left, edge label=\(S_0 (\sigma)\),insertion={[size=3pt]0.9999}] (j3) -- [scalar, quarter left, edge label=\(S_0 (\sigma)\)] (i3),
};
\vertex[label=above:\((IV)\)] (l4) at (8.2,-2);
\vertex[label=left:\(L_i\)] (a4) at (8.2,-4.7);
\vertex[label=right:\(L_j\)] (b4) at (14.6,-4.7);
\vertex (c4) at (10.3,-4.7);
\vertex[label=above:\(N_R (\rho)\)] (m4) at (11.4,-4.7);
\vertex (d4) at (12.5,-4.7);
\vertex (e4) at (11.4,-3.6);
\vertex[label=above:\(\langle H \rangle\)] (f4) at (11,-3.2);
\vertex[label=above:\(\langle H \rangle\)] (g4) at (11.8,-3.2);
\vertex (h4) at (8.7,-4.7);
\vertex (i4) at (14.1,-4.7);
\vertex (j4) at (11.4,-2);
\diagram* {
(a4) -- (h4) -- [edge label'=\(L'_R\)] (c4) -- (d4) -- [insertion={[size=2.5pt]0.5}, edge label'=\(L'_L ~~~ L'_R\)] (i4) -- (b4),
(c4) -- [scalar, quarter left, edge label=\(\eta (\epsilon)\)] (e4) -- [scalar, quarter left, edge label=\(\eta (\epsilon)\)] (d4),
(f4) -- [scalar] (e4),
(e4) -- [scalar] (g4),
(h4) -- [scalar, quarter left, edge label=\(S_0 (\sigma)\),insertion={[size=3pt]0.9999}] (j4) -- [scalar, quarter left, edge label=\(S_0 (\sigma)\)] (i4),
};
\end{feynman}
\end{tikzpicture}
\caption{Two-loop diagrams for the neutrino mass matrix. }
\label{fig:neut}
\end{figure}

The neutrino mass matrix consists of four kinds of two-loop diagrams as shown in Fig.~\ref{fig:neut}. 
Each contribution to the mass matrix can be estimated as
\begin{align}
(m_{\nu}^{(I)})_{ij} &\simeq \frac{2}{(4 \pi)^4} f_{ia} h^{\dag}_{\alpha a} M_{R_{\alpha}} h^{\dag}_{\alpha b} f^T_{bj}(m_{\eta_R}^2 - m_{\eta_I}^2) F_D^{(a,b,\alpha)} \, , \\
(m_{\nu}^{(II)})_{ij} &\simeq \frac{1}{(4 \pi)^4} f_{ia} M'_a g_{a\alpha} M_{R_{\alpha}} g^T_{\alpha b}M'_b f^T_{bj}(m_{\eta_R}^2 - m_{\eta_I}^2) F_T^{(a,b,\alpha)} \, , \\
(m_{\nu}^{(III)})_{ij} &\simeq \frac{2}{(4 \pi)^4} f_{ia} M'_a g_{a\alpha } h^{\dag}_{\alpha b} f^T_{bj}(m_{\eta_R}^2 - m_{\eta_I}^2) F_S^{(a,b,\alpha)} \, , \\
(m_{\nu}^{(IV)})_{ij} &\simeq \frac{2}{(4 \pi)^4} f_{ia} h^{\dag}_{\alpha a} g^T_{\alpha b}M'_b f^T_{bj}(m_{\eta_R}^2 - m_{\eta_I}^2) F_S^{(a,b,\alpha)} \, ,
\end{align}
where loop functions $F_{D, S, T}^{(a,b,\alpha)}$ are respectively found as
\begin{align}
F_D^{(a,b,\alpha)} &\equiv \int \! \frac{[dx]_3}{(x_2 + x_3)(x_2 + x_3 - 1)} \int \! \frac{[dy]_4}{y_1 M'^2_a + y_2 M'^2_b + y_3 m^2_S + y_4 {\cal M}_{\alpha}^2} \, , \\
F_S^{(a,b,\alpha)} &\equiv \int \! \frac{[dx]_3}{x_2 + x_3 - 1} \int \! \frac{[dy]_4}{y_1 M'^2_a + y_2 M'^2_b + y_3 m^2_S + y_4 {\cal M}_{\alpha}^2} \, , \\
F_T^{(a,b,\alpha)} &\equiv \int \! \frac{[dx]_3}{(x_2 + x_3)(x_2 + x_3 - 1)} \int \! \frac{[dy]_4}{(y_1 M'^2_a + y_2 M'^2_b + y_3 m^2_S + y_4 {\cal M}^2_{\alpha})^2} \, , \\[0.5ex]
\text{with } ~~ {\cal M}_{\alpha}^2 &\equiv \frac{x_1 M^2_{R_{\alpha}} + x_2 m_{\eta_R}^2 + x_3 m_{\eta_I}^2}{(x_2 + x_3)(x_2 + x_3 - 1)} \, ,
\end{align}
with $[dx]_3 \equiv dx_1 dx_2 dx_3 \delta (1 - x_1 - x_2 - x_3)$ and $[dy]_4 \equiv dy_1 dy_2 dy_3 dy_4 \delta (1 - y_1 - y_2 - y_3 - y_4)$. 
Before formulating the neutrino mass eigenvalues and its mixing, it is worthwhile discussing the dominance of each mass matrix $m_{\nu}^{(I-IV)}$. 
Since we assume the lightest mode of neutral components in vector-like fermions $n'$ to be an intriguing DM candidate, we suppose to have the following mass hierarchies:
\begin{align}
10^2 \, {\rm GeV} \le M' \lesssim 10^4 \, {\rm GeV} \, , \quad \{M_{R_{\alpha}}, m_{\eta_R}, m_{\eta_I}, m_S\} = [1.2 \times M', 10^5 \, {\rm GeV}] \, , \label{eq:dm-cond}
\end{align}
where $M'$ is mass eigenvalue of $e'$. 
The lower mass bound $1.2 \times M'$ above is determined so that our DM neither have relevant effects of co-annihilations nor semi-annihilations~\cite{Griest:1990kh,DEramo:2010keq}. 
Under these mass hierarchies, we numerically evaluate the mass scale of $m_{\nu}^{(I-IV)}$, and obtain the following tendency:
\begin{align}
\left| \frac{m_{\nu}^{(III, IV)}}{m_{\nu}^{(I)}} \right| \lesssim 0.4 \, , \quad \left| \frac{m_{\nu}^{(II)}}{m_{\nu}^{(I)}} \right| \lesssim 0.2 \, ,
\end{align}
which implies that $m_{\nu}^{(I)}$ is dominant contribution to the active neutrino mass matrix.\footnote{{\it For simplicity, we impose $h \sim g$, and this assumption is applied to our numerical analysis below.} 
Precisely speaking, $m_{\nu}^{(I)}$ minimally dominates by 50\% in total neutrino mass matrix, and we might have to consider the other contributions in case of the minimum value of $m_{\nu}^{(I)}$. 
However, in our numerical analysis, we assume that this possibility is avoided by choosing proper structures for couplings $f$ and $h \sim g$.} 

Now that we fix $m_{\nu} \approx m_{\nu}^{(I)}$, which would be a numerically appropriate hypothesis, we proceed our analysis of the neutrino mass matrix. 
At first we redefine $m_{\nu}$ to be
\begin{align}
(m_{\nu})_{ij} &\equiv f_{ia} h^*_{a \alpha} F_{\nu}^{(a,b,\alpha)} h^{\dag}_{\alpha b} f^T_{bj} = f_{ia} M_{S_{ab}} f^T_{bj} \, , \label{eq:ourmnuMat} \\
F_{\nu}^{(a,b,\alpha)} &\equiv \frac{2 (m_{\eta_R}^2 - m_{\eta_I}^2)}{(4 \pi)^4} M_{R_{\alpha}} F_D^{(a,b,\alpha)} \, .
\end{align}
Then, $M_{S_{ab}} \equiv h^*_{a \alpha} F_{\nu}^{(a,b,\alpha)} h^{\dag}_{\alpha b}$ (no summation for $a/b$), which is complex symmetric $3 \times 3$ matrix, can be diagonalized by a single unitary matrix $U_S$ as $D_S \equiv U_S^T M_S U_S$. 
Here, $U_S$ is determined so that $D_S$ is the real and positive mass eigenvalue. 
Using this fact, $m_{\nu}$ can be written in terms of $D_S$ and $U_S$ as well as $f$ by
\begin{align}
m_{\nu} = f U_S^* D_S U_S^{\dag} f^T \, .
\label{eq:ourmnuMat2}
\end{align}
On the other hand, $m_{\nu}$ is a symmetric matrix as well (see Eqs.~\eqref{eq:ourmnuMat} and \eqref{eq:ourmnuMat2}), and hence, $m_{\nu}$ is also diagonalized by a single unitary matrix $U_{\nu}$ via $D_{\nu} \equiv U_{\nu}^{\dag} m_{\nu} U_{\nu}^*$.
Therefore, we can parametrize unknown Yukawa coupling $f$ in terms of neutrino experimental results and some degrees of freedom in our model as
\begin{align}
f = U_{\nu} D^{1/2}_{\nu} {\cal O}_f D_S^{-1/2} U_S^T \, ,
\label{eq:fExp}
\end{align}
where ${\cal O}_f$ is a $3 \times 3$ complex orthogonal matrix; ${\cal O}_f^T {\cal O}_f = {\cal O}_f {\cal O}_f^T = \mathbbm{1}$, and $U_{\nu}$ can be an observed mixing matrix, since the charged-lepton mass matrix is diagonal. 

\subsubsection{Constraints on neutrino masses and mixing angles}
\label{sec:const-nu}

In the neutrino sector, there are several experimental constraints on the model. 
First, the sum of neutrino masses, which is denoted by $\sum D_{\nu} = D_{\nu_1} + D_{\nu_2}+ D_{\nu_3}$, is estimated to be its upper bound by the minimal standard cosmological model with CMB data~\cite{Planck:2018vyg}:\footnote{Here, we do not consider the stringent constraint $\sum D_{\nu} \le 71 \, {\rm meV}$ at 95\% CL via DESI Collaboration~\cite{DESI:2024hhd}, since our numerical result does not suggest this region.}
\begin{align}
\sum D_{\nu} \le 120 \, {\rm meV} \, .
\label{eq:sum1}
\end{align}
However extend cosmological models provide different upper bounds, e.g., in ref.~\cite{Shao:2024mag}
\begin{align}
\sum D_{\nu} \le [110-190] \, {\rm meV} \, ,
\label{eq:sum2}
\end{align}
and for specific neutrino mass hierarchy, we have~\cite{Pang:2023joc}
\begin{align}
{\rm NH} &:\ \sum D_{\nu} \le [135-152] \, {\rm meV} \, , \label{eq:sum3_nh} \\
{\rm IH} &:\ \sum D_{\nu} \le [167-178] \, {\rm meV} \, . \label{eq:sum3_ih}
\end{align}

Next, the effective mass for neutrinoless double beta decay $m_{ee} \equiv |\sum_{i} D_{\nu_i} U_{\nu_{ei}}^2|$ in our model is given by
\begin{align}
m_{ee} = \left| D_{\nu_1} c^2_{12} c^2_{13} + D_{\nu_2} s^2_{12} c^2_{13} e^{i \alpha_{21}} + D_{\nu_3} s^2_{13}e^{i (\alpha_{31} - 2 \delta_{CP})} \right| \, ,
\label{eq:0nu2beta}
\end{align}
where $s_{12,23,13} \, (c_{12,23,13})$, which are short-hand notations $\sin\theta_{12,23,13} \, (\cos\theta_{12,23,13})$, are lepton mixing of $U_\nu$,\footnote{Since the charged-lepton mass matrix is diagonal in our model, the neutrino mixing matrix $U_{\nu}$ is identical to the observed lepton mixing matrix $U_{\rm PMNS}$~\cite{Maki:1962mu}.} $\delta_{CP}$ is the Dirac phase, and $\alpha_{21, 31}$ are the Majorana phases. 
The KamLAND-Zen collaboration gives an upper bound on $m_{ee}$ at the 90\% confidence level (CL)~\cite{KamLAND-Zen:2024eml} as
\begin{align}
m_{ee} < (28-122) \, {\rm meV} \, .
\label{eq:mee_current}
\end{align}
Future experiments provide us more stringent constraints as follows:
\begin{align}
m_{ee} < \begin{cases}
(9-21) \, {\rm meV} & \text{from LEGEND-1000~\cite{LEGEND:2021bnm}} \, , \\[0.3ex]
(4.7-20.3) \, {\rm meV} & \text{from nEXO~\cite{nEXO:2021ujk}} \, .
\end{cases}
\label{eq:mee_future}
\end{align}

Moreover, $m_{\nu_e}\equiv \sum_{i}D_{\nu_i}^2 |U_{\nu_{ei}}|^2 $, which is model independent observable, is given by
\begin{align}
m_{\nu_e} = \sqrt{D_{\nu_1}^2 c^2_{13} c^2_{12} + D_{\nu_2}^2 c^2_{13} s^2_{12} + D_{\nu_3}^2 s^2_{13}} \, .
\label{eq:mnue-modelindep}
\end{align}
Its upper bound at 90\% CL from KATRIN~\cite{KATRIN:2024cdt} is given by
\begin{align}
m_{\nu_e} \le 450 \, {\rm meV} \, ,
\end{align}
which is weaker than the other constraints.

\subsection{Other constraints: LFVs and muon $g-2$}
\label{sec:leptonpheno}

In addition to the constraints related with the neutrino sector, we should care about the experimental bounds on the charged lepton sector. 
In this work, we focus on $\ell_i \to \ell_j \gamma$ and charged lepton $g-2$; $\Delta a_{\ell_i}$, since these are arisen from $f$. 
The resultant branching ratios for $\ell_i \to \ell_j \gamma$ and new physics contributions to $\Delta a_{\ell_i}$ are found as
\begin{align}
{\rm BR} (\ell_i \to \ell_j \gamma) &\approx \frac{48 \pi^3 \alpha_{\rm em} C_{ij}}{G_F^2 m_{\ell_i}} \Bigl[ |(a_L)_{ij}|^2 + |(a_R)_{ij}|^2 \Bigr] \, , \\
\Delta a_{\ell_i} &\approx - m_{\ell_i} \Bigl[ (a_L)_{ii} + (a_R)_{ii} \Bigr] \, ,
\end{align}
where $a_L$ and $a_R$ are respectively given by
\begin{align}
(a_L)_{ij} &\approx - \frac{f_{ja} f^{\dag}_{ai}}{(4 \pi)^2} m_{\ell_j} 
\int [dx]_3 \frac{x_1 x_3}{x_1 m_S^2 + (x_2 + x_3) M'^2_a} \, , \label{eq:aL_2lp} \\
(a_R)_{ij} &\approx - \frac{f_{ja} f^{\dag}_{ai}}{(4 \pi)^2} m_{\ell_i} 
\int [dx]_3\frac{x_1 x_2 }{x_1 m_S^2 + (x_2 + x_3) M'^2_a} \, , \label{eq:aR_2lp}
\end{align}
$\alpha_{em}\approx1/137$ is the fine structure constant, $C_{21, 31, 32} \approx (1, 0.1784, 0.1736)$, $G_F \approx 1.17 \times 10^{-5} \, {\rm GeV}^{-2}$ is the Fermi constant, and $m_{\ell_{1, 2, 3}} \equiv m_{e, \mu, \tau} \, (\ll m_S, \ M'_a)$. 
The current upper bounds on ${\rm BR} (\ell_i \to \ell_j \gamma)$ are~\cite{MEGII:2023ltw,BaBar:2009hkt,Belle:2021ysv}
\begin{align}
{\rm BR} (\mu \to e \gamma) < 3.1 \times 10^{-13} \, , \quad {\rm BR} (\tau \to e \gamma) < 3.3 \times 10^{-8} \, , \quad {\rm BR} (\tau \to \mu \gamma) < 4.2 \times 10^{-8} \, ,
\end{align}
at 90\% CL. 
Among these bounds, ${\rm BR} (\mu \to e \gamma)$ is relevant to our discussion. 
For the lepton g-2, we have following experimental measurements~\cite{Fan:2022oyb,Fan:2022eto,ParticleDataGroup:2024cfk,Morel:2020dww}:
\begin{align}
\Delta a_e = (3.41 \pm 1.64) \times 10^{-13} \, , \quad \Delta a_{\mu} = (39 \pm 64) \times 10^{-11} \, ,
\label{eq:g-2ell}
\end{align}
while $\Delta a_{\tau}$ is less precise compared with $\Delta a_{e, \mu}$. 
Since our contributions to $\Delta a_{\ell_i}$ are proportional to $m_{\ell_i}^2 / m_S^2$ and roughly follow the minimal flavor violation relation $\Delta a_{\ell_i} / \Delta a_{\ell_j} \sim m_{\ell_i}^2 / m_{\ell_j}^2$, it is enough to check $\Delta a_{\mu}$ for our numerical analysis. 
Note that our prediction of $\Delta a_{\mu}$ is always positive, and the typical value is $\Delta a_{\mu} \sim 10^{-9}$ for $m_S, M' \sim 10^2$ with ${\cal O} (1)$ values for $f_{ia}$.

\section{Dark matter candidates}
\label{sec:III}

In our model, we have four kinds of DM candidates: the lightest mode of $N_R$, the lightest neutral component of $L'$, neutral component of inert boson $\eta$, and a singlet inert boson $S_0$. 
In our study, we concentrate on the case of the lightest neutral component of $L'$, since the other cases have already been discussed in our previous paper ref.~\cite{Okada:2025kfm}. 
We can evade the stringent constraints from the direct detection searches, because our DM candidate does not have spin independent cross section due to its Majorana nature! 
In this sense, Majorana mass terms generated at one-loop level are crucially important. 

In addition, the Majorana mass terms $\delta m_{R, L}$ play another important role for the DM phenomenology in our model. 
Due to one-loop contributions to $n'_{R, L}$, it is natural to consider co-annihilation processes among $\psi_R$, in order to explain the current observed relic density. 
In fact, we have checked that there are no solutions to satisfy the observed relic density of DM in the case with only annihilation process via the Yukawa coupling.\footnote{Even in case of gauge interactions, such as $\overline{\psi_{R_1}} \psi_{R_1} \to \bar{f} f$ via $Z$ boson mediated $s$-channel, its annihilation cross section is ${\cal O} (10^{-12}) \, {\rm GeV}^{-2}$ at most in the DM mass is ${\cal O} (10^3) \, {\rm GeV}$. 
Here, $f$ runs over all fermions in the SM.} 
Therefore, we focus on the co-annihilation processes only among $\psi_R$ for correct relic density.\footnote{If we also consider the co-annihilation with $e'$, we have to include a lot of processes via the SM gauge bosons. 
However, in our setup and allowed region we find below, the mass difference between $\psi_{R_1}$ and $e'$ is enough large to suppress its co-annihilation processes. 
Therefore, we only consider the co-annihilation with $\psi_{R_{2, 3}}$.} 

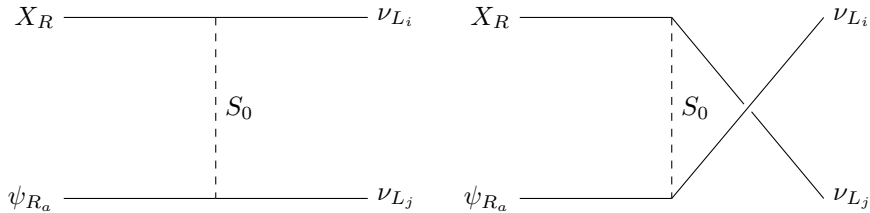
\begin{figure}
\centering
\begin{tikzpicture}
\begin{feynman}
\vertex[label=left:\(X_R\)] (at) at (0,2.4);
\vertex[label=left:\(\psi_{R_a}\)] (bt) at (0,0);
\vertex[label=right:\(\nu_{L_i}\)] (ct) at (4,2.4);
\vertex[label=right:\(\nu_{L_j}\)] (dt) at (4,0);
\vertex (et) at (2,2.4);
\vertex (ft) at (2,0);
\diagram* {
(at) -- (et) -- (ct),
(bt) -- (ft) -- (dt),
(et) -- [scalar, edge label=\(S_0\)] (ft),
};
\vertex[label=left:\(X_R\)] (au) at (6,2.4);
\vertex[label=left:\(\psi_{R_a}\)] (bu) at (6,0);
\vertex[label=right:\(\nu_{L_i}\)] (cu) at (10,2.4);
\vertex[label=right:\(\nu_{L_j}\)] (du) at (10,0);
\vertex (eu) at (8,2.4);
\vertex (fu) at (8,0);
\vertex (gu) at (8.95,1.25);
\vertex (hu) at (9.05,1.15);
\diagram* {
(au) -- (eu) -- (gu),
(hu) -- (du),
(bu) -- (fu) -- (cu),
(eu) -- [scalar, edge label=\(S_0\)] (fu),
};
\end{feynman}
\end{tikzpicture}
\caption{The main annihilation and co-annihilation processes in our model. 
$X_R \equiv \psi_{R_1}$ is our DM, and $a/i/j$ run $1, 2, 3$. }
\label{fig:anni}
\end{figure}

The dominant term to explain the relic density arises from
\begin{align}
- \frac{i}{\sqrt2} f_{ik} \left( 1 + \frac{1}{2} M'^{-1} m_- \right)_{ka} \overline{\nu_{L_i}} \psi_{R_a} S_0 + {\rm h.c.} \equiv i \tilde{f}_{ia} \overline{\nu_{L_i}} \psi_{R_a} S_0 + {\rm h.c.} \, ,
\end{align}
where $\psi_{R_1} \equiv X_R$ is our DM candidate, and its mass is defined by $D_{R_1} \equiv D_X$. 
This term leads to the annihilation and co-annihilation processes shown in Fig.~\ref{fig:anni}. 
The relic density is given by~\cite{Kong:2005hn} 
\begin{align}
\Omega h^2 \approx \frac{1.04 \times 10^9}{M_P \sqrt{g_*} J} \, , \quad J \approx \int_{x_f}^\infty \! dx \int_0^{\pi} \! d\theta \frac{da_{\rm eff} + \frac{6}{x} \left( db_{\rm eff} - \frac{da_{\rm eff}}{4} \right)}{x^2},
\end{align}
where $M_P \approx 1.22 \times 10^{19} \, {\rm GeV}$ is the reduced Planck mass, $g_* \approx 100$ is relativistic degrees of freedom, and $x_f \approx 25$. 
$da_{\rm eff}$ and $db_{\rm eff}$ are respectively given by 
\begin{align}
da_{\rm eff} &\approx \frac{4}{g^2_{\rm eff}} \sum_{a, b = 1}^3 da_{ab} (1 + \Delta_a)^{3/2} (1 + \Delta_b)^{3/2} e^{-x (\Delta_a + \Delta_b)} \, , \\
db_{\rm eff} &\approx \frac{4}{g^2_{\rm eff}} \sum_{a, b = 1}^3 db_{ab} (1 + \Delta_a)^{3/2} (1 + \Delta_b)^{3/2} e^{-x (\Delta_a + \Delta_b)} \, , \\
g_{\rm eff} &\approx 2 \left[ 1 + (1 + \Delta_a)^{3/2} e^{-x \Delta_a} + (1 + \Delta_b)^{3/2} e^{-x \Delta_b} \right] \, ,
\end{align}
where $\Delta_a \equiv (D_{R_a} - D_X) / D_X$. 
We fixed our co-annihilation condition in the neutrino sector that we consider all the processes satisfying the mass difference be within 20\% from $D_X$: $\Delta_a \le 20\%$. 
When we use $x_f = 25$, the Boltzmann suppression factor is about $e^{- x_f \Delta} = e^{-25 \times 0.2} \approx 6.74 \times 10^{-3} \sim {\cal O} (0.01)$. 
Therefore, we expect that this setup would lead us to appropriate evaluation of relic density in our co-annihilation system. 
Furthermore, $da$ and $db$ are found by expanding form of the differential cross section $(d\sigma v_{\rm rel})_{ab} \approx da + db v^2_{\rm rel} + {\cal O} (v^4_{\rm rel})$ in terms of relative velocity $v_{\rm rel}$, and $(d\sigma v_{\rm rel})_{ab}$ in our model is estimated as
\begin{align}
(d\sigma v_{\rm rel})_{ab} &\approx \sin \theta \sum_{i, j = 1}^3 |\tilde{f}_{ia} \tilde{f}_{bj}^{\dag}|^2 \frac{|\overline{M_{ab}}|^2}{16 \pi s_{ab}} (1 + \Delta_a)^{3/2} (1 + \Delta_b)^{3/2} e^{-x (\Delta_a + \Delta_b)} \, , \\
|\overline {M_{ab}}|^2 &\approx \frac{1}{T^2_{ab}} (p_1 \cdot k_1)_{ab} (p_2 \cdot k_2)_{ab} + \frac{1}{U^2_{ab}} (p_2 \cdot k_1)_{ab} (p_2 \cdot k_1)_{ab} - \frac{D_{R_a} D_{R_b}}{T_{ab} U_{ab}} (k_1 \cdot k_2)_{ab} \, ,
\end{align}
where $s_{ab} \approx (D_{R_a} + D_{R_b})^2 + D_{R_a} D_{R_b} v_{\rm rel}^2$ is the Mandelstam variable expanded by $v_{\rm rel}$, $p_1, p_2$ are momenta for initial state particles $\psi_R$, and $k_1, k_2$ are those for final state particles $\nu_L$ which are supposed to be massless compared with $D_R$. 
$T_{ab} 
= D_{R_a}^2 - m_S^2 - 2 (p_1 \cdot k_1)_{ab}$ and
$U_{ab} 
= D_{R_a}^2 - m_S^2 - 2 (p_1 \cdot k_2)_{ab}$ are propagators for $t$-channel and $u$-channel, respectively. 
$v_{\rm rel}$ expanded form of scalar products for each combination of momenta are given in Appendix~\ref{app:DMkine}.

\section{Numerical results}
\label{sec:IV}

In this section, we perform the numerical analysis taking into account all the constraints discussed above. 
Our input parameters are randomly selected by the following regions:
\begin{align}
&10^2 \, {\rm GeV} \le M' \lesssim 10^4 \, {\rm GeV} \, , \quad \{M_{R_{\alpha}}, m_{\eta_R}, m_S\} = [1.2 \times M', 10^5 \, {\rm GeV}] \, , \\
&|\theta_{12, 23, 13}, \alpha_{21, 31}| = [0 - 2 \pi] \, , \quad |h| = [10^{-3} - \sqrt{4 \pi}] \, ,
\end{align}
where $\theta_{12, 23, 13}$ are mixing angles of ${\cal O}_f$ in $f$ of Eq.~\eqref{eq:fExp}, and the hierarchy of $M_{R_{\alpha}}$ is set to be $M_{R_1} \le M_{R_2} \le M_{R_3}$. 
$m_{\eta_I}$ is taken by the range of $[0.7 m_{\eta_R} - 1.3 m_{\eta_R}]$ so that we can simply avoid constraints of electroweak precision tests~\cite{Barbieri:2006dq}.

\subsubsection{\rm NH case}

\begin{figure}[tb]\begin{center}
\includegraphics[width=85mm]{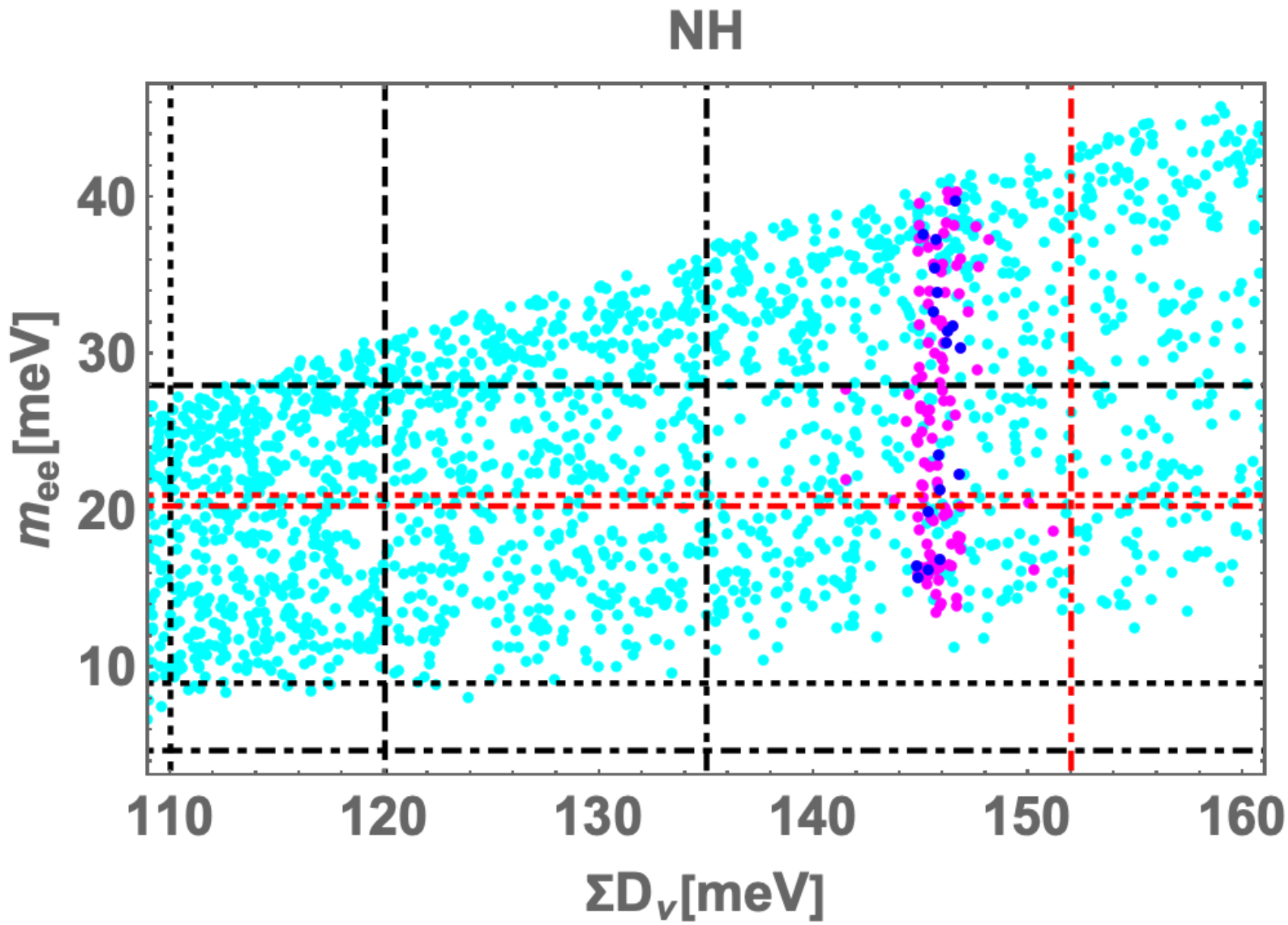}
\caption{Allowed region of $m_{ee}$ in terms of $\sum D_{\nu}$ in meV unit, for the case of NH. 
The cyan points are obtained by experimental result from Nufit6.0~\cite{Esteban:2024eli}. 
Satisfying relevant bounds for our model, magenta and blue points correspond to the DM relic density of $0.05 \le \Omega h^2 \le 0.20$ and $0.11 \le \Omega h^2 \le 0.13$, respectively. 
The lines represent experimental results for the neutrino sector, discussed in Sec.~\ref{sec:const-nu}. 
See the main text for more details. }
\label{fig:nh1_neut}
\end{center}\end{figure}
In Fig.~\ref{fig:nh1_neut}, we show the allowed region of $m_{ee}$ in terms of $\sum D_{\nu}$ in meV unit. 
The cyan points are purely obtained by experimental result from Nufit6.0~\cite{Esteban:2024eli}, with satisfying following requirements: safe for $\ell_i \to \ell_j \gamma$ processes, small contributions to $\Delta a_{\mu} \lesssim 10^{-9}$ (within current $2\sigma$ error in Eq.~\eqref{eq:g-2ell}), $\sum D_{\nu} < 200 \, {\rm meV}$, $\sigma v_{\rm rel} \geq 10^9 \, {\rm GeV}^{-2}$ and perturbatively safe value for $f_{ia} \leq 4 \pi$ as well as consistency with the neutrino oscillation data. 
The magenta points satisfy the relic density of DM of $0.05 \le \Omega h^2 \le 0.20$, while the blue ones $0.11 \le \Omega h^2 \le 0.13$. 
The vertical dashed black line represents upper bound by the minimal standard cosmological model with CMB data, $\sum D_{\nu} \leq 120 \, {\rm meV}$ in Eq.~\eqref{eq:sum1}. 
We also show upper bounds from extended cosmological models as different vertical lines, dotted black one for $\sum D_{\nu} \leq 110 \, {\rm meV}$ (lowest bound in Eq.~\eqref{eq:sum2}) and dot-dashed black (red) one for $\sum D_{\nu} \leq 135 \, (152) \, {\rm meV}$ (in Eq.~\eqref{eq:sum3_nh}). 
The current lowest upper bound of the neutrinoless double beta decay which is $m_{ee} < 28 \, {\rm meV}$ in Eq.~\eqref{eq:mee_current} is shown by the horizontal dashed black line. 
The horizontal dotted and dot-dashed black (red) lines correspond to the future lowest (highest) upper bounds of LEGEND-1000, $m_{ee} < 9 \, (21) \, {\rm meV}$, and nEXO experiments, $m_{ee} < 4.7 \, (20.3) \, {\rm meV}$, respectively, which are shown in Eq.~\eqref{eq:mee_future}. 

The result suggests that our allowed region is localized at nearby $141 \, {\rm meV}$ - $151 \, {\rm meV}$ for $\sum D_{\nu}$ and $13.5 \, {\rm meV}$ - $40 \, {\rm meV}$ for $m_{ee}$. 
Mostly half of allowed region satisfies the lowest upper bound of $m_{ee}$ in Eq.~\eqref{eq:mee_current}, but all of them are excluded by upper bound of $\sum D_{\nu}$ from the minimal standard cosmological model with CMB data in Eq.~\eqref{eq:sum1}. 
However, they are still allowed by highest upper bounds from extended cosmological models in Eqs.~\eqref{eq:sum2} and \eqref{eq:sum3_nh}. 
It is emphasized that future experiments for $m_{ee}$, which are for example LEGEND-1000 and nEXO, have a potential to test our allowed region, and we expect that there will be some signature of our model at these experiments. 
Below we list predicted ranges of the specific parameters:
\begin{enumerate} 
\item the lightest neutrino mass is $38.5 \, {\rm meV} \le D_{\nu_1} \le 41 \, {\rm meV}$,

\item the model independent observable in Eq.~\eqref{eq:mnue-modelindep} is $39.5 \, {\rm meV} \le m_{\nu_e} \le 42.1 \, {\rm meV}$,

\item two Majorana phases $\alpha_{21, 31}$ run whole the regions.
\end{enumerate}

\begin{figure}[tb]\begin{center}
\includegraphics[width=78mm]{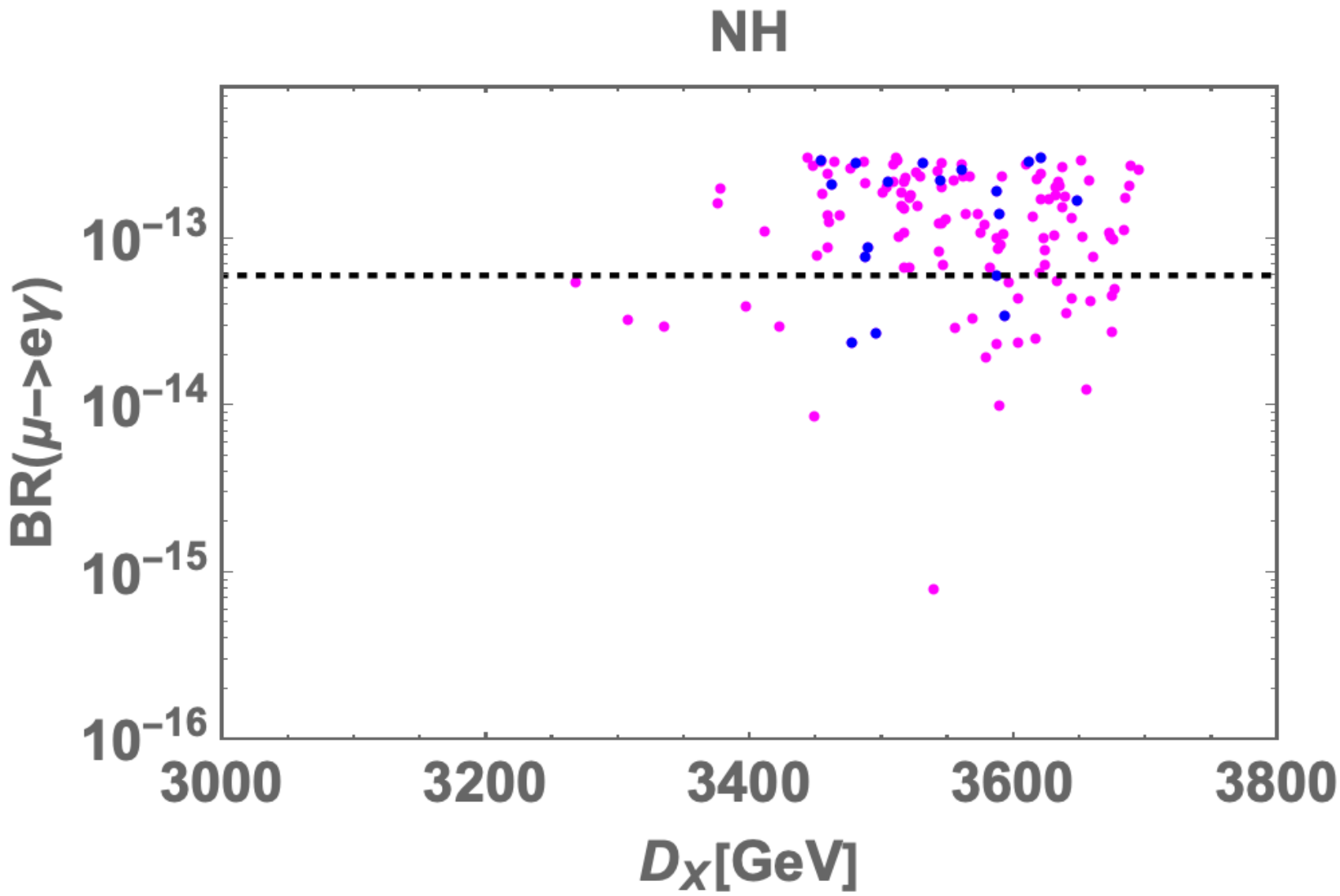}
\includegraphics[width=84mm]{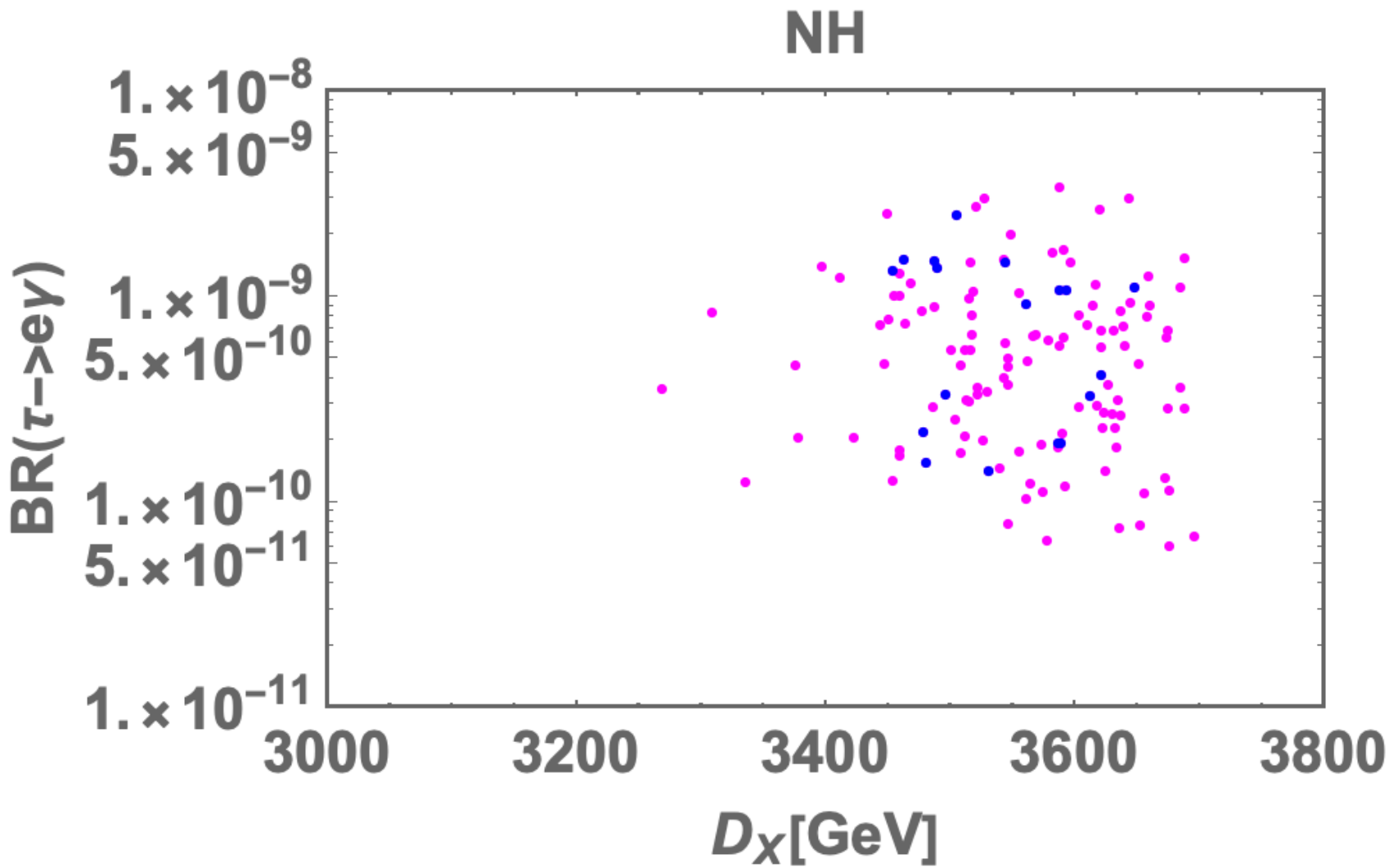} \\[1.2ex]
\includegraphics[width=78mm]{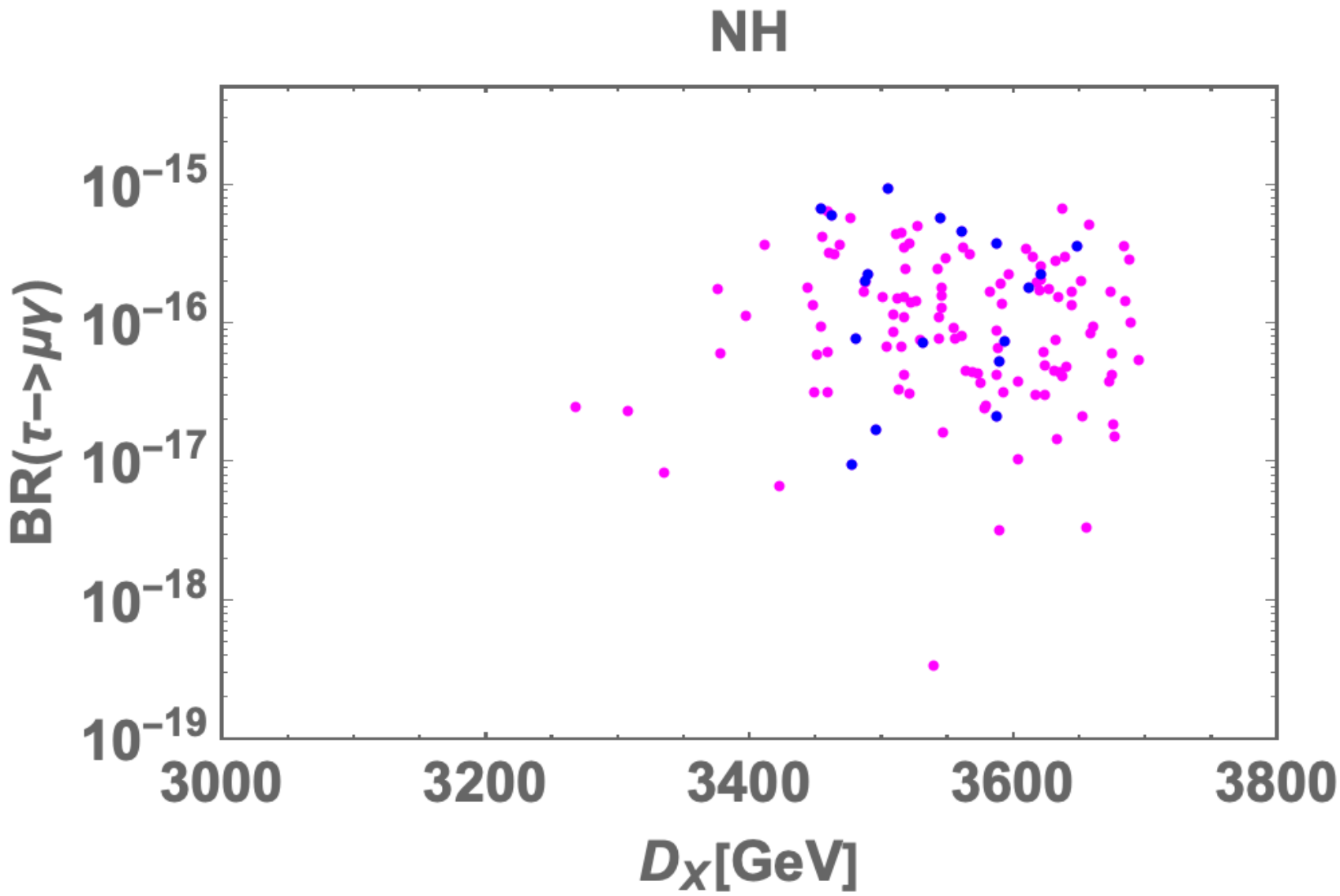} ~~~~
\includegraphics[width=78mm]{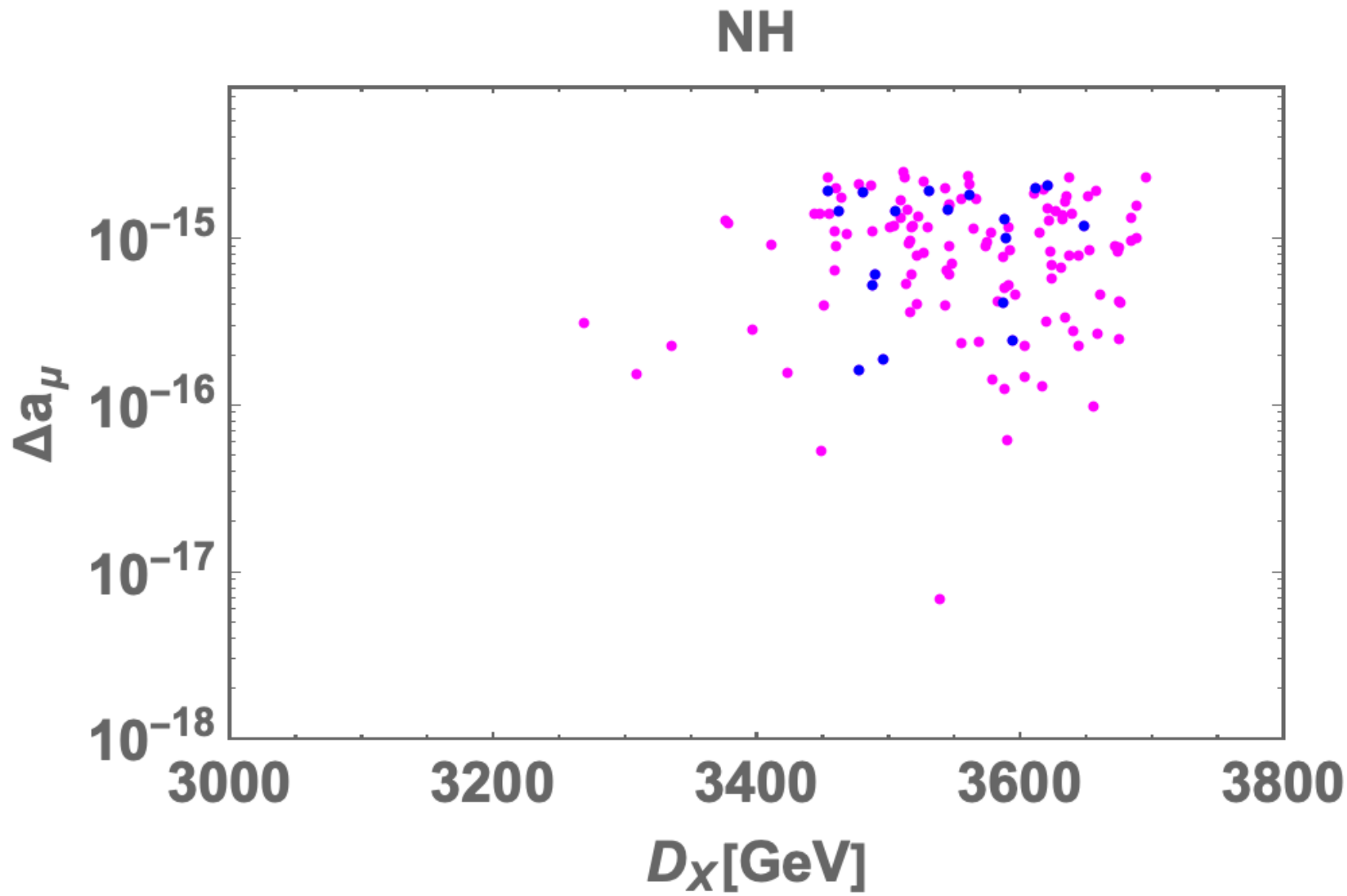}
\caption{Our predictions for LFVs and muon $g-2$ in terms of DM mass $D_X$, for the case of NH. 
We only show the points of $0.05 \leq \Omega h^2 \leq 0.2$, with same color legends as in Fig.~\ref{fig:nh1_neut}. 
For ${\rm BR} (\mu \to e \gamma)$ plot, we also show the future sensitivity as dotted black line. }
\label{fig:nh1_s}
\end{center}\end{figure}
In Fig.~\ref{fig:nh1_s}, we show LFVs and muon $g-2$ predictions in terms of DM mass $D_X$, for the case of NH. 
Here, we only show the magenta and blue points in Fig.~\ref{fig:nh1_neut} with same color legends. 
These figures tell us that the DM mass range is 3250 GeV - 3700 GeV, and there is no correlation between the relic density and predictions of LFVs and $\Delta a_{\mu}$. 
For $\mu \to e \gamma$ process, the allowed points are located at nearby the experimental upper limits, and therefore, our allowed region is also tested by the future experiment, e.g., MEG II~\cite{MEGII:2023ltw} whose goal sensitivity is $6 \times 10^{-14}$, as shown by the dotted black line in the plot. 
In case of $\tau$ LFV decays, $\tau \to e \gamma$ is below the experimental upper bound by one order of magnitude, while $\tau \to \mu \gamma$ is far below the current experimental bound. 
Therefore, we also have a possibility to test our model by future $\tau \to e \gamma$ measurement, e.g., at the Belle II~\cite{Bodrov:2024wrw}. 
$\Delta a_{\mu}$ is, on the other hand, much smaller than the current deviation and totally within the range of the experimental region.

\subsubsection{\rm IH case}

\begin{figure}[tb]\begin{center}
\includegraphics[width=78mm]{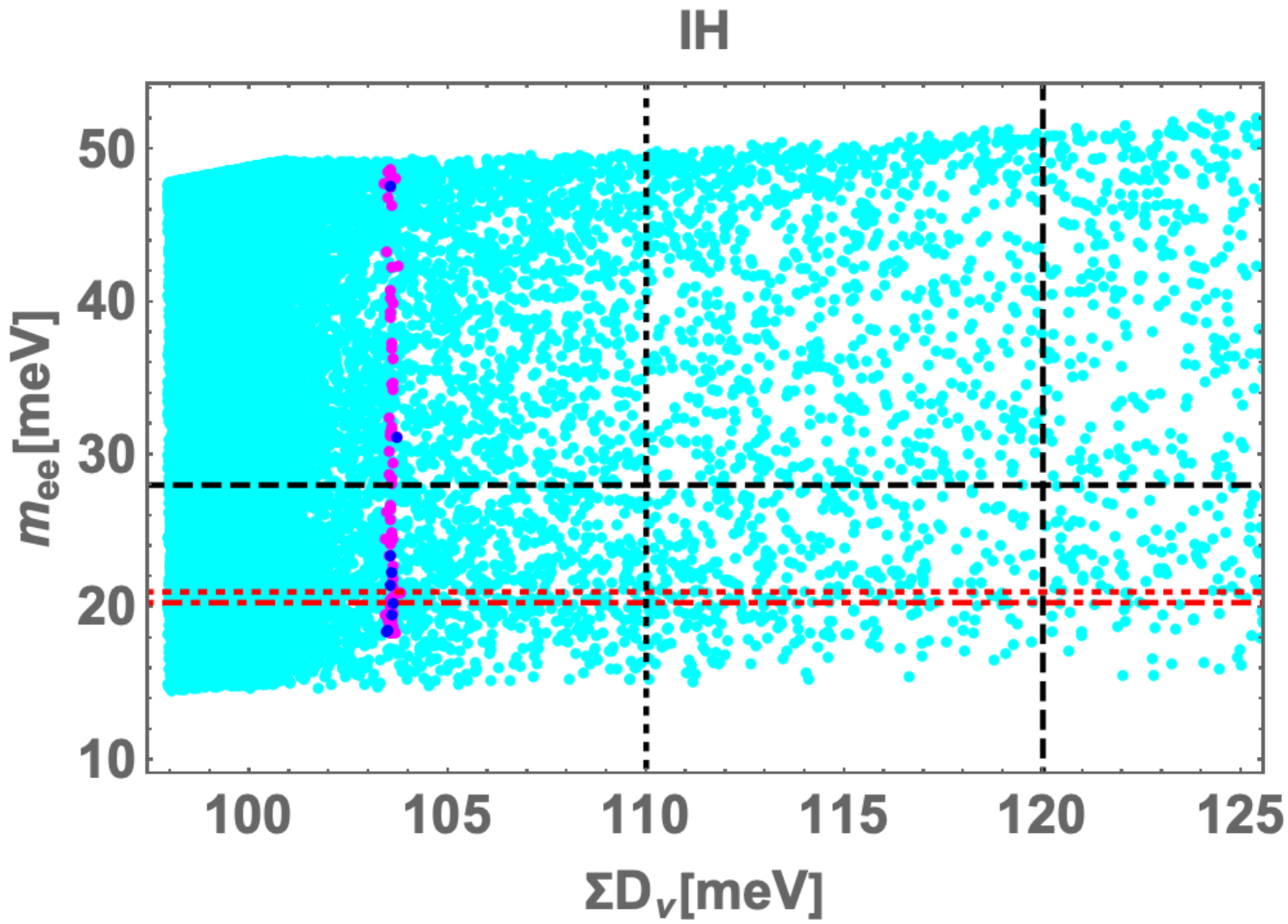}
\caption{Allowed region of $m_{ee}$ in terms of $\sum D_{\nu}$ in meV unit, for the case of IH. 
The color legends are same as in Fig.~\ref{fig:nh1_neut}. }
\label{fig:ih1_neut}
\end{center}\end{figure}
In Fig.~\ref{fig:ih1_neut}, we show the allowed region of $m_{ee}$ in terms of $\sum D_{\nu}$ in meV unit. 
The color legends as well as meaning of all lines are the same as those in Fig.~\ref{fig:nh1_neut}. 
The result for $\sum D_{\nu}$ has different feature and tendency from the case of NH, and we found that the allowed region is localized at nearby $103 \, {\rm meV}$ - $104 \, {\rm meV}$ for $\sum D_{\nu}$. 
This is allowed and favored by upper bound from the minimal standard cosmological model with CMB data in Eq.~\eqref{eq:sum1}. 
In contrast, $m_{ee}$ has similar (but a bit larger) range to that in the NH case, and we have $18 \, {\rm meV}$ - $49 \, {\rm meV}$. 
These values will be able to be tested by future experiments, LEGEND-1000 and nEXO. 
Therefore, if there is no signal for $\sum D_{\nu} \geq 110 \, {\rm meV}$ in some experiments, the IH for the neutrino mass hierarchy is strongly favored in our model. 
Below, we list predicted ranges and favored values for the specific parameters:
\begin{enumerate} 
\item the lightest neutrino mass is $3.66 \, {\rm meV} \le D_{\nu_3} \le 4.4 \, {\rm meV}$,

\item the model independent observable in Eq.~\eqref{eq:mnue-modelindep} is $m_{\nu_e} \approx 49.2 \, {\rm meV}$,

\item two Majorana phases $\alpha_{21, 31}$ run whole the regions.
\end{enumerate}

\begin{figure}[tb]\begin{center}
\includegraphics[width=84mm]{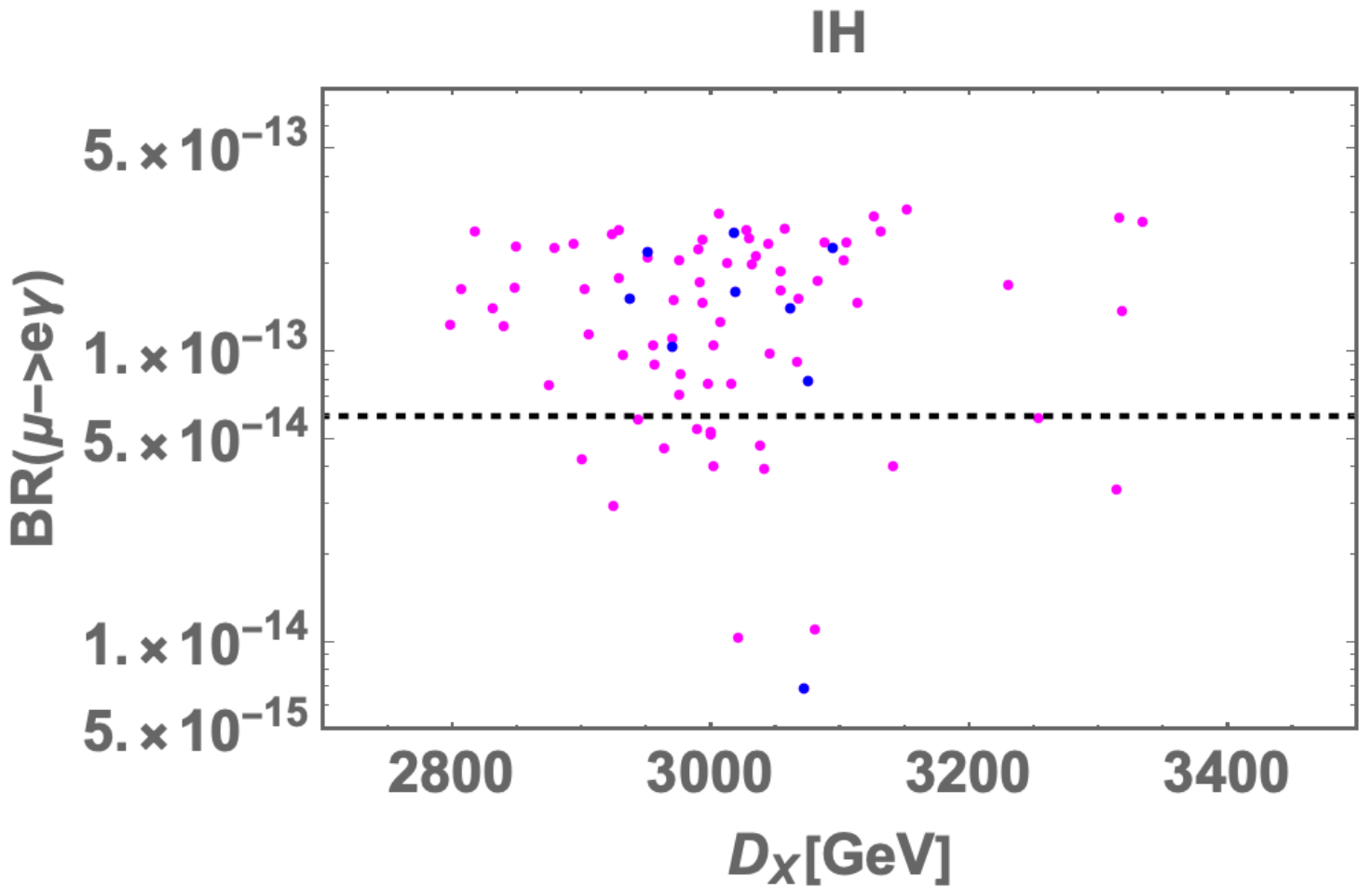}
\includegraphics[width=78mm]{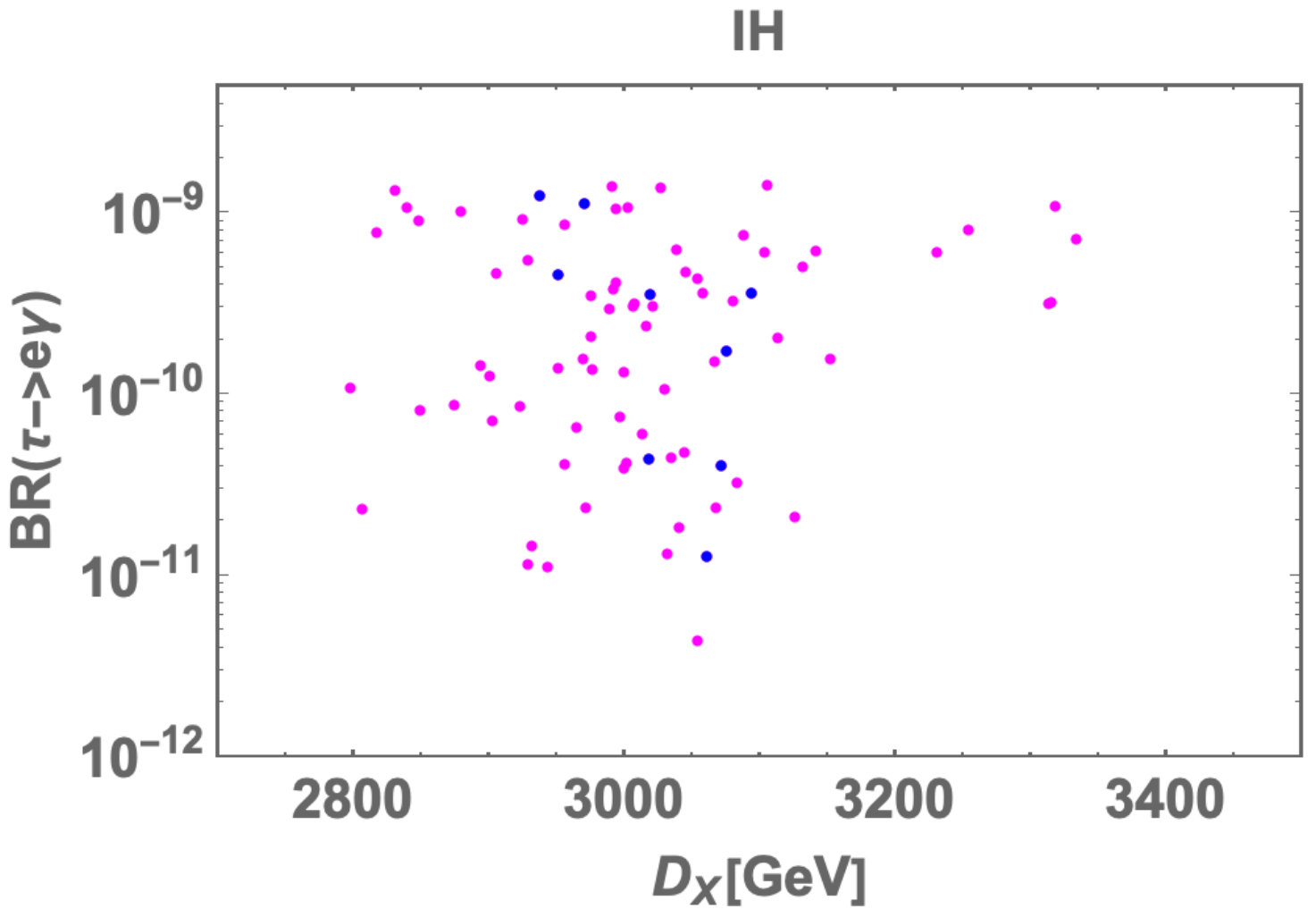}
\includegraphics[width=78mm]{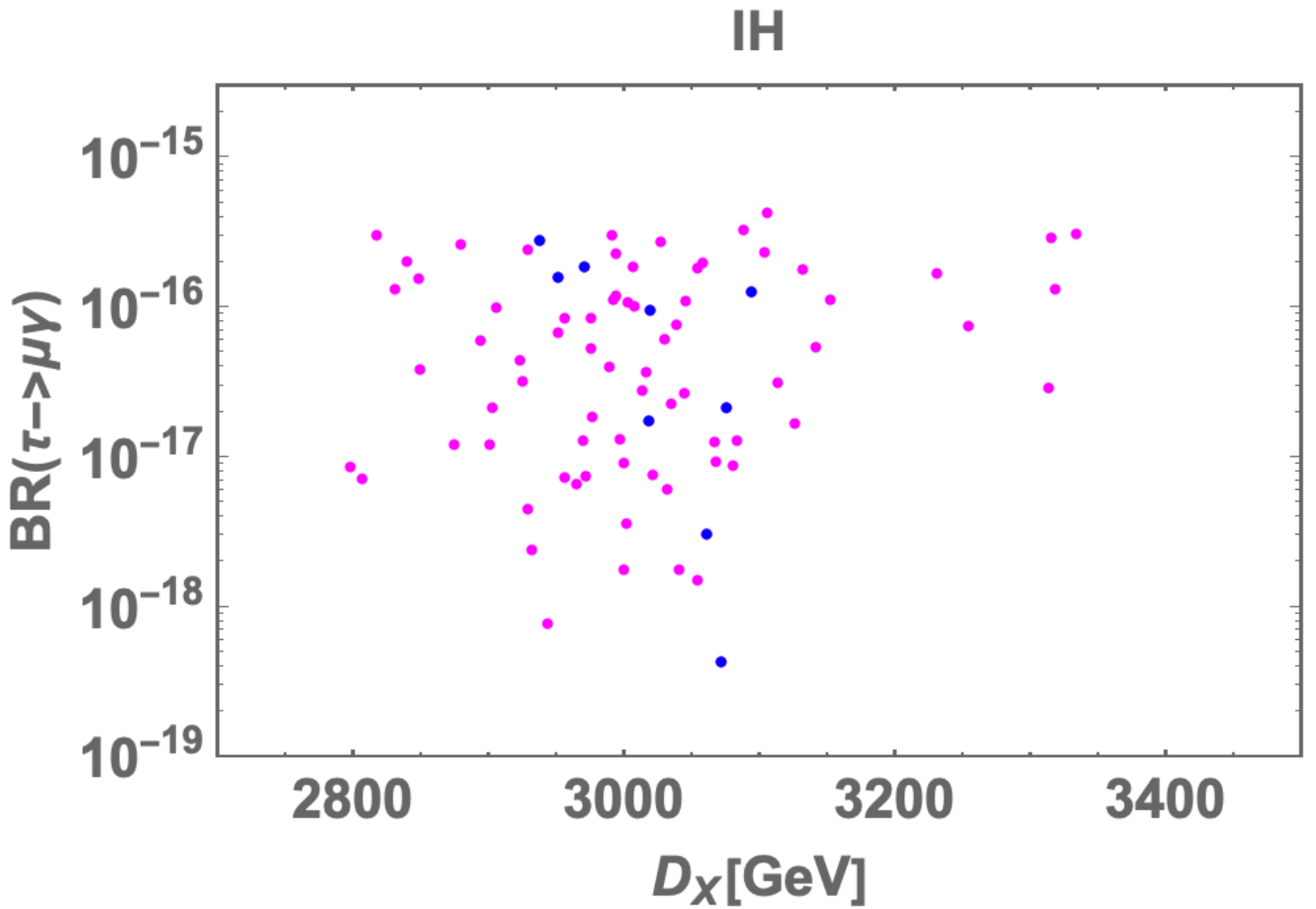}
\includegraphics[width=84mm]{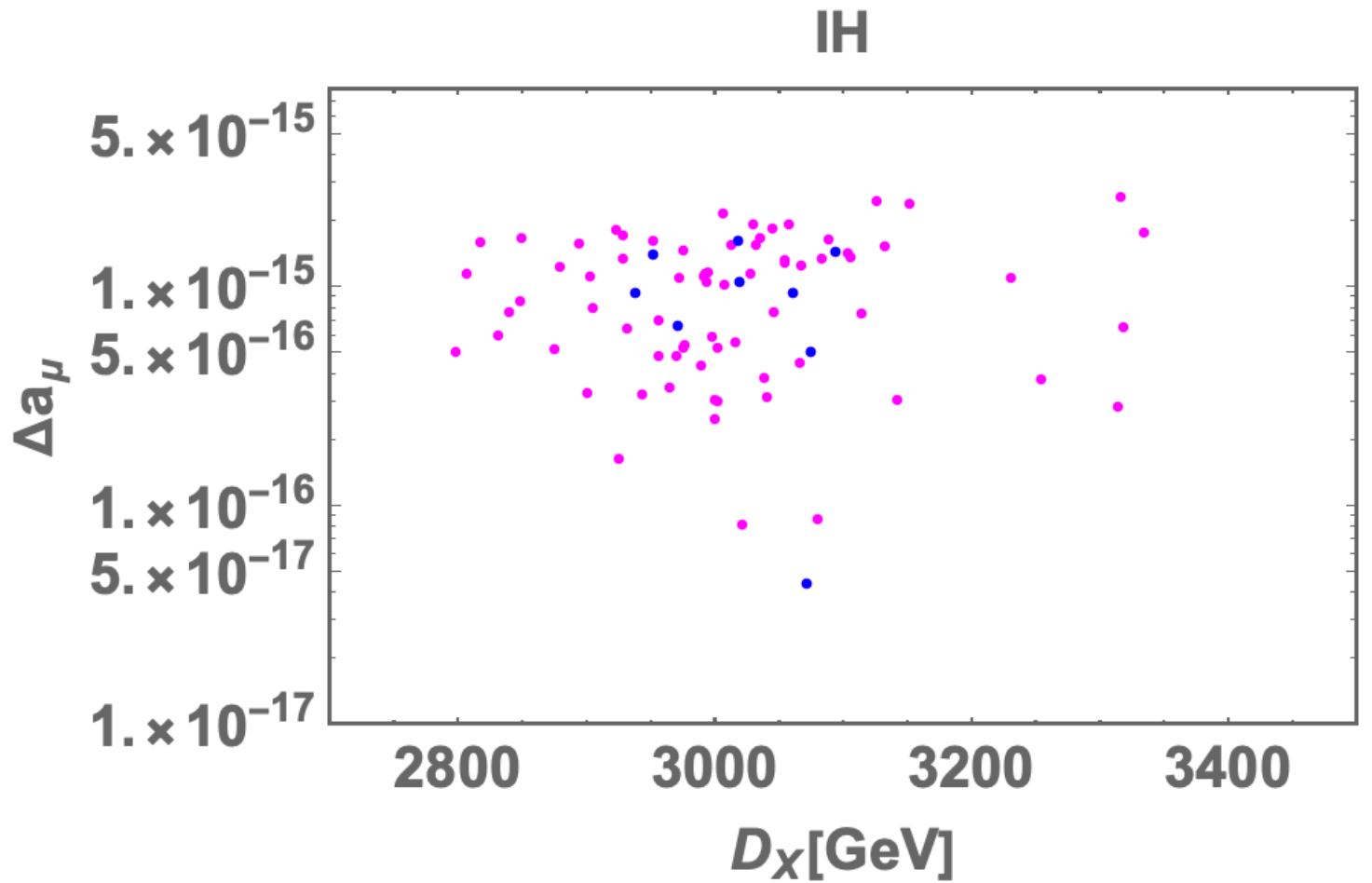}
\caption{Our predictions for LFVs and muon $g-2$ in terms of DM mass $D_X$, for the case of IH, with same color legends as in Figs.~\ref{fig:nh1_neut} and \ref{fig:nh1_s}. }
\label{fig:ih1_s}
\end{center}\end{figure}
In Fig.~\ref{fig:ih1_s}, we show our predictions of LFVs and muon $g-2$ in terms of DM mass $D_X$, for the case of IH. 
Same as in Fig.~\ref{fig:nh1_s}, we only show the magenta and blue points in Fig.~\ref{fig:ih1_neut} with same color legends. 
These figures indicate that each prediction has same tendency as those in the case of NH, for the DM mass range of $2800 \, {\rm GeV}$ - $3350 \, {\rm GeV}$, which results in slightly smaller mass for our DM. 
Then, we can conclude that the IH case can also be tested by the future measurement of $\mu \to e \gamma$ experiments.

\section{Summary and discussion}
\label{sec:summary}

We have proposed a two-loop neutrino mass model via $\mathbbm{Z}_2$ gauging $\mathbbm{Z}_6$ non-invertible symmetry in which we have introduced three families of isospin doublet vector fermions, heavy right-handed neutrinos, isospin doublet and singlet bosons. 
All these new fields have nonzero charges under our non-invertible symmetry, and its neutral components can be DM candidates, because of the remnant $\mathbbm{Z}_2$ symmetry in the non-invertible symmetry, which plays a role to stabilize our DM candidates. 
The non-invertible symmetry in our model is dynamically broken at one-loop level, but its violation does not affect our scenario. 
We have considered the lightest mode of the neutral components in the doublet vector-like fermions as our main DM candidate. 
Even with degenerated bare masses for them, one-loop corrections to their masses give tiny hierarchy, and then, it is natural to discuss the DM relic density via co-annihilation processes among them. 
Although the other new particles can have similar masses to the DM, its co-annihilation effects are suppressed by enough mass difference between the DM and their masses. 
As a result, we can discuss the relic density in our setup with analytical estimation. 

Different from our previous work~\cite{Okada:2025kfm}, the neutrino mass matrix in this model is dominated by two-loop rainbow type diagrams, and therefore, distinct features for relevant observable will be expected. 
Considering all the constraints of neutrino oscillation data, LFVs, muon $g-2$ and the relic density of DM candidate, we have performed the numerical analysis and found some allowed regions, both for NH and IH cases. 
Remarkably, through our numerical analysis, the sum of neutrino masses in the case of NH is larger than that in the case of IH, which is opposite situation compared to typical active neutrino models as well as our previous work in ref.~\cite{Okada:2025kfm}, due to our DM phenomenology. 
We have also found that the future measurement of $\mu \to e \gamma$ has a potential to test our model with both NH and IH cases, because our prediction for its branching ratio is just below the current upper bound. 
On the other hand, our predictions of the tau LFV processes $\tau \to e/\mu + \gamma$ and the muon $g-2$ are smaller than the current experimental bound and measurement, although there will be a chance to have some signature in $\tau \to e \gamma$ process at the future experiments.

\section*{Acknowledgments}
HO is supported by Zhongyuan Talent (Talent Recruitment Series) Foreign Experts Project. 
YS is supported by Natural Science Foundation of China under grant No. W2433006. 

\appendix

\section{Fusion rule of ${\mathbb Z}_2$ gauging of ${\mathbb Z}_6$}
\label{app:FusionRule}

Here, we write down multiplication rules of ${\mathbb Z}_2$ gauging of ${\mathbb Z}_6$ symmetry:\footnote{For studies of ${\mathbb Z}_2$ gauging of ${\mathbb Z}_N$ symmetries, see, e.g., refs.~\cite{Kobayashi:2024yqq,Kobayashi:2024cvp,Funakoshi:2024uvy}.}
\begin{align}
\begin{array}{lll}
\epsilon \otimes \epsilon = \mathbbm{1} \oplus \sigma \, , ~~~~~ & \sigma \otimes \sigma = \mathbbm{1} \oplus \sigma \, , ~~~~~ & \rho \otimes \rho = \mathbbm{1} \, , \\[0.3ex]
\epsilon \otimes \sigma = \epsilon \oplus \rho \, , ~~~~~ & \epsilon \otimes \rho = \sigma \, , ~~~~~ & \sigma \otimes \rho = \epsilon \, .
\end{array}
\label{eq:fusionrule}
\end{align}
Note here that all elements are commutable each other. 
With these rules, we can investigate all invariant combinations. 
We have
\begin{align}
\epsilon \otimes \epsilon \otimes \sigma \, , \quad \sigma \otimes \sigma \otimes \sigma \, , \quad \epsilon \otimes \sigma \otimes \rho \, ,
\label{eq:singletin3F}
\end{align}
for three fields and
\begin{align}
\begin{array}{llll}
\epsilon \otimes \epsilon \otimes \epsilon \otimes \epsilon \, , ~~~~ & \epsilon \otimes \epsilon \otimes \epsilon \otimes \rho \, , ~~~~ & \epsilon \otimes \epsilon \otimes \sigma \otimes \sigma \, , ~~~~ & \epsilon \otimes \epsilon \otimes \rho \otimes \rho \, , \\[0.3ex]
\epsilon \otimes \sigma \otimes \sigma \otimes \rho \, , ~~~~ & \sigma \otimes \sigma \otimes \sigma \otimes \sigma \, , ~~~~ & \sigma \otimes \sigma \otimes \rho \otimes \rho \, , ~~~~ & \rho \otimes \rho \otimes \rho \otimes \rho \, ,
\end{array}
\label{eq:singletin4F}
\end{align}
for four fields, which are invariant under the symmetry. 
The former is applied to the Yukawa and scalar trilinear couplings, and the latter to the scalar quartic couplings.

\section{Kinematics for DM computation}
\label{app:DMkine}

Below, we display formulas of scalar products in terms of $v_{\rm rel}$ expansion: 
\begin{align}
&(k_1 \cdot k_2)_{ab} = \frac{s_{ab}}{2} \, , \\
&(p_1 \cdot k_1)_{ab} = \frac{m_a (m_a + m_b)}{2} - \frac{m_a m_b \cos \theta}{2} v_{\rm rel} + \frac{m_a m_b}{4} v_{\rm rel}^2 \, , \\
&(p_1 \cdot k_2)_{ab} = \frac{m_a (m_a + m_b)}{2} + \frac{m_a m_b \cos \theta}{2} v_{\rm rel} + \frac{m_a m_b} {4} v_{\rm rel}^2 \, , \\
&(p_2 \cdot k_1)_{ab} = \frac{m_b (m_a + m_b)}{2} + \frac{m_a m_b \cos \theta}{2} v_{\rm rel} + \frac{m_a m_b}{4} v_{\rm rel}^2 \, , \\
&(p_2 \cdot k_2)_{ab} = \frac{m_b (m_a + m_b)}{2} - \frac{m_a m_b \cos \theta}{2} v_{\rm rel} + \frac{m_a m_b}{4} v_{\rm rel}^2 \, ,
\end{align}
where masses in the final state are supposed to be zero due to neutrinos in our annihilation processes, which is justified by the fact that $m_{a, b}$ correspond to $D_{R_{a, b}} (\gg m_{\nu})$ in our analysis.

\bibliography{ctma4}

\begin{thebibliography}{204}
\expandafter\ifx\csname natexlab\endcsname\relax\def\natexlab#1{#1}\fi
\expandafter\ifx\csname bibnamefont\endcsname\relax
  \def\bibnamefont#1{#1}\fi
\expandafter\ifx\csname bibfnamefont\endcsname\relax
  \def\bibfnamefont#1{#1}\fi
\expandafter\ifx\csname citenamefont\endcsname\relax
  \def\citenamefont#1{#1}\fi
\expandafter\ifx\csname url\endcsname\relax
  \def\url#1{\texttt{#1}}\fi
\expandafter\ifx\csname urlprefix\endcsname\relax\def\urlprefix{URL }\fi
\providecommand{\bibinfo}[2]{#2}
\providecommand{\eprint}[2][]{\url{#2}}

\bibitem[{\citenamefont{Ma}(2006)}]{Ma:2006km}
\bibinfo{author}{\bibfnamefont{E.}~\bibnamefont{Ma}}, \bibinfo{journal}{Phys.
  Rev. D} \textbf{\bibinfo{volume}{73}}, \bibinfo{pages}{077301}
  (\bibinfo{year}{2006}), \eprint{hep-ph/0601225}.

\bibitem[{\citenamefont{Zee}(1980)}]{Zee:1980ai}
\bibinfo{author}{\bibfnamefont{A.}~\bibnamefont{Zee}}, \bibinfo{journal}{Phys.
  Lett. B} \textbf{\bibinfo{volume}{93}}, \bibinfo{pages}{389}
  (\bibinfo{year}{1980}), \bibinfo{note}{[Erratum: Phys.Lett.B 95, 461
  (1980)]}.

\bibitem[{\citenamefont{Zee}(1986)}]{Zee:1985id}
\bibinfo{author}{\bibfnamefont{A.}~\bibnamefont{Zee}}, \bibinfo{journal}{Nucl.
  Phys. B} \textbf{\bibinfo{volume}{264}}, \bibinfo{pages}{99}
  (\bibinfo{year}{1986}).

\bibitem[{\citenamefont{Babu}(1988)}]{Babu:1988ki}
\bibinfo{author}{\bibfnamefont{K.~S.} \bibnamefont{Babu}},
  \bibinfo{journal}{Phys. Lett. B} \textbf{\bibinfo{volume}{203}},
  \bibinfo{pages}{132} (\bibinfo{year}{1988}).

\bibitem[{\citenamefont{Kajiyama
  et~al.}(2013{\natexlab{a}})\citenamefont{Kajiyama, Okada, and
  Toma}}]{Kajiyama:2013rla}
\bibinfo{author}{\bibfnamefont{Y.}~\bibnamefont{Kajiyama}},
  \bibinfo{author}{\bibfnamefont{H.}~\bibnamefont{Okada}}, \bibnamefont{and}
  \bibinfo{author}{\bibfnamefont{T.}~\bibnamefont{Toma}},
  \bibinfo{journal}{Phys. Rev. D} \textbf{\bibinfo{volume}{88}},
  \bibinfo{pages}{015029} (\bibinfo{year}{2013}{\natexlab{a}}),
  \eprint{1303.7356}.

\bibitem[{\citenamefont{Krauss et~al.}(2003)\citenamefont{Krauss, Nasri, and
  Trodden}}]{Krauss:2002px}
\bibinfo{author}{\bibfnamefont{L.~M.} \bibnamefont{Krauss}},
  \bibinfo{author}{\bibfnamefont{S.}~\bibnamefont{Nasri}}, \bibnamefont{and}
  \bibinfo{author}{\bibfnamefont{M.}~\bibnamefont{Trodden}},
  \bibinfo{journal}{Phys. Rev. D} \textbf{\bibinfo{volume}{67}},
  \bibinfo{pages}{085002} (\bibinfo{year}{2003}), \eprint{hep-ph/0210389}.

\bibitem[{\citenamefont{Aoki et~al.}(2009)\citenamefont{Aoki, Kanemura, and
  Seto}}]{Aoki:2008av}
\bibinfo{author}{\bibfnamefont{M.}~\bibnamefont{Aoki}},
  \bibinfo{author}{\bibfnamefont{S.}~\bibnamefont{Kanemura}}, \bibnamefont{and}
  \bibinfo{author}{\bibfnamefont{O.}~\bibnamefont{Seto}},
  \bibinfo{journal}{Phys. Rev. Lett.} \textbf{\bibinfo{volume}{102}},
  \bibinfo{pages}{051805} (\bibinfo{year}{2009}), \eprint{0807.0361}.

\bibitem[{\citenamefont{Gustafsson et~al.}(2013)\citenamefont{Gustafsson, No,
  and Rivera}}]{Gustafsson:2012vj}
\bibinfo{author}{\bibfnamefont{M.}~\bibnamefont{Gustafsson}},
  \bibinfo{author}{\bibfnamefont{J.~M.} \bibnamefont{No}}, \bibnamefont{and}
  \bibinfo{author}{\bibfnamefont{M.~A.} \bibnamefont{Rivera}},
  \bibinfo{journal}{Phys. Rev. Lett.} \textbf{\bibinfo{volume}{110}},
  \bibinfo{pages}{211802} (\bibinfo{year}{2013}), \bibinfo{note}{[Erratum:
  Phys.Rev.Lett. 112, 259902 (2014)]}, \eprint{1212.4806}.

\bibitem[{\citenamefont{Nishiwaki et~al.}(2015)\citenamefont{Nishiwaki, Okada,
  and Orikasa}}]{Nishiwaki:2015iqa}
\bibinfo{author}{\bibfnamefont{K.}~\bibnamefont{Nishiwaki}},
  \bibinfo{author}{\bibfnamefont{H.}~\bibnamefont{Okada}}, \bibnamefont{and}
  \bibinfo{author}{\bibfnamefont{Y.}~\bibnamefont{Orikasa}},
  \bibinfo{journal}{Phys. Rev. D} \textbf{\bibinfo{volume}{92}},
  \bibinfo{pages}{093013} (\bibinfo{year}{2015}), \eprint{1507.02412}.

\bibitem[{\citenamefont{Bertone et~al.}(2005)\citenamefont{Bertone, Hooper, and
  Silk}}]{Bertone:2004pz}
\bibinfo{author}{\bibfnamefont{G.}~\bibnamefont{Bertone}},
  \bibinfo{author}{\bibfnamefont{D.}~\bibnamefont{Hooper}}, \bibnamefont{and}
  \bibinfo{author}{\bibfnamefont{J.}~\bibnamefont{Silk}},
  \bibinfo{journal}{Phys. Rept.} \textbf{\bibinfo{volume}{405}},
  \bibinfo{pages}{279} (\bibinfo{year}{2005}), \eprint{hep-ph/0404175}.

\bibitem[{\citenamefont{Cirelli et~al.}(2024)\citenamefont{Cirelli, Strumia,
  and Zupan}}]{Cirelli:2024ssz}
\bibinfo{author}{\bibfnamefont{M.}~\bibnamefont{Cirelli}},
  \bibinfo{author}{\bibfnamefont{A.}~\bibnamefont{Strumia}}, \bibnamefont{and}
  \bibinfo{author}{\bibfnamefont{J.}~\bibnamefont{Zupan}}
  (\bibinfo{year}{2024}), \eprint{2406.01705}.

\bibitem[{\citenamefont{Lopez~Honorez et~al.}(2007)\citenamefont{Lopez~Honorez,
  Nezri, Oliver, and Tytgat}}]{LopezHonorez:2006gr}
\bibinfo{author}{\bibfnamefont{L.}~\bibnamefont{Lopez~Honorez}},
  \bibinfo{author}{\bibfnamefont{E.}~\bibnamefont{Nezri}},
  \bibinfo{author}{\bibfnamefont{J.~F.} \bibnamefont{Oliver}},
  \bibnamefont{and} \bibinfo{author}{\bibfnamefont{M.~H.~G.}
  \bibnamefont{Tytgat}}, \bibinfo{journal}{JCAP} \textbf{\bibinfo{volume}{02}},
  \bibinfo{pages}{028} (\bibinfo{year}{2007}), \eprint{hep-ph/0612275}.

\bibitem[{\citenamefont{Pierce and Thaler}(2007)}]{Pierce:2007ut}
\bibinfo{author}{\bibfnamefont{A.}~\bibnamefont{Pierce}} \bibnamefont{and}
  \bibinfo{author}{\bibfnamefont{J.}~\bibnamefont{Thaler}},
  \bibinfo{journal}{JHEP} \textbf{\bibinfo{volume}{08}}, \bibinfo{pages}{026}
  (\bibinfo{year}{2007}), \eprint{hep-ph/0703056}.

\bibitem[{\citenamefont{Chen et~al.}(2009)\citenamefont{Chen, Cline, and
  Frey}}]{Chen:2009ab}
\bibinfo{author}{\bibfnamefont{F.}~\bibnamefont{Chen}},
  \bibinfo{author}{\bibfnamefont{J.~M.} \bibnamefont{Cline}}, \bibnamefont{and}
  \bibinfo{author}{\bibfnamefont{A.~R.} \bibnamefont{Frey}},
  \bibinfo{journal}{Phys. Rev. D} \textbf{\bibinfo{volume}{80}},
  \bibinfo{pages}{083516} (\bibinfo{year}{2009}), \eprint{0907.4746}.

\bibitem[{\citenamefont{Okada and Seto}(2010)}]{Okada:2010wd}
\bibinfo{author}{\bibfnamefont{N.}~\bibnamefont{Okada}} \bibnamefont{and}
  \bibinfo{author}{\bibfnamefont{O.}~\bibnamefont{Seto}},
  \bibinfo{journal}{Phys. Rev. D} \textbf{\bibinfo{volume}{82}},
  \bibinfo{pages}{023507} (\bibinfo{year}{2010}), \eprint{1002.2525}.

\bibitem[{\citenamefont{Lopez~Honorez and Yaguna}(2010)}]{LopezHonorez:2010eeh}
\bibinfo{author}{\bibfnamefont{L.}~\bibnamefont{Lopez~Honorez}}
  \bibnamefont{and} \bibinfo{author}{\bibfnamefont{C.~E.}
  \bibnamefont{Yaguna}}, \bibinfo{journal}{JHEP} \textbf{\bibinfo{volume}{09}},
  \bibinfo{pages}{046} (\bibinfo{year}{2010}), \eprint{1003.3125}.

\bibitem[{\citenamefont{Eby and Frampton}(2012)}]{Eby:2011qa}
\bibinfo{author}{\bibfnamefont{D.~A.} \bibnamefont{Eby}} \bibnamefont{and}
  \bibinfo{author}{\bibfnamefont{P.~H.} \bibnamefont{Frampton}},
  \bibinfo{journal}{Phys. Lett. B} \textbf{\bibinfo{volume}{713}},
  \bibinfo{pages}{249} (\bibinfo{year}{2012}), \eprint{1111.4938}.

\bibitem[{\citenamefont{Chao}(2015)}]{Chao:2012sz}
\bibinfo{author}{\bibfnamefont{W.}~\bibnamefont{Chao}}, \bibinfo{journal}{Int.
  J. Mod. Phys. A} \textbf{\bibinfo{volume}{30}}, \bibinfo{pages}{1550007}
  (\bibinfo{year}{2015}), \eprint{1202.6394}.

\bibitem[{\citenamefont{Farzan and Akbarieh}(2012)}]{Farzan:2012hh}
\bibinfo{author}{\bibfnamefont{Y.}~\bibnamefont{Farzan}} \bibnamefont{and}
  \bibinfo{author}{\bibfnamefont{A.~R.} \bibnamefont{Akbarieh}},
  \bibinfo{journal}{JCAP} \textbf{\bibinfo{volume}{10}}, \bibinfo{pages}{026}
  (\bibinfo{year}{2012}), \eprint{1207.4272}.

\bibitem[{\citenamefont{Kashiwase and Suematsu}(2013)}]{Kashiwase:2013uy}
\bibinfo{author}{\bibfnamefont{S.}~\bibnamefont{Kashiwase}} \bibnamefont{and}
  \bibinfo{author}{\bibfnamefont{D.}~\bibnamefont{Suematsu}},
  \bibinfo{journal}{Eur. Phys. J. C} \textbf{\bibinfo{volume}{73}},
  \bibinfo{pages}{2484} (\bibinfo{year}{2013}), \eprint{1301.2087}.

\bibitem[{\citenamefont{Biswas et~al.}(2013)\citenamefont{Biswas, Majumdar,
  Sil, and Bhattacharjee}}]{Biswas:2013nn}
\bibinfo{author}{\bibfnamefont{A.}~\bibnamefont{Biswas}},
  \bibinfo{author}{\bibfnamefont{D.}~\bibnamefont{Majumdar}},
  \bibinfo{author}{\bibfnamefont{A.}~\bibnamefont{Sil}}, \bibnamefont{and}
  \bibinfo{author}{\bibfnamefont{P.}~\bibnamefont{Bhattacharjee}},
  \bibinfo{journal}{JCAP} \textbf{\bibinfo{volume}{12}}, \bibinfo{pages}{049}
  (\bibinfo{year}{2013}), \eprint{1301.3668}.

\bibitem[{\citenamefont{Aoki et~al.}(2013)\citenamefont{Aoki, Kubo, and
  Takano}}]{Aoki:2013gzs}
\bibinfo{author}{\bibfnamefont{M.}~\bibnamefont{Aoki}},
  \bibinfo{author}{\bibfnamefont{J.}~\bibnamefont{Kubo}}, \bibnamefont{and}
  \bibinfo{author}{\bibfnamefont{H.}~\bibnamefont{Takano}},
  \bibinfo{journal}{Phys. Rev. D} \textbf{\bibinfo{volume}{87}},
  \bibinfo{pages}{116001} (\bibinfo{year}{2013}), \eprint{1302.3936}.

\bibitem[{\citenamefont{Klasen et~al.}(2013)\citenamefont{Klasen, Yaguna,
  Ruiz-Alvarez, Restrepo, and Zapata}}]{Klasen:2013jpa}
\bibinfo{author}{\bibfnamefont{M.}~\bibnamefont{Klasen}},
  \bibinfo{author}{\bibfnamefont{C.~E.} \bibnamefont{Yaguna}},
  \bibinfo{author}{\bibfnamefont{J.~D.} \bibnamefont{Ruiz-Alvarez}},
  \bibinfo{author}{\bibfnamefont{D.}~\bibnamefont{Restrepo}}, \bibnamefont{and}
  \bibinfo{author}{\bibfnamefont{O.}~\bibnamefont{Zapata}},
  \bibinfo{journal}{JCAP} \textbf{\bibinfo{volume}{04}}, \bibinfo{pages}{044}
  (\bibinfo{year}{2013}), \eprint{1302.5298}.

\bibitem[{\citenamefont{Goudelis et~al.}(2013)\citenamefont{Goudelis, Herrmann,
  and St{\r{a}}l}}]{Goudelis:2013uca}
\bibinfo{author}{\bibfnamefont{A.}~\bibnamefont{Goudelis}},
  \bibinfo{author}{\bibfnamefont{B.}~\bibnamefont{Herrmann}}, \bibnamefont{and}
  \bibinfo{author}{\bibfnamefont{O.}~\bibnamefont{St{\r{a}}l}},
  \bibinfo{journal}{JHEP} \textbf{\bibinfo{volume}{09}}, \bibinfo{pages}{106}
  (\bibinfo{year}{2013}), \eprint{1303.3010}.

\bibitem[{\citenamefont{Basak and Mondal}(2014)}]{Basak:2013cga}
\bibinfo{author}{\bibfnamefont{T.}~\bibnamefont{Basak}} \bibnamefont{and}
  \bibinfo{author}{\bibfnamefont{T.}~\bibnamefont{Mondal}},
  \bibinfo{journal}{Phys. Rev. D} \textbf{\bibinfo{volume}{89}},
  \bibinfo{pages}{063527} (\bibinfo{year}{2014}), \eprint{1308.0023}.

\bibitem[{\citenamefont{Ibarra et~al.}(2014)\citenamefont{Ibarra, Toma,
  Totzauer, and Wild}}]{Ibarra:2014qma}
\bibinfo{author}{\bibfnamefont{A.}~\bibnamefont{Ibarra}},
  \bibinfo{author}{\bibfnamefont{T.}~\bibnamefont{Toma}},
  \bibinfo{author}{\bibfnamefont{M.}~\bibnamefont{Totzauer}}, \bibnamefont{and}
  \bibinfo{author}{\bibfnamefont{S.}~\bibnamefont{Wild}},
  \bibinfo{journal}{Phys. Rev. D} \textbf{\bibinfo{volume}{90}},
  \bibinfo{pages}{043526} (\bibinfo{year}{2014}), \eprint{1405.6917}.

\bibitem[{\citenamefont{Alanne et~al.}(2014)\citenamefont{Alanne, Tuominen, and
  Vaskonen}}]{Alanne:2014bra}
\bibinfo{author}{\bibfnamefont{T.}~\bibnamefont{Alanne}},
  \bibinfo{author}{\bibfnamefont{K.}~\bibnamefont{Tuominen}}, \bibnamefont{and}
  \bibinfo{author}{\bibfnamefont{V.}~\bibnamefont{Vaskonen}},
  \bibinfo{journal}{Nucl. Phys. B} \textbf{\bibinfo{volume}{889}},
  \bibinfo{pages}{692} (\bibinfo{year}{2014}), \eprint{1407.0688}.

\bibitem[{\citenamefont{Keus et~al.}(2014)\citenamefont{Keus, King, Moretti,
  and Sokolowska}}]{Keus:2014jha}
\bibinfo{author}{\bibfnamefont{V.}~\bibnamefont{Keus}},
  \bibinfo{author}{\bibfnamefont{S.~F.} \bibnamefont{King}},
  \bibinfo{author}{\bibfnamefont{S.}~\bibnamefont{Moretti}}, \bibnamefont{and}
  \bibinfo{author}{\bibfnamefont{D.}~\bibnamefont{Sokolowska}},
  \bibinfo{journal}{JHEP} \textbf{\bibinfo{volume}{11}}, \bibinfo{pages}{016}
  (\bibinfo{year}{2014}), \eprint{1407.7859}.

\bibitem[{\citenamefont{Okada et~al.}(2014)\citenamefont{Okada, Toma, and
  Yagyu}}]{Okada:2014qsa}
\bibinfo{author}{\bibfnamefont{H.}~\bibnamefont{Okada}},
  \bibinfo{author}{\bibfnamefont{T.}~\bibnamefont{Toma}}, \bibnamefont{and}
  \bibinfo{author}{\bibfnamefont{K.}~\bibnamefont{Yagyu}},
  \bibinfo{journal}{Phys. Rev. D} \textbf{\bibinfo{volume}{90}},
  \bibinfo{pages}{095005} (\bibinfo{year}{2014}), \eprint{1408.0961}.

\bibitem[{\citenamefont{Vicente and Yaguna}(2015)}]{Vicente:2014wga}
\bibinfo{author}{\bibfnamefont{A.}~\bibnamefont{Vicente}} \bibnamefont{and}
  \bibinfo{author}{\bibfnamefont{C.~E.} \bibnamefont{Yaguna}},
  \bibinfo{journal}{JHEP} \textbf{\bibinfo{volume}{02}}, \bibinfo{pages}{144}
  (\bibinfo{year}{2015}), \eprint{1412.2545}.

\bibitem[{\citenamefont{Bonilla et~al.}(2016)\citenamefont{Bonilla, Sokolowska,
  Darvishi, Diaz-Cruz, and Krawczyk}}]{Bonilla:2014xba}
\bibinfo{author}{\bibfnamefont{C.}~\bibnamefont{Bonilla}},
  \bibinfo{author}{\bibfnamefont{D.}~\bibnamefont{Sokolowska}},
  \bibinfo{author}{\bibfnamefont{N.}~\bibnamefont{Darvishi}},
  \bibinfo{author}{\bibfnamefont{J.~L.} \bibnamefont{Diaz-Cruz}},
  \bibnamefont{and} \bibinfo{author}{\bibfnamefont{M.}~\bibnamefont{Krawczyk}},
  \bibinfo{journal}{J. Phys. G} \textbf{\bibinfo{volume}{43}},
  \bibinfo{pages}{065001} (\bibinfo{year}{2016}), \eprint{1412.8730}.

\bibitem[{\citenamefont{Jin et~al.}(2015)\citenamefont{Jin, Tang, and
  Zhang}}]{Jin:2014glp}
\bibinfo{author}{\bibfnamefont{L.-G.} \bibnamefont{Jin}},
  \bibinfo{author}{\bibfnamefont{R.}~\bibnamefont{Tang}}, \bibnamefont{and}
  \bibinfo{author}{\bibfnamefont{F.}~\bibnamefont{Zhang}},
  \bibinfo{journal}{Phys. Lett. B} \textbf{\bibinfo{volume}{741}},
  \bibinfo{pages}{163} (\bibinfo{year}{2015}), \eprint{1501.02020}.

\bibitem[{\citenamefont{Chakrabarty et~al.}(2015)\citenamefont{Chakrabarty,
  Ghosh, Mukhopadhyaya, and Saha}}]{Chakrabarty:2015yia}
\bibinfo{author}{\bibfnamefont{N.}~\bibnamefont{Chakrabarty}},
  \bibinfo{author}{\bibfnamefont{D.~K.} \bibnamefont{Ghosh}},
  \bibinfo{author}{\bibfnamefont{B.}~\bibnamefont{Mukhopadhyaya}},
  \bibnamefont{and} \bibinfo{author}{\bibfnamefont{I.}~\bibnamefont{Saha}},
  \bibinfo{journal}{Phys. Rev. D} \textbf{\bibinfo{volume}{92}},
  \bibinfo{pages}{015002} (\bibinfo{year}{2015}), \eprint{1501.03700}.

\bibitem[{\citenamefont{Ibarra and Wild}(2015)}]{Ibarra:2015fqa}
\bibinfo{author}{\bibfnamefont{A.}~\bibnamefont{Ibarra}} \bibnamefont{and}
  \bibinfo{author}{\bibfnamefont{S.}~\bibnamefont{Wild}},
  \bibinfo{journal}{JCAP} \textbf{\bibinfo{volume}{05}}, \bibinfo{pages}{047}
  (\bibinfo{year}{2015}), \eprint{1503.03382}.

\bibitem[{\citenamefont{Blinov et~al.}(2016)\citenamefont{Blinov, Kozaczuk,
  Morrissey, and de~la Puente}}]{Blinov:2015qva}
\bibinfo{author}{\bibfnamefont{N.}~\bibnamefont{Blinov}},
  \bibinfo{author}{\bibfnamefont{J.}~\bibnamefont{Kozaczuk}},
  \bibinfo{author}{\bibfnamefont{D.~E.} \bibnamefont{Morrissey}},
  \bibnamefont{and} \bibinfo{author}{\bibfnamefont{A.}~\bibnamefont{de~la
  Puente}}, \bibinfo{journal}{Phys. Rev. D} \textbf{\bibinfo{volume}{93}},
  \bibinfo{pages}{035020} (\bibinfo{year}{2016}), \eprint{1510.08069}.

\bibitem[{\citenamefont{Okada and Okada}(2016)}]{Okada:2016gsh}
\bibinfo{author}{\bibfnamefont{N.}~\bibnamefont{Okada}} \bibnamefont{and}
  \bibinfo{author}{\bibfnamefont{S.}~\bibnamefont{Okada}},
  \bibinfo{journal}{Phys. Rev. D} \textbf{\bibinfo{volume}{93}},
  \bibinfo{pages}{075003} (\bibinfo{year}{2016}), \eprint{1601.07526}.

\bibitem[{\citenamefont{Kanemura et~al.}(2016)\citenamefont{Kanemura, Kikuchi,
  and Sakurai}}]{Kanemura:2016sos}
\bibinfo{author}{\bibfnamefont{S.}~\bibnamefont{Kanemura}},
  \bibinfo{author}{\bibfnamefont{M.}~\bibnamefont{Kikuchi}}, \bibnamefont{and}
  \bibinfo{author}{\bibfnamefont{K.}~\bibnamefont{Sakurai}},
  \bibinfo{journal}{Phys. Rev. D} \textbf{\bibinfo{volume}{94}},
  \bibinfo{pages}{115011} (\bibinfo{year}{2016}), \eprint{1605.08520}.

\bibitem[{\citenamefont{Singirala}(2017)}]{Singirala:2016kam}
\bibinfo{author}{\bibfnamefont{S.}~\bibnamefont{Singirala}},
  \bibinfo{journal}{Chin. Phys. C} \textbf{\bibinfo{volume}{41}},
  \bibinfo{pages}{043102} (\bibinfo{year}{2017}), \eprint{1607.03309}.

\bibitem[{\citenamefont{Arcadi et~al.}(2016)\citenamefont{Arcadi, Gross,
  Lebedev, Mambrini, Pokorski, and Toma}}]{Arcadi:2016kmk}
\bibinfo{author}{\bibfnamefont{G.}~\bibnamefont{Arcadi}},
  \bibinfo{author}{\bibfnamefont{C.}~\bibnamefont{Gross}},
  \bibinfo{author}{\bibfnamefont{O.}~\bibnamefont{Lebedev}},
  \bibinfo{author}{\bibfnamefont{Y.}~\bibnamefont{Mambrini}},
  \bibinfo{author}{\bibfnamefont{S.}~\bibnamefont{Pokorski}}, \bibnamefont{and}
  \bibinfo{author}{\bibfnamefont{T.}~\bibnamefont{Toma}},
  \bibinfo{journal}{JHEP} \textbf{\bibinfo{volume}{12}}, \bibinfo{pages}{081}
  (\bibinfo{year}{2016}), \eprint{1611.00365}.

\bibitem[{\citenamefont{Okada and Okada}(2017)}]{Okada:2016tci}
\bibinfo{author}{\bibfnamefont{N.}~\bibnamefont{Okada}} \bibnamefont{and}
  \bibinfo{author}{\bibfnamefont{S.}~\bibnamefont{Okada}},
  \bibinfo{journal}{Phys. Rev. D} \textbf{\bibinfo{volume}{95}},
  \bibinfo{pages}{035025} (\bibinfo{year}{2017}), \eprint{1611.02672}.

\bibitem[{\citenamefont{Dutta~Banik et~al.}(2017)\citenamefont{Dutta~Banik,
  Pandey, Majumdar, and Biswas}}]{DuttaBanik:2016jzv}
\bibinfo{author}{\bibfnamefont{A.}~\bibnamefont{Dutta~Banik}},
  \bibinfo{author}{\bibfnamefont{M.}~\bibnamefont{Pandey}},
  \bibinfo{author}{\bibfnamefont{D.}~\bibnamefont{Majumdar}}, \bibnamefont{and}
  \bibinfo{author}{\bibfnamefont{A.}~\bibnamefont{Biswas}},
  \bibinfo{journal}{Eur. Phys. J. C} \textbf{\bibinfo{volume}{77}},
  \bibinfo{pages}{657} (\bibinfo{year}{2017}), \eprint{1612.08621}.

\bibitem[{\citenamefont{Oda et~al.}(2017)\citenamefont{Oda, Okada, and
  Takahashi}}]{Oda:2017kwl}
\bibinfo{author}{\bibfnamefont{S.}~\bibnamefont{Oda}},
  \bibinfo{author}{\bibfnamefont{N.}~\bibnamefont{Okada}}, \bibnamefont{and}
  \bibinfo{author}{\bibfnamefont{D.-s.} \bibnamefont{Takahashi}},
  \bibinfo{journal}{Phys. Rev. D} \textbf{\bibinfo{volume}{96}},
  \bibinfo{pages}{095032} (\bibinfo{year}{2017}), \eprint{1704.05023}.

\bibitem[{\citenamefont{Chao}(2018)}]{Chao:2017rwv}
\bibinfo{author}{\bibfnamefont{W.}~\bibnamefont{Chao}}, \bibinfo{journal}{Eur.
  Phys. J. C} \textbf{\bibinfo{volume}{78}}, \bibinfo{pages}{103}
  (\bibinfo{year}{2018}), \eprint{1707.07858}.

\bibitem[{\citenamefont{Cox et~al.}(2018)\citenamefont{Cox, Han, and
  Yanagida}}]{Cox:2017rgn}
\bibinfo{author}{\bibfnamefont{P.}~\bibnamefont{Cox}},
  \bibinfo{author}{\bibfnamefont{C.}~\bibnamefont{Han}}, \bibnamefont{and}
  \bibinfo{author}{\bibfnamefont{T.~T.} \bibnamefont{Yanagida}},
  \bibinfo{journal}{JCAP} \textbf{\bibinfo{volume}{01}}, \bibinfo{pages}{029}
  (\bibinfo{year}{2018}), \eprint{1710.01585}.

\bibitem[{\citenamefont{Okada}(2018)}]{Okada:2018ktp}
\bibinfo{author}{\bibfnamefont{S.}~\bibnamefont{Okada}}, \bibinfo{journal}{Adv.
  High Energy Phys.} \textbf{\bibinfo{volume}{2018}}, \bibinfo{pages}{5340935}
  (\bibinfo{year}{2018}), \eprint{1803.06793}.

\bibitem[{\citenamefont{Escudero et~al.}(2018)\citenamefont{Escudero, Witte,
  and Rius}}]{Escudero:2018fwn}
\bibinfo{author}{\bibfnamefont{M.}~\bibnamefont{Escudero}},
  \bibinfo{author}{\bibfnamefont{S.~J.} \bibnamefont{Witte}}, \bibnamefont{and}
  \bibinfo{author}{\bibfnamefont{N.}~\bibnamefont{Rius}},
  \bibinfo{journal}{JHEP} \textbf{\bibinfo{volume}{08}}, \bibinfo{pages}{190}
  (\bibinfo{year}{2018}), \eprint{1806.02823}.

\bibitem[{\citenamefont{Bandyopadhyay et~al.}(2019)\citenamefont{Bandyopadhyay,
  Chun, Mandal, and Queiroz}}]{Bandyopadhyay:2018qcv}
\bibinfo{author}{\bibfnamefont{P.}~\bibnamefont{Bandyopadhyay}},
  \bibinfo{author}{\bibfnamefont{E.~J.} \bibnamefont{Chun}},
  \bibinfo{author}{\bibfnamefont{R.}~\bibnamefont{Mandal}}, \bibnamefont{and}
  \bibinfo{author}{\bibfnamefont{F.~S.} \bibnamefont{Queiroz}},
  \bibinfo{journal}{Phys. Lett. B} \textbf{\bibinfo{volume}{788}},
  \bibinfo{pages}{530} (\bibinfo{year}{2019}), \eprint{1807.05122}.

\bibitem[{\citenamefont{Bhattacharya et~al.}(2019)\citenamefont{Bhattacharya,
  Ghosh, and Sahu}}]{Bhattacharya:2018cgx}
\bibinfo{author}{\bibfnamefont{S.}~\bibnamefont{Bhattacharya}},
  \bibinfo{author}{\bibfnamefont{P.}~\bibnamefont{Ghosh}}, \bibnamefont{and}
  \bibinfo{author}{\bibfnamefont{N.}~\bibnamefont{Sahu}},
  \bibinfo{journal}{JHEP} \textbf{\bibinfo{volume}{02}}, \bibinfo{pages}{059}
  (\bibinfo{year}{2019}), \eprint{1809.07474}.

\bibitem[{\citenamefont{Borah et~al.}(2020)\citenamefont{Borah, Nanda,
  Narendra, and Sahu}}]{Borah:2018smz}
\bibinfo{author}{\bibfnamefont{D.}~\bibnamefont{Borah}},
  \bibinfo{author}{\bibfnamefont{D.}~\bibnamefont{Nanda}},
  \bibinfo{author}{\bibfnamefont{N.}~\bibnamefont{Narendra}}, \bibnamefont{and}
  \bibinfo{author}{\bibfnamefont{N.}~\bibnamefont{Sahu}},
  \bibinfo{journal}{Nucl. Phys. B} \textbf{\bibinfo{volume}{950}},
  \bibinfo{pages}{114841} (\bibinfo{year}{2020}), \eprint{1810.12920}.

\bibitem[{\citenamefont{Das et~al.}(2020)\citenamefont{Das, Goswami,
  Vishnudath, and Nomura}}]{Das:2019pua}
\bibinfo{author}{\bibfnamefont{A.}~\bibnamefont{Das}},
  \bibinfo{author}{\bibfnamefont{S.}~\bibnamefont{Goswami}},
  \bibinfo{author}{\bibfnamefont{K.~N.} \bibnamefont{Vishnudath}},
  \bibnamefont{and} \bibinfo{author}{\bibfnamefont{T.}~\bibnamefont{Nomura}},
  \bibinfo{journal}{Phys. Rev. D} \textbf{\bibinfo{volume}{101}},
  \bibinfo{pages}{055026} (\bibinfo{year}{2020}), \eprint{1905.00201}.

\bibitem[{\citenamefont{Dutta et~al.}(2021)\citenamefont{Dutta, Bhattacharya,
  Ghosh, and Sahu}}]{Dutta:2020xwn}
\bibinfo{author}{\bibfnamefont{M.}~\bibnamefont{Dutta}},
  \bibinfo{author}{\bibfnamefont{S.}~\bibnamefont{Bhattacharya}},
  \bibinfo{author}{\bibfnamefont{P.}~\bibnamefont{Ghosh}}, \bibnamefont{and}
  \bibinfo{author}{\bibfnamefont{N.}~\bibnamefont{Sahu}},
  \bibinfo{journal}{JCAP} \textbf{\bibinfo{volume}{03}}, \bibinfo{pages}{008}
  (\bibinfo{year}{2021}), \eprint{2009.00885}.

\bibitem[{\citenamefont{Barman et~al.}(2020{\natexlab{a}})\citenamefont{Barman,
  Bhattacharya, and Grzadkowski}}]{Barman:2020ifq}
\bibinfo{author}{\bibfnamefont{B.}~\bibnamefont{Barman}},
  \bibinfo{author}{\bibfnamefont{S.}~\bibnamefont{Bhattacharya}},
  \bibnamefont{and}
  \bibinfo{author}{\bibfnamefont{B.}~\bibnamefont{Grzadkowski}},
  \bibinfo{journal}{JHEP} \textbf{\bibinfo{volume}{12}}, \bibinfo{pages}{162}
  (\bibinfo{year}{2020}{\natexlab{a}}), \eprint{2009.07438}.

\bibitem[{\citenamefont{Nam}(2020)}]{Nam:2020byw}
\bibinfo{author}{\bibfnamefont{C.~H.} \bibnamefont{Nam}},
  \bibinfo{journal}{Eur. Phys. J. C} \textbf{\bibinfo{volume}{80}},
  \bibinfo{pages}{1114} (\bibinfo{year}{2020}), \eprint{2011.11207}.

\bibitem[{\citenamefont{Chen et~al.}(2021)\citenamefont{Chen, Dutta~Banik, and
  Liu}}]{Chen:2020ark}
\bibinfo{author}{\bibfnamefont{S.-L.} \bibnamefont{Chen}},
  \bibinfo{author}{\bibfnamefont{A.}~\bibnamefont{Dutta~Banik}},
  \bibnamefont{and} \bibinfo{author}{\bibfnamefont{Z.-K.} \bibnamefont{Liu}},
  \bibinfo{journal}{Nucl. Phys. B} \textbf{\bibinfo{volume}{966}},
  \bibinfo{pages}{115394} (\bibinfo{year}{2021}), \eprint{2011.13551}.

\bibitem[{\citenamefont{Arcadi et~al.}(2021)\citenamefont{Arcadi, Djouadi, and
  Kado}}]{Arcadi:2021mag}
\bibinfo{author}{\bibfnamefont{G.}~\bibnamefont{Arcadi}},
  \bibinfo{author}{\bibfnamefont{A.}~\bibnamefont{Djouadi}}, \bibnamefont{and}
  \bibinfo{author}{\bibfnamefont{M.}~\bibnamefont{Kado}},
  \bibinfo{journal}{Eur. Phys. J. C} \textbf{\bibinfo{volume}{81}},
  \bibinfo{pages}{653} (\bibinfo{year}{2021}), \eprint{2101.02507}.

\bibitem[{\citenamefont{Sarazin et~al.}(2021)\citenamefont{Sarazin, Bernigaud,
  and Herrmann}}]{Sarazin:2021nwo}
\bibinfo{author}{\bibfnamefont{M.}~\bibnamefont{Sarazin}},
  \bibinfo{author}{\bibfnamefont{J.}~\bibnamefont{Bernigaud}},
  \bibnamefont{and} \bibinfo{author}{\bibfnamefont{B.}~\bibnamefont{Herrmann}},
  \bibinfo{journal}{JHEP} \textbf{\bibinfo{volume}{12}}, \bibinfo{pages}{116}
  (\bibinfo{year}{2021}), \eprint{2107.04613}.

\bibitem[{\citenamefont{Datta et~al.}(2022)\citenamefont{Datta, Roshan, and
  Sil}}]{Datta:2021gyi}
\bibinfo{author}{\bibfnamefont{A.}~\bibnamefont{Datta}},
  \bibinfo{author}{\bibfnamefont{R.}~\bibnamefont{Roshan}}, \bibnamefont{and}
  \bibinfo{author}{\bibfnamefont{A.}~\bibnamefont{Sil}},
  \bibinfo{journal}{Phys. Rev. D} \textbf{\bibinfo{volume}{105}},
  \bibinfo{pages}{095032} (\bibinfo{year}{2022}), \eprint{2110.03914}.

\bibitem[{\citenamefont{Bandyopadhyay et~al.}(2022)\citenamefont{Bandyopadhyay,
  Mitra, Padhan, Roy, and Spannowsky}}]{Bandyopadhyay:2022xlp}
\bibinfo{author}{\bibfnamefont{P.}~\bibnamefont{Bandyopadhyay}},
  \bibinfo{author}{\bibfnamefont{M.}~\bibnamefont{Mitra}},
  \bibinfo{author}{\bibfnamefont{R.}~\bibnamefont{Padhan}},
  \bibinfo{author}{\bibfnamefont{A.}~\bibnamefont{Roy}}, \bibnamefont{and}
  \bibinfo{author}{\bibfnamefont{M.}~\bibnamefont{Spannowsky}},
  \bibinfo{journal}{JHEP} \textbf{\bibinfo{volume}{05}}, \bibinfo{pages}{182}
  (\bibinfo{year}{2022}), \eprint{2201.09203}.

\bibitem[{\citenamefont{Khaw et~al.}(2023)\citenamefont{Khaw, Nakai, Sato,
  Shigekami, and Zhang}}]{Khaw:2022qxh}
\bibinfo{author}{\bibfnamefont{K.~S.} \bibnamefont{Khaw}},
  \bibinfo{author}{\bibfnamefont{Y.}~\bibnamefont{Nakai}},
  \bibinfo{author}{\bibfnamefont{R.}~\bibnamefont{Sato}},
  \bibinfo{author}{\bibfnamefont{Y.}~\bibnamefont{Shigekami}},
  \bibnamefont{and} \bibinfo{author}{\bibfnamefont{Z.}~\bibnamefont{Zhang}},
  \bibinfo{journal}{JHEP} \textbf{\bibinfo{volume}{02}}, \bibinfo{pages}{234}
  (\bibinfo{year}{2023}), \eprint{2212.02891}.

\bibitem[{\citenamefont{Van~Loi et~al.}(2023)\citenamefont{Van~Loi, Nam, and
  Van~Dong}}]{VanLoi:2023pkt}
\bibinfo{author}{\bibfnamefont{D.}~\bibnamefont{Van~Loi}},
  \bibinfo{author}{\bibfnamefont{C.~H.} \bibnamefont{Nam}}, \bibnamefont{and}
  \bibinfo{author}{\bibfnamefont{P.}~\bibnamefont{Van~Dong}},
  \bibinfo{journal}{Phys. Rev. D} \textbf{\bibinfo{volume}{108}},
  \bibinfo{pages}{095018} (\bibinfo{year}{2023}), \eprint{2305.04681}.

\bibitem[{\citenamefont{Zhang et~al.}(2025)\citenamefont{Zhang, Han, Liu, and
  Shao}}]{Zhang:2024sox}
\bibinfo{author}{\bibfnamefont{J.-J.} \bibnamefont{Zhang}},
  \bibinfo{author}{\bibfnamefont{Z.-L.} \bibnamefont{Han}},
  \bibinfo{author}{\bibfnamefont{A.}~\bibnamefont{Liu}}, \bibnamefont{and}
  \bibinfo{author}{\bibfnamefont{F.-L.} \bibnamefont{Shao}},
  \bibinfo{journal}{Nucl. Phys. B} \textbf{\bibinfo{volume}{1014}},
  \bibinfo{pages}{116864} (\bibinfo{year}{2025}), \eprint{2411.06744}.

\bibitem[{\citenamefont{Lu et~al.}(2025)\citenamefont{Lu, Wu, and
  Xu}}]{Lu:2025vif}
\bibinfo{author}{\bibfnamefont{C.-T.} \bibnamefont{Lu}},
  \bibinfo{author}{\bibfnamefont{Y.}~\bibnamefont{Wu}}, \bibnamefont{and}
  \bibinfo{author}{\bibfnamefont{S.}~\bibnamefont{Xu}}, \bibinfo{journal}{JHEP}
  \textbf{\bibinfo{volume}{12}}, \bibinfo{pages}{155} (\bibinfo{year}{2025}),
  \eprint{2504.10930}.

\bibitem[{\citenamefont{Liu et~al.}(2025)\citenamefont{Liu, Han, Huang, Shao,
  and Wang}}]{Liu:2025swd}
\bibinfo{author}{\bibfnamefont{A.}~\bibnamefont{Liu}},
  \bibinfo{author}{\bibfnamefont{Z.-L.} \bibnamefont{Han}},
  \bibinfo{author}{\bibfnamefont{F.}~\bibnamefont{Huang}},
  \bibinfo{author}{\bibfnamefont{F.-L.} \bibnamefont{Shao}}, \bibnamefont{and}
  \bibinfo{author}{\bibfnamefont{W.}~\bibnamefont{Wang}}
  (\bibinfo{year}{2025}), \eprint{2510.13231}.

\bibitem[{\citenamefont{Khalil and Seto}(2008)}]{Khalil:2008kp}
\bibinfo{author}{\bibfnamefont{S.}~\bibnamefont{Khalil}} \bibnamefont{and}
  \bibinfo{author}{\bibfnamefont{O.}~\bibnamefont{Seto}},
  \bibinfo{journal}{JCAP} \textbf{\bibinfo{volume}{10}}, \bibinfo{pages}{024}
  (\bibinfo{year}{2008}), \eprint{0804.0336}.

\bibitem[{\citenamefont{Mambrini}(2010)}]{Mambrini:2010dq}
\bibinfo{author}{\bibfnamefont{Y.}~\bibnamefont{Mambrini}},
  \bibinfo{journal}{JCAP} \textbf{\bibinfo{volume}{09}}, \bibinfo{pages}{022}
  (\bibinfo{year}{2010}), \eprint{1006.3318}.

\bibitem[{\citenamefont{Kanemura et~al.}(2011)\citenamefont{Kanemura, Seto, and
  Shimomura}}]{Kanemura:2011vm}
\bibinfo{author}{\bibfnamefont{S.}~\bibnamefont{Kanemura}},
  \bibinfo{author}{\bibfnamefont{O.}~\bibnamefont{Seto}}, \bibnamefont{and}
  \bibinfo{author}{\bibfnamefont{T.}~\bibnamefont{Shimomura}},
  \bibinfo{journal}{Phys. Rev. D} \textbf{\bibinfo{volume}{84}},
  \bibinfo{pages}{016004} (\bibinfo{year}{2011}), \eprint{1101.5713}.

\bibitem[{\citenamefont{Lindner et~al.}(2011)\citenamefont{Lindner, Schmidt,
  and Schwetz}}]{Lindner:2011it}
\bibinfo{author}{\bibfnamefont{M.}~\bibnamefont{Lindner}},
  \bibinfo{author}{\bibfnamefont{D.}~\bibnamefont{Schmidt}}, \bibnamefont{and}
  \bibinfo{author}{\bibfnamefont{T.}~\bibnamefont{Schwetz}},
  \bibinfo{journal}{Phys. Lett. B} \textbf{\bibinfo{volume}{705}},
  \bibinfo{pages}{324} (\bibinfo{year}{2011}), \eprint{1105.4626}.

\bibitem[{\citenamefont{Okada and Orikasa}(2012)}]{Okada:2012sg}
\bibinfo{author}{\bibfnamefont{N.}~\bibnamefont{Okada}} \bibnamefont{and}
  \bibinfo{author}{\bibfnamefont{Y.}~\bibnamefont{Orikasa}},
  \bibinfo{journal}{Phys. Rev. D} \textbf{\bibinfo{volume}{85}},
  \bibinfo{pages}{115006} (\bibinfo{year}{2012}), \eprint{1202.1405}.

\bibitem[{\citenamefont{Lindner et~al.}(2014)\citenamefont{Lindner, Schmidt,
  and Watanabe}}]{Lindner:2013awa}
\bibinfo{author}{\bibfnamefont{M.}~\bibnamefont{Lindner}},
  \bibinfo{author}{\bibfnamefont{D.}~\bibnamefont{Schmidt}}, \bibnamefont{and}
  \bibinfo{author}{\bibfnamefont{A.}~\bibnamefont{Watanabe}},
  \bibinfo{journal}{Phys. Rev. D} \textbf{\bibinfo{volume}{89}},
  \bibinfo{pages}{013007} (\bibinfo{year}{2014}), \eprint{1310.6582}.

\bibitem[{\citenamefont{Kanemura et~al.}(2014)\citenamefont{Kanemura, Matsui,
  and Sugiyama}}]{Kanemura:2014rpa}
\bibinfo{author}{\bibfnamefont{S.}~\bibnamefont{Kanemura}},
  \bibinfo{author}{\bibfnamefont{T.}~\bibnamefont{Matsui}}, \bibnamefont{and}
  \bibinfo{author}{\bibfnamefont{H.}~\bibnamefont{Sugiyama}},
  \bibinfo{journal}{Phys. Rev. D} \textbf{\bibinfo{volume}{90}},
  \bibinfo{pages}{013001} (\bibinfo{year}{2014}), \eprint{1405.1935}.

\bibitem[{\citenamefont{Ko et~al.}(2014)\citenamefont{Ko, Omura, and
  Yu}}]{Ko:2014uka}
\bibinfo{author}{\bibfnamefont{P.}~\bibnamefont{Ko}},
  \bibinfo{author}{\bibfnamefont{Y.}~\bibnamefont{Omura}}, \bibnamefont{and}
  \bibinfo{author}{\bibfnamefont{C.}~\bibnamefont{Yu}}, \bibinfo{journal}{JHEP}
  \textbf{\bibinfo{volume}{11}}, \bibinfo{pages}{054} (\bibinfo{year}{2014}),
  \eprint{1405.2138}.

\bibitem[{\citenamefont{Rodejohann and Yaguna}(2015)}]{Rodejohann:2015lca}
\bibinfo{author}{\bibfnamefont{W.}~\bibnamefont{Rodejohann}} \bibnamefont{and}
  \bibinfo{author}{\bibfnamefont{C.~E.} \bibnamefont{Yaguna}},
  \bibinfo{journal}{JCAP} \textbf{\bibinfo{volume}{12}}, \bibinfo{pages}{032}
  (\bibinfo{year}{2015}), \eprint{1509.04036}.

\bibitem[{\citenamefont{Biswas et~al.}(2016)\citenamefont{Biswas, Choubey, and
  Khan}}]{Biswas:2016ewm}
\bibinfo{author}{\bibfnamefont{A.}~\bibnamefont{Biswas}},
  \bibinfo{author}{\bibfnamefont{S.}~\bibnamefont{Choubey}}, \bibnamefont{and}
  \bibinfo{author}{\bibfnamefont{S.}~\bibnamefont{Khan}},
  \bibinfo{journal}{JHEP} \textbf{\bibinfo{volume}{08}}, \bibinfo{pages}{114}
  (\bibinfo{year}{2016}), \eprint{1604.06566}.

\bibitem[{\citenamefont{Escudero et~al.}(2017)\citenamefont{Escudero, Rius, and
  Sanz}}]{Escudero:2016tzx}
\bibinfo{author}{\bibfnamefont{M.}~\bibnamefont{Escudero}},
  \bibinfo{author}{\bibfnamefont{N.}~\bibnamefont{Rius}}, \bibnamefont{and}
  \bibinfo{author}{\bibfnamefont{V.}~\bibnamefont{Sanz}},
  \bibinfo{journal}{JHEP} \textbf{\bibinfo{volume}{02}}, \bibinfo{pages}{045}
  (\bibinfo{year}{2017}), \eprint{1606.01258}.

\bibitem[{\citenamefont{Patra et~al.}(2016)\citenamefont{Patra, Rodejohann, and
  Yaguna}}]{Patra:2016ofq}
\bibinfo{author}{\bibfnamefont{S.}~\bibnamefont{Patra}},
  \bibinfo{author}{\bibfnamefont{W.}~\bibnamefont{Rodejohann}},
  \bibnamefont{and} \bibinfo{author}{\bibfnamefont{C.~E.}
  \bibnamefont{Yaguna}}, \bibinfo{journal}{JHEP} \textbf{\bibinfo{volume}{09}},
  \bibinfo{pages}{076} (\bibinfo{year}{2016}), \eprint{1607.04029}.

\bibitem[{\citenamefont{Singirala
  et~al.}(2018{\natexlab{a}})\citenamefont{Singirala, Mohanta, and
  Patra}}]{Singirala:2017see}
\bibinfo{author}{\bibfnamefont{S.}~\bibnamefont{Singirala}},
  \bibinfo{author}{\bibfnamefont{R.}~\bibnamefont{Mohanta}}, \bibnamefont{and}
  \bibinfo{author}{\bibfnamefont{S.}~\bibnamefont{Patra}},
  \bibinfo{journal}{Eur. Phys. J. Plus} \textbf{\bibinfo{volume}{133}},
  \bibinfo{pages}{477} (\bibinfo{year}{2018}{\natexlab{a}}),
  \eprint{1704.01107}.

\bibitem[{\citenamefont{Nomura and Okada}(2019{\natexlab{a}})}]{Nomura:2017vzp}
\bibinfo{author}{\bibfnamefont{T.}~\bibnamefont{Nomura}} \bibnamefont{and}
  \bibinfo{author}{\bibfnamefont{H.}~\bibnamefont{Okada}},
  \bibinfo{journal}{Nucl. Phys. B} \textbf{\bibinfo{volume}{941}},
  \bibinfo{pages}{586} (\bibinfo{year}{2019}{\natexlab{a}}),
  \eprint{1705.08309}.

\bibitem[{\citenamefont{Bandyopadhyay et~al.}(2018)\citenamefont{Bandyopadhyay,
  Chun, and Mandal}}]{Bandyopadhyay:2017bgh}
\bibinfo{author}{\bibfnamefont{P.}~\bibnamefont{Bandyopadhyay}},
  \bibinfo{author}{\bibfnamefont{E.~J.} \bibnamefont{Chun}}, \bibnamefont{and}
  \bibinfo{author}{\bibfnamefont{R.}~\bibnamefont{Mandal}},
  \bibinfo{journal}{Phys. Rev. D} \textbf{\bibinfo{volume}{97}},
  \bibinfo{pages}{015001} (\bibinfo{year}{2018}), \eprint{1707.00874}.

\bibitem[{\citenamefont{De~Romeri et~al.}(2017)\citenamefont{De~Romeri,
  Fernandez-Martinez, Gehrlein, Machado, and Niro}}]{DeRomeri:2017oxa}
\bibinfo{author}{\bibfnamefont{V.}~\bibnamefont{De~Romeri}},
  \bibinfo{author}{\bibfnamefont{E.}~\bibnamefont{Fernandez-Martinez}},
  \bibinfo{author}{\bibfnamefont{J.}~\bibnamefont{Gehrlein}},
  \bibinfo{author}{\bibfnamefont{P.~A.~N.} \bibnamefont{Machado}},
  \bibnamefont{and} \bibinfo{author}{\bibfnamefont{V.}~\bibnamefont{Niro}},
  \bibinfo{journal}{JHEP} \textbf{\bibinfo{volume}{10}}, \bibinfo{pages}{169}
  (\bibinfo{year}{2017}), \eprint{1707.08606}.

\bibitem[{\citenamefont{Nomura and Okada}(2018{\natexlab{a}})}]{Nomura:2017jxb}
\bibinfo{author}{\bibfnamefont{T.}~\bibnamefont{Nomura}} \bibnamefont{and}
  \bibinfo{author}{\bibfnamefont{H.}~\bibnamefont{Okada}},
  \bibinfo{journal}{Eur. Phys. J. C} \textbf{\bibinfo{volume}{78}},
  \bibinfo{pages}{189} (\bibinfo{year}{2018}{\natexlab{a}}),
  \eprint{1708.08737}.

\bibitem[{\citenamefont{Nomura and Okada}(2018{\natexlab{b}})}]{Nomura:2017wxf}
\bibinfo{author}{\bibfnamefont{T.}~\bibnamefont{Nomura}} \bibnamefont{and}
  \bibinfo{author}{\bibfnamefont{H.}~\bibnamefont{Okada}},
  \bibinfo{journal}{Phys. Rev. D} \textbf{\bibinfo{volume}{97}},
  \bibinfo{pages}{075038} (\bibinfo{year}{2018}{\natexlab{b}}),
  \eprint{1709.06406}.

\bibitem[{\citenamefont{Nanda and Borah}(2017)}]{Nanda:2017bmi}
\bibinfo{author}{\bibfnamefont{D.}~\bibnamefont{Nanda}} \bibnamefont{and}
  \bibinfo{author}{\bibfnamefont{D.}~\bibnamefont{Borah}},
  \bibinfo{journal}{Phys. Rev. D} \textbf{\bibinfo{volume}{96}},
  \bibinfo{pages}{115014} (\bibinfo{year}{2017}), \eprint{1709.08417}.

\bibitem[{\citenamefont{Singirala
  et~al.}(2018{\natexlab{b}})\citenamefont{Singirala, Mohanta, Patra, and
  Rao}}]{Singirala:2017cch}
\bibinfo{author}{\bibfnamefont{S.}~\bibnamefont{Singirala}},
  \bibinfo{author}{\bibfnamefont{R.}~\bibnamefont{Mohanta}},
  \bibinfo{author}{\bibfnamefont{S.}~\bibnamefont{Patra}}, \bibnamefont{and}
  \bibinfo{author}{\bibfnamefont{S.}~\bibnamefont{Rao}},
  \bibinfo{journal}{JCAP} \textbf{\bibinfo{volume}{11}}, \bibinfo{pages}{026}
  (\bibinfo{year}{2018}{\natexlab{b}}), \eprint{1710.05775}.

\bibitem[{\citenamefont{Okada et~al.}(2019)\citenamefont{Okada, Okada, and
  Raut}}]{Okada:2018tgy}
\bibinfo{author}{\bibfnamefont{N.}~\bibnamefont{Okada}},
  \bibinfo{author}{\bibfnamefont{S.}~\bibnamefont{Okada}}, \bibnamefont{and}
  \bibinfo{author}{\bibfnamefont{D.}~\bibnamefont{Raut}},
  \bibinfo{journal}{Phys. Rev. D} \textbf{\bibinfo{volume}{100}},
  \bibinfo{pages}{035022} (\bibinfo{year}{2019}), \eprint{1811.11927}.

\bibitem[{\citenamefont{Das et~al.}(2019)\citenamefont{Das, Okada, Okada, and
  Raut}}]{Das:2018tbd}
\bibinfo{author}{\bibfnamefont{A.}~\bibnamefont{Das}},
  \bibinfo{author}{\bibfnamefont{N.}~\bibnamefont{Okada}},
  \bibinfo{author}{\bibfnamefont{S.}~\bibnamefont{Okada}}, \bibnamefont{and}
  \bibinfo{author}{\bibfnamefont{D.}~\bibnamefont{Raut}},
  \bibinfo{journal}{Phys. Lett. B} \textbf{\bibinfo{volume}{797}},
  \bibinfo{pages}{134849} (\bibinfo{year}{2019}), \eprint{1812.11931}.

\bibitem[{\citenamefont{Biswas et~al.}(2019)\citenamefont{Biswas, Borah, and
  Nanda}}]{Biswas:2019ygr}
\bibinfo{author}{\bibfnamefont{A.}~\bibnamefont{Biswas}},
  \bibinfo{author}{\bibfnamefont{D.}~\bibnamefont{Borah}}, \bibnamefont{and}
  \bibinfo{author}{\bibfnamefont{D.}~\bibnamefont{Nanda}},
  \bibinfo{journal}{JHEP} \textbf{\bibinfo{volume}{12}}, \bibinfo{pages}{109}
  (\bibinfo{year}{2019}), \eprint{1908.04308}.

\bibitem[{\citenamefont{Mohapatra and Okada}(2020)}]{Mohapatra:2019ysk}
\bibinfo{author}{\bibfnamefont{R.~N.} \bibnamefont{Mohapatra}}
  \bibnamefont{and} \bibinfo{author}{\bibfnamefont{N.}~\bibnamefont{Okada}},
  \bibinfo{journal}{Phys. Rev. D} \textbf{\bibinfo{volume}{102}},
  \bibinfo{pages}{035028} (\bibinfo{year}{2020}), \eprint{1908.11325}.

\bibitem[{\citenamefont{Okada et~al.}(2020)\citenamefont{Okada, Okada, Raut,
  and Shafi}}]{Okada:2020evk}
\bibinfo{author}{\bibfnamefont{N.}~\bibnamefont{Okada}},
  \bibinfo{author}{\bibfnamefont{S.}~\bibnamefont{Okada}},
  \bibinfo{author}{\bibfnamefont{D.}~\bibnamefont{Raut}}, \bibnamefont{and}
  \bibinfo{author}{\bibfnamefont{Q.}~\bibnamefont{Shafi}},
  \bibinfo{journal}{Phys. Lett. B} \textbf{\bibinfo{volume}{810}},
  \bibinfo{pages}{135785} (\bibinfo{year}{2020}), \eprint{2007.02898}.

\bibitem[{\citenamefont{Seto and Shimomura}(2020)}]{Seto:2020udg}
\bibinfo{author}{\bibfnamefont{O.}~\bibnamefont{Seto}} \bibnamefont{and}
  \bibinfo{author}{\bibfnamefont{T.}~\bibnamefont{Shimomura}},
  \bibinfo{journal}{Phys. Lett. B} \textbf{\bibinfo{volume}{811}},
  \bibinfo{pages}{135880} (\bibinfo{year}{2020}), \eprint{2007.14605}.

\bibitem[{\citenamefont{Nagao and Okada}(2021)}]{Nagao:2020azf}
\bibinfo{author}{\bibfnamefont{K.~I.} \bibnamefont{Nagao}} \bibnamefont{and}
  \bibinfo{author}{\bibfnamefont{H.}~\bibnamefont{Okada}},
  \bibinfo{journal}{JCAP} \textbf{\bibinfo{volume}{05}}, \bibinfo{pages}{063}
  (\bibinfo{year}{2021}), \eprint{2008.13686}.

\bibitem[{\citenamefont{Asai et~al.}(2021)\citenamefont{Asai, Okawa, and
  Tsumura}}]{Asai:2020qlp}
\bibinfo{author}{\bibfnamefont{K.}~\bibnamefont{Asai}},
  \bibinfo{author}{\bibfnamefont{S.}~\bibnamefont{Okawa}}, \bibnamefont{and}
  \bibinfo{author}{\bibfnamefont{K.}~\bibnamefont{Tsumura}},
  \bibinfo{journal}{JHEP} \textbf{\bibinfo{volume}{03}}, \bibinfo{pages}{047}
  (\bibinfo{year}{2021}), \eprint{2011.03165}.

\bibitem[{\citenamefont{Ghosh et~al.}(2022)\citenamefont{Ghosh, Mahapatra,
  Narendra, and Sahu}}]{Ghosh:2021khk}
\bibinfo{author}{\bibfnamefont{P.}~\bibnamefont{Ghosh}},
  \bibinfo{author}{\bibfnamefont{S.}~\bibnamefont{Mahapatra}},
  \bibinfo{author}{\bibfnamefont{N.}~\bibnamefont{Narendra}}, \bibnamefont{and}
  \bibinfo{author}{\bibfnamefont{N.}~\bibnamefont{Sahu}},
  \bibinfo{journal}{Phys. Rev. D} \textbf{\bibinfo{volume}{106}},
  \bibinfo{pages}{015001} (\bibinfo{year}{2022}), \eprint{2107.11951}.

\bibitem[{\citenamefont{Nath et~al.}(2022)\citenamefont{Nath, Okada, Okada,
  Raut, and Shafi}}]{Nath:2021uqb}
\bibinfo{author}{\bibfnamefont{N.}~\bibnamefont{Nath}},
  \bibinfo{author}{\bibfnamefont{N.}~\bibnamefont{Okada}},
  \bibinfo{author}{\bibfnamefont{S.}~\bibnamefont{Okada}},
  \bibinfo{author}{\bibfnamefont{D.}~\bibnamefont{Raut}}, \bibnamefont{and}
  \bibinfo{author}{\bibfnamefont{Q.}~\bibnamefont{Shafi}},
  \bibinfo{journal}{Eur. Phys. J. C} \textbf{\bibinfo{volume}{82}},
  \bibinfo{pages}{864} (\bibinfo{year}{2022}), \eprint{2112.08960}.

\bibitem[{\citenamefont{Barman et~al.}(2022)\citenamefont{Barman, Ghosh,
  Ghoshal, and Mukherjee}}]{Barman:2021yaz}
\bibinfo{author}{\bibfnamefont{B.}~\bibnamefont{Barman}},
  \bibinfo{author}{\bibfnamefont{P.}~\bibnamefont{Ghosh}},
  \bibinfo{author}{\bibfnamefont{A.}~\bibnamefont{Ghoshal}}, \bibnamefont{and}
  \bibinfo{author}{\bibfnamefont{L.}~\bibnamefont{Mukherjee}},
  \bibinfo{journal}{JCAP} \textbf{\bibinfo{volume}{08}}, \bibinfo{pages}{049}
  (\bibinfo{year}{2022}), \eprint{2112.12798}.

\bibitem[{\citenamefont{Okada and Seto}(2022)}]{Okada:2022cby}
\bibinfo{author}{\bibfnamefont{N.}~\bibnamefont{Okada}} \bibnamefont{and}
  \bibinfo{author}{\bibfnamefont{O.}~\bibnamefont{Seto}},
  \bibinfo{journal}{Phys. Rev. D} \textbf{\bibinfo{volume}{105}},
  \bibinfo{pages}{123512} (\bibinfo{year}{2022}), \eprint{2202.08508}.

\bibitem[{\citenamefont{Okada and Seto}(2023)}]{Okada:2023mdv}
\bibinfo{author}{\bibfnamefont{N.}~\bibnamefont{Okada}} \bibnamefont{and}
  \bibinfo{author}{\bibfnamefont{O.}~\bibnamefont{Seto}},
  \bibinfo{journal}{Phys. Rev. D} \textbf{\bibinfo{volume}{108}},
  \bibinfo{pages}{083504} (\bibinfo{year}{2023}), \eprint{2307.14053}.

\bibitem[{\citenamefont{de~Boer et~al.}(2024)\citenamefont{de~Boer, Klasen, and
  Zeinstra}}]{deBoer:2023phz}
\bibinfo{author}{\bibfnamefont{T.}~\bibnamefont{de~Boer}},
  \bibinfo{author}{\bibfnamefont{M.}~\bibnamefont{Klasen}}, \bibnamefont{and}
  \bibinfo{author}{\bibfnamefont{S.}~\bibnamefont{Zeinstra}},
  \bibinfo{journal}{JHEP} \textbf{\bibinfo{volume}{01}}, \bibinfo{pages}{013}
  (\bibinfo{year}{2024}), \eprint{2309.06920}.

\bibitem[{\citenamefont{Figueroa et~al.}(2024)\citenamefont{Figueroa, Herrera,
  and Ochoa}}]{Figueroa:2024tmn}
\bibinfo{author}{\bibfnamefont{P.}~\bibnamefont{Figueroa}},
  \bibinfo{author}{\bibfnamefont{G.}~\bibnamefont{Herrera}}, \bibnamefont{and}
  \bibinfo{author}{\bibfnamefont{F.}~\bibnamefont{Ochoa}},
  \bibinfo{journal}{Phys. Rev. D} \textbf{\bibinfo{volume}{110}},
  \bibinfo{pages}{095018} (\bibinfo{year}{2024}), \eprint{2404.03090}.

\bibitem[{\citenamefont{Babu et~al.}(2025)\citenamefont{Babu, Chakdar, and
  P.~K.}}]{Babu:2024zoe}
\bibinfo{author}{\bibfnamefont{K.~S.} \bibnamefont{Babu}},
  \bibinfo{author}{\bibfnamefont{S.}~\bibnamefont{Chakdar}}, \bibnamefont{and}
  \bibinfo{author}{\bibfnamefont{V.}~\bibnamefont{P.~K.}},
  \bibinfo{journal}{JCAP} \textbf{\bibinfo{volume}{02}}, \bibinfo{pages}{010}
  (\bibinfo{year}{2025}), \eprint{2409.09008}.

\bibitem[{\citenamefont{He et~al.}(2025)\citenamefont{He, Nomura, and
  Yokozaki}}]{He:2025uvm}
\bibinfo{author}{\bibfnamefont{X.}~\bibnamefont{He}},
  \bibinfo{author}{\bibfnamefont{T.}~\bibnamefont{Nomura}}, \bibnamefont{and}
  \bibinfo{author}{\bibfnamefont{N.}~\bibnamefont{Yokozaki}}
  (\bibinfo{year}{2025}), \eprint{2506.04718}.

\bibitem[{\citenamefont{Abdelrahim et~al.}(2025)\citenamefont{Abdelrahim,
  Batell, Berger, McKeen, and Shams Es~Haghi}}]{Abdelrahim:2025fiz}
\bibinfo{author}{\bibfnamefont{A.~E.~B.} \bibnamefont{Abdelrahim}},
  \bibinfo{author}{\bibfnamefont{B.}~\bibnamefont{Batell}},
  \bibinfo{author}{\bibfnamefont{J.}~\bibnamefont{Berger}},
  \bibinfo{author}{\bibfnamefont{D.}~\bibnamefont{McKeen}}, \bibnamefont{and}
  \bibinfo{author}{\bibfnamefont{B.}~\bibnamefont{Shams Es~Haghi}}
  (\bibinfo{year}{2025}), \eprint{2506.09137}.

\bibitem[{\citenamefont{Hirsch et~al.}(2010)\citenamefont{Hirsch, Morisi,
  Peinado, and Valle}}]{Hirsch:2010ru}
\bibinfo{author}{\bibfnamefont{M.}~\bibnamefont{Hirsch}},
  \bibinfo{author}{\bibfnamefont{S.}~\bibnamefont{Morisi}},
  \bibinfo{author}{\bibfnamefont{E.}~\bibnamefont{Peinado}}, \bibnamefont{and}
  \bibinfo{author}{\bibfnamefont{J.~W.~F.} \bibnamefont{Valle}},
  \bibinfo{journal}{Phys. Rev. D} \textbf{\bibinfo{volume}{82}},
  \bibinfo{pages}{116003} (\bibinfo{year}{2010}), \eprint{1007.0871}.

\bibitem[{\citenamefont{Meloni et~al.}(2011)\citenamefont{Meloni, Morisi, and
  Peinado}}]{Meloni:2010sk}
\bibinfo{author}{\bibfnamefont{D.}~\bibnamefont{Meloni}},
  \bibinfo{author}{\bibfnamefont{S.}~\bibnamefont{Morisi}}, \bibnamefont{and}
  \bibinfo{author}{\bibfnamefont{E.}~\bibnamefont{Peinado}},
  \bibinfo{journal}{Phys. Lett. B} \textbf{\bibinfo{volume}{697}},
  \bibinfo{pages}{339} (\bibinfo{year}{2011}), \eprint{1011.1371}.

\bibitem[{\citenamefont{Boucenna et~al.}(2011)\citenamefont{Boucenna, Hirsch,
  Morisi, Peinado, Taoso, and Valle}}]{Boucenna:2011tj}
\bibinfo{author}{\bibfnamefont{M.~S.} \bibnamefont{Boucenna}},
  \bibinfo{author}{\bibfnamefont{M.}~\bibnamefont{Hirsch}},
  \bibinfo{author}{\bibfnamefont{S.}~\bibnamefont{Morisi}},
  \bibinfo{author}{\bibfnamefont{E.}~\bibnamefont{Peinado}},
  \bibinfo{author}{\bibfnamefont{M.}~\bibnamefont{Taoso}}, \bibnamefont{and}
  \bibinfo{author}{\bibfnamefont{J.~W.~F.} \bibnamefont{Valle}},
  \bibinfo{journal}{JHEP} \textbf{\bibinfo{volume}{05}}, \bibinfo{pages}{037}
  (\bibinfo{year}{2011}), \eprint{1101.2874}.

\bibitem[{\citenamefont{Hamada et~al.}(2014)\citenamefont{Hamada, Kobayashi,
  Ogasahara, Omura, Takayama, and Yasuhara}}]{Hamada:2014xha}
\bibinfo{author}{\bibfnamefont{Y.}~\bibnamefont{Hamada}},
  \bibinfo{author}{\bibfnamefont{T.}~\bibnamefont{Kobayashi}},
  \bibinfo{author}{\bibfnamefont{A.}~\bibnamefont{Ogasahara}},
  \bibinfo{author}{\bibfnamefont{Y.}~\bibnamefont{Omura}},
  \bibinfo{author}{\bibfnamefont{F.}~\bibnamefont{Takayama}}, \bibnamefont{and}
  \bibinfo{author}{\bibfnamefont{D.}~\bibnamefont{Yasuhara}},
  \bibinfo{journal}{JHEP} \textbf{\bibinfo{volume}{10}}, \bibinfo{pages}{183}
  (\bibinfo{year}{2014}), \eprint{1405.3592}.

\bibitem[{\citenamefont{de~Medeiros~Varzielas
  et~al.}(2015)\citenamefont{de~Medeiros~Varzielas, Fischer, and
  Maurer}}]{deMedeirosVarzielas:2015ybd}
\bibinfo{author}{\bibfnamefont{I.}~\bibnamefont{de~Medeiros~Varzielas}},
  \bibinfo{author}{\bibfnamefont{O.}~\bibnamefont{Fischer}}, \bibnamefont{and}
  \bibinfo{author}{\bibfnamefont{V.}~\bibnamefont{Maurer}},
  \bibinfo{journal}{JHEP} \textbf{\bibinfo{volume}{08}}, \bibinfo{pages}{080}
  (\bibinfo{year}{2015}), \eprint{1504.03955}.

\bibitem[{\citenamefont{de~Medeiros~Varzielas and
  Fischer}(2016)}]{deMedeirosVarzielas:2015lmh}
\bibinfo{author}{\bibfnamefont{I.}~\bibnamefont{de~Medeiros~Varzielas}}
  \bibnamefont{and} \bibinfo{author}{\bibfnamefont{O.}~\bibnamefont{Fischer}},
  \bibinfo{journal}{JHEP} \textbf{\bibinfo{volume}{01}}, \bibinfo{pages}{160}
  (\bibinfo{year}{2016}), \eprint{1512.00869}.

\bibitem[{\citenamefont{Lamprea and Peinado}(2016)}]{Lamprea:2016egz}
\bibinfo{author}{\bibfnamefont{J.~M.} \bibnamefont{Lamprea}} \bibnamefont{and}
  \bibinfo{author}{\bibfnamefont{E.}~\bibnamefont{Peinado}},
  \bibinfo{journal}{Phys. Rev. D} \textbf{\bibinfo{volume}{94}},
  \bibinfo{pages}{055007} (\bibinfo{year}{2016}), \eprint{1603.02190}.

\bibitem[{\citenamefont{De~La~Vega et~al.}(2019)\citenamefont{De~La~Vega,
  Ferro-Hernandez, and Peinado}}]{DeLaVega:2018bkp}
\bibinfo{author}{\bibfnamefont{L.~M.~G.} \bibnamefont{De~La~Vega}},
  \bibinfo{author}{\bibfnamefont{R.}~\bibnamefont{Ferro-Hernandez}},
  \bibnamefont{and} \bibinfo{author}{\bibfnamefont{E.}~\bibnamefont{Peinado}},
  \bibinfo{journal}{Phys. Rev. D} \textbf{\bibinfo{volume}{99}},
  \bibinfo{pages}{055044} (\bibinfo{year}{2019}), \eprint{1811.10619}.

\bibitem[{\citenamefont{Nomura and Okada}(2019{\natexlab{b}})}]{Nomura:2019jxj}
\bibinfo{author}{\bibfnamefont{T.}~\bibnamefont{Nomura}} \bibnamefont{and}
  \bibinfo{author}{\bibfnamefont{H.}~\bibnamefont{Okada}},
  \bibinfo{journal}{Phys. Lett. B} \textbf{\bibinfo{volume}{797}},
  \bibinfo{pages}{134799} (\bibinfo{year}{2019}{\natexlab{b}}),
  \eprint{1904.03937}.

\bibitem[{\citenamefont{Nomura and Okada}(2021)}]{Nomura:2019yft}
\bibinfo{author}{\bibfnamefont{T.}~\bibnamefont{Nomura}} \bibnamefont{and}
  \bibinfo{author}{\bibfnamefont{H.}~\bibnamefont{Okada}},
  \bibinfo{journal}{Nucl. Phys. B} \textbf{\bibinfo{volume}{966}},
  \bibinfo{pages}{115372} (\bibinfo{year}{2021}), \eprint{1906.03927}.

\bibitem[{\citenamefont{Nomura et~al.}(2020)\citenamefont{Nomura, Okada, and
  Popov}}]{Nomura:2019lnr}
\bibinfo{author}{\bibfnamefont{T.}~\bibnamefont{Nomura}},
  \bibinfo{author}{\bibfnamefont{H.}~\bibnamefont{Okada}}, \bibnamefont{and}
  \bibinfo{author}{\bibfnamefont{O.}~\bibnamefont{Popov}},
  \bibinfo{journal}{Phys. Lett. B} \textbf{\bibinfo{volume}{803}},
  \bibinfo{pages}{135294} (\bibinfo{year}{2020}), \eprint{1908.07457}.

\bibitem[{\citenamefont{Okada and Shoji}(2020)}]{Okada:2020dmb}
\bibinfo{author}{\bibfnamefont{H.}~\bibnamefont{Okada}} \bibnamefont{and}
  \bibinfo{author}{\bibfnamefont{Y.}~\bibnamefont{Shoji}},
  \bibinfo{journal}{Nucl. Phys. B} \textbf{\bibinfo{volume}{961}},
  \bibinfo{pages}{115216} (\bibinfo{year}{2020}), \eprint{2003.13219}.

\bibitem[{\citenamefont{Behera et~al.}(2022)\citenamefont{Behera, Singirala,
  Mishra, and Mohanta}}]{Behera:2020lpd}
\bibinfo{author}{\bibfnamefont{M.~K.} \bibnamefont{Behera}},
  \bibinfo{author}{\bibfnamefont{S.}~\bibnamefont{Singirala}},
  \bibinfo{author}{\bibfnamefont{S.}~\bibnamefont{Mishra}}, \bibnamefont{and}
  \bibinfo{author}{\bibfnamefont{R.}~\bibnamefont{Mohanta}},
  \bibinfo{journal}{J. Phys. G} \textbf{\bibinfo{volume}{49}},
  \bibinfo{pages}{035002} (\bibinfo{year}{2022}), \eprint{2009.01806}.

\bibitem[{\citenamefont{Hutauruk et~al.}(2020)\citenamefont{Hutauruk, Kang,
  Kim, and Okada}}]{Hutauruk:2020xtk}
\bibinfo{author}{\bibfnamefont{P.~T.~P.} \bibnamefont{Hutauruk}},
  \bibinfo{author}{\bibfnamefont{D.~W.} \bibnamefont{Kang}},
  \bibinfo{author}{\bibfnamefont{J.}~\bibnamefont{Kim}}, \bibnamefont{and}
  \bibinfo{author}{\bibfnamefont{H.}~\bibnamefont{Okada}}
  (\bibinfo{year}{2020}), \eprint{2012.11156}.

\bibitem[{\citenamefont{Nagao and Okada}(2022)}]{Nagao:2021rio}
\bibinfo{author}{\bibfnamefont{K.~I.} \bibnamefont{Nagao}} \bibnamefont{and}
  \bibinfo{author}{\bibfnamefont{H.}~\bibnamefont{Okada}},
  \bibinfo{journal}{Phys. Dark Univ.} \textbf{\bibinfo{volume}{36}},
  \bibinfo{pages}{101039} (\bibinfo{year}{2022}), \eprint{2108.09984}.

\bibitem[{\citenamefont{Kobayashi et~al.}(2022)\citenamefont{Kobayashi, Okada,
  and Orikasa}}]{Kobayashi:2021ajl}
\bibinfo{author}{\bibfnamefont{T.}~\bibnamefont{Kobayashi}},
  \bibinfo{author}{\bibfnamefont{H.}~\bibnamefont{Okada}}, \bibnamefont{and}
  \bibinfo{author}{\bibfnamefont{Y.}~\bibnamefont{Orikasa}},
  \bibinfo{journal}{Phys. Dark Univ.} \textbf{\bibinfo{volume}{37}},
  \bibinfo{pages}{101080} (\bibinfo{year}{2022}), \eprint{2111.05674}.

\bibitem[{\citenamefont{Dasgupta et~al.}(2021)\citenamefont{Dasgupta, Nomura,
  Okada, Popov, and Tanimoto}}]{Dasgupta:2021ggp}
\bibinfo{author}{\bibfnamefont{A.}~\bibnamefont{Dasgupta}},
  \bibinfo{author}{\bibfnamefont{T.}~\bibnamefont{Nomura}},
  \bibinfo{author}{\bibfnamefont{H.}~\bibnamefont{Okada}},
  \bibinfo{author}{\bibfnamefont{O.}~\bibnamefont{Popov}}, \bibnamefont{and}
  \bibinfo{author}{\bibfnamefont{M.}~\bibnamefont{Tanimoto}}
  (\bibinfo{year}{2021}), \eprint{2111.06898}.

\bibitem[{\citenamefont{Otsuka and Okada}(2022)}]{Otsuka:2022rak}
\bibinfo{author}{\bibfnamefont{H.}~\bibnamefont{Otsuka}} \bibnamefont{and}
  \bibinfo{author}{\bibfnamefont{H.}~\bibnamefont{Okada}}
  (\bibinfo{year}{2022}), \eprint{2202.10089}.

\bibitem[{\citenamefont{Kang et~al.}(2022)\citenamefont{Kang, Kim, Nomura, and
  Okada}}]{Kang:2022psa}
\bibinfo{author}{\bibfnamefont{D.~W.} \bibnamefont{Kang}},
  \bibinfo{author}{\bibfnamefont{J.}~\bibnamefont{Kim}},
  \bibinfo{author}{\bibfnamefont{T.}~\bibnamefont{Nomura}}, \bibnamefont{and}
  \bibinfo{author}{\bibfnamefont{H.}~\bibnamefont{Okada}},
  \bibinfo{journal}{JHEP} \textbf{\bibinfo{volume}{07}}, \bibinfo{pages}{050}
  (\bibinfo{year}{2022}), \eprint{2205.08269}.

\bibitem[{\citenamefont{Bonilla et~al.}(2023)\citenamefont{Bonilla, Herms,
  Medina, and Peinado}}]{Bonilla:2023pna}
\bibinfo{author}{\bibfnamefont{C.}~\bibnamefont{Bonilla}},
  \bibinfo{author}{\bibfnamefont{J.}~\bibnamefont{Herms}},
  \bibinfo{author}{\bibfnamefont{O.}~\bibnamefont{Medina}}, \bibnamefont{and}
  \bibinfo{author}{\bibfnamefont{E.}~\bibnamefont{Peinado}},
  \bibinfo{journal}{JHEP} \textbf{\bibinfo{volume}{06}}, \bibinfo{pages}{078}
  (\bibinfo{year}{2023}), \eprint{2301.10811}.

\bibitem[{\citenamefont{Kim and Okada}(2023)}]{Kim:2023jto}
\bibinfo{author}{\bibfnamefont{J.}~\bibnamefont{Kim}} \bibnamefont{and}
  \bibinfo{author}{\bibfnamefont{H.}~\bibnamefont{Okada}}
  (\bibinfo{year}{2023}), \eprint{2302.09747}.

\bibitem[{\citenamefont{Nomura and Okada}(2025{\natexlab{a}})}]{Nomura:2023kwz}
\bibinfo{author}{\bibfnamefont{T.}~\bibnamefont{Nomura}} \bibnamefont{and}
  \bibinfo{author}{\bibfnamefont{H.}~\bibnamefont{Okada}},
  \bibinfo{journal}{Phys. Dark Univ.} \textbf{\bibinfo{volume}{49}},
  \bibinfo{pages}{101986} (\bibinfo{year}{2025}{\natexlab{a}}),
  \eprint{2304.13361}.

\bibitem[{\citenamefont{Kumar et~al.}(2024)\citenamefont{Kumar, Nath, and
  Srivastava}}]{Kumar:2024zfb}
\bibinfo{author}{\bibfnamefont{R.}~\bibnamefont{Kumar}},
  \bibinfo{author}{\bibfnamefont{N.}~\bibnamefont{Nath}}, \bibnamefont{and}
  \bibinfo{author}{\bibfnamefont{R.}~\bibnamefont{Srivastava}},
  \bibinfo{journal}{JHEP} \textbf{\bibinfo{volume}{12}}, \bibinfo{pages}{036}
  (\bibinfo{year}{2024}), \eprint{2406.00188}.

\bibitem[{\citenamefont{Borah et~al.}(2025)\citenamefont{Borah, Das, Karmakar,
  and Mahapatra}}]{Borah:2024gql}
\bibinfo{author}{\bibfnamefont{D.}~\bibnamefont{Borah}},
  \bibinfo{author}{\bibfnamefont{P.}~\bibnamefont{Das}},
  \bibinfo{author}{\bibfnamefont{B.}~\bibnamefont{Karmakar}}, \bibnamefont{and}
  \bibinfo{author}{\bibfnamefont{S.}~\bibnamefont{Mahapatra}},
  \bibinfo{journal}{Phys. Rev. D} \textbf{\bibinfo{volume}{111}},
  \bibinfo{pages}{035032} (\bibinfo{year}{2025}), \eprint{2406.17861}.

\bibitem[{\citenamefont{Kumar et~al.}(2025)\citenamefont{Kumar, Prajapati,
  Srivastava, and Yadav}}]{Kumar:2025zvv}
\bibinfo{author}{\bibfnamefont{R.}~\bibnamefont{Kumar}},
  \bibinfo{author}{\bibfnamefont{H.~K.} \bibnamefont{Prajapati}},
  \bibinfo{author}{\bibfnamefont{R.}~\bibnamefont{Srivastava}},
  \bibnamefont{and} \bibinfo{author}{\bibfnamefont{S.}~\bibnamefont{Yadav}},
  \bibinfo{journal}{JHEP} \textbf{\bibinfo{volume}{11}}, \bibinfo{pages}{094}
  (\bibinfo{year}{2025}), \eprint{2510.02972}.

\bibitem[{\citenamefont{Hambye}(2009)}]{Hambye:2008bq}
\bibinfo{author}{\bibfnamefont{T.}~\bibnamefont{Hambye}},
  \bibinfo{journal}{JHEP} \textbf{\bibinfo{volume}{01}}, \bibinfo{pages}{028}
  (\bibinfo{year}{2009}), \eprint{0811.0172}.

\bibitem[{\citenamefont{Walker}(2009)}]{Walker:2009en}
\bibinfo{author}{\bibfnamefont{D.~G.~E.} \bibnamefont{Walker}}
  (\bibinfo{year}{2009}), \eprint{0907.3146}.

\bibitem[{\citenamefont{Diaz-Cruz and Ma}(2011)}]{Diaz-Cruz:2010czr}
\bibinfo{author}{\bibfnamefont{J.~L.} \bibnamefont{Diaz-Cruz}}
  \bibnamefont{and} \bibinfo{author}{\bibfnamefont{E.}~\bibnamefont{Ma}},
  \bibinfo{journal}{Phys. Lett. B} \textbf{\bibinfo{volume}{695}},
  \bibinfo{pages}{264} (\bibinfo{year}{2011}), \eprint{1007.2631}.

\bibitem[{\citenamefont{Chiang et~al.}(2014)\citenamefont{Chiang, Nomura, and
  Tandean}}]{Chiang:2013kqa}
\bibinfo{author}{\bibfnamefont{C.-W.} \bibnamefont{Chiang}},
  \bibinfo{author}{\bibfnamefont{T.}~\bibnamefont{Nomura}}, \bibnamefont{and}
  \bibinfo{author}{\bibfnamefont{J.}~\bibnamefont{Tandean}},
  \bibinfo{journal}{JHEP} \textbf{\bibinfo{volume}{01}}, \bibinfo{pages}{183}
  (\bibinfo{year}{2014}), \eprint{1306.0882}.

\bibitem[{\citenamefont{Baek et~al.}(2014)\citenamefont{Baek, Ko, and
  Park}}]{Baek:2013dwa}
\bibinfo{author}{\bibfnamefont{S.}~\bibnamefont{Baek}},
  \bibinfo{author}{\bibfnamefont{P.}~\bibnamefont{Ko}}, \bibnamefont{and}
  \bibinfo{author}{\bibfnamefont{W.-I.} \bibnamefont{Park}},
  \bibinfo{journal}{JCAP} \textbf{\bibinfo{volume}{10}}, \bibinfo{pages}{067}
  (\bibinfo{year}{2014}), \eprint{1311.1035}.

\bibitem[{\citenamefont{Khoze and Ro}(2014)}]{Khoze:2014woa}
\bibinfo{author}{\bibfnamefont{V.~V.} \bibnamefont{Khoze}} \bibnamefont{and}
  \bibinfo{author}{\bibfnamefont{G.}~\bibnamefont{Ro}}, \bibinfo{journal}{JHEP}
  \textbf{\bibinfo{volume}{10}}, \bibinfo{pages}{061} (\bibinfo{year}{2014}),
  \eprint{1406.2291}.

\bibitem[{\citenamefont{Chen and Nomura}(2015)}]{Chen:2015nea}
\bibinfo{author}{\bibfnamefont{C.-H.} \bibnamefont{Chen}} \bibnamefont{and}
  \bibinfo{author}{\bibfnamefont{T.}~\bibnamefont{Nomura}},
  \bibinfo{journal}{Phys. Lett. B} \textbf{\bibinfo{volume}{746}},
  \bibinfo{pages}{351} (\bibinfo{year}{2015}), \eprint{1501.07413}.

\bibitem[{\citenamefont{Chen et~al.}(2015)\citenamefont{Chen, Chiang, and
  Nomura}}]{Chen:2015cqa}
\bibinfo{author}{\bibfnamefont{C.-H.} \bibnamefont{Chen}},
  \bibinfo{author}{\bibfnamefont{C.-W.} \bibnamefont{Chiang}},
  \bibnamefont{and} \bibinfo{author}{\bibfnamefont{T.}~\bibnamefont{Nomura}},
  \bibinfo{journal}{Phys. Lett. B} \textbf{\bibinfo{volume}{747}},
  \bibinfo{pages}{495} (\bibinfo{year}{2015}), \eprint{1504.07848}.

\bibitem[{\citenamefont{Gross et~al.}(2015)\citenamefont{Gross, Lebedev, and
  Mambrini}}]{Gross:2015cwa}
\bibinfo{author}{\bibfnamefont{C.}~\bibnamefont{Gross}},
  \bibinfo{author}{\bibfnamefont{O.}~\bibnamefont{Lebedev}}, \bibnamefont{and}
  \bibinfo{author}{\bibfnamefont{Y.}~\bibnamefont{Mambrini}},
  \bibinfo{journal}{JHEP} \textbf{\bibinfo{volume}{08}}, \bibinfo{pages}{158}
  (\bibinfo{year}{2015}), \eprint{1505.07480}.

\bibitem[{\citenamefont{Chen and Nomura}(2016)}]{Chen:2015dea}
\bibinfo{author}{\bibfnamefont{C.-H.} \bibnamefont{Chen}} \bibnamefont{and}
  \bibinfo{author}{\bibfnamefont{T.}~\bibnamefont{Nomura}},
  \bibinfo{journal}{Phys. Rev. D} \textbf{\bibinfo{volume}{93}},
  \bibinfo{pages}{074019} (\bibinfo{year}{2016}), \eprint{1507.00886}.

\bibitem[{\citenamefont{Karam and Tamvakis}(2015)}]{Karam:2015jta}
\bibinfo{author}{\bibfnamefont{A.}~\bibnamefont{Karam}} \bibnamefont{and}
  \bibinfo{author}{\bibfnamefont{K.}~\bibnamefont{Tamvakis}},
  \bibinfo{journal}{Phys. Rev. D} \textbf{\bibinfo{volume}{92}},
  \bibinfo{pages}{075010} (\bibinfo{year}{2015}), \eprint{1508.03031}.

\bibitem[{\citenamefont{Ko}(2016)}]{Ko:2016yfb}
\bibinfo{author}{\bibfnamefont{P.}~\bibnamefont{Ko}}, \bibinfo{journal}{New
  Phys. Sae Mulli} \textbf{\bibinfo{volume}{66}}, \bibinfo{pages}{966}
  (\bibinfo{year}{2016}).

\bibitem[{\citenamefont{Choi et~al.}(2017)\citenamefont{Choi, Hochberg, Kuflik,
  Lee, Mambrini, Murayama, and Pierre}}]{Choi:2017zww}
\bibinfo{author}{\bibfnamefont{S.-M.} \bibnamefont{Choi}},
  \bibinfo{author}{\bibfnamefont{Y.}~\bibnamefont{Hochberg}},
  \bibinfo{author}{\bibfnamefont{E.}~\bibnamefont{Kuflik}},
  \bibinfo{author}{\bibfnamefont{H.~M.} \bibnamefont{Lee}},
  \bibinfo{author}{\bibfnamefont{Y.}~\bibnamefont{Mambrini}},
  \bibinfo{author}{\bibfnamefont{H.}~\bibnamefont{Murayama}}, \bibnamefont{and}
  \bibinfo{author}{\bibfnamefont{M.}~\bibnamefont{Pierre}},
  \bibinfo{journal}{JHEP} \textbf{\bibinfo{volume}{10}}, \bibinfo{pages}{162}
  (\bibinfo{year}{2017}), \eprint{1707.01434}.

\bibitem[{\citenamefont{Barman et~al.}(2018)\citenamefont{Barman, Bhattacharya,
  and Zakeri}}]{Barman:2018esi}
\bibinfo{author}{\bibfnamefont{B.}~\bibnamefont{Barman}},
  \bibinfo{author}{\bibfnamefont{S.}~\bibnamefont{Bhattacharya}},
  \bibnamefont{and} \bibinfo{author}{\bibfnamefont{M.}~\bibnamefont{Zakeri}},
  \bibinfo{journal}{JCAP} \textbf{\bibinfo{volume}{09}}, \bibinfo{pages}{023}
  (\bibinfo{year}{2018}), \eprint{1806.01129}.

\bibitem[{\citenamefont{Choi et~al.}(2019)\citenamefont{Choi, Lee, Mambrini,
  and Pierre}}]{Choi:2019zeb}
\bibinfo{author}{\bibfnamefont{S.-M.} \bibnamefont{Choi}},
  \bibinfo{author}{\bibfnamefont{H.~M.} \bibnamefont{Lee}},
  \bibinfo{author}{\bibfnamefont{Y.}~\bibnamefont{Mambrini}}, \bibnamefont{and}
  \bibinfo{author}{\bibfnamefont{M.}~\bibnamefont{Pierre}},
  \bibinfo{journal}{JHEP} \textbf{\bibinfo{volume}{07}}, \bibinfo{pages}{049}
  (\bibinfo{year}{2019}), \eprint{1904.04109}.

\bibitem[{\citenamefont{Barman et~al.}(2020{\natexlab{b}})\citenamefont{Barman,
  Bhattacharya, and Zakeri}}]{Barman:2019lvm}
\bibinfo{author}{\bibfnamefont{B.}~\bibnamefont{Barman}},
  \bibinfo{author}{\bibfnamefont{S.}~\bibnamefont{Bhattacharya}},
  \bibnamefont{and} \bibinfo{author}{\bibfnamefont{M.}~\bibnamefont{Zakeri}},
  \bibinfo{journal}{JCAP} \textbf{\bibinfo{volume}{02}}, \bibinfo{pages}{029}
  (\bibinfo{year}{2020}{\natexlab{b}}), \eprint{1905.07236}.

\bibitem[{\citenamefont{Abe et~al.}(2020)\citenamefont{Abe, Fujiwara, Hisano,
  and Matsushita}}]{Abe:2020mph}
\bibinfo{author}{\bibfnamefont{T.}~\bibnamefont{Abe}},
  \bibinfo{author}{\bibfnamefont{M.}~\bibnamefont{Fujiwara}},
  \bibinfo{author}{\bibfnamefont{J.}~\bibnamefont{Hisano}}, \bibnamefont{and}
  \bibinfo{author}{\bibfnamefont{K.}~\bibnamefont{Matsushita}},
  \bibinfo{journal}{JHEP} \textbf{\bibinfo{volume}{07}}, \bibinfo{pages}{136}
  (\bibinfo{year}{2020}), \eprint{2004.00884}.

\bibitem[{\citenamefont{Ko et~al.}(2021)\citenamefont{Ko, Nomura, and
  Okada}}]{Ko:2020qlt}
\bibinfo{author}{\bibfnamefont{P.}~\bibnamefont{Ko}},
  \bibinfo{author}{\bibfnamefont{T.}~\bibnamefont{Nomura}}, \bibnamefont{and}
  \bibinfo{author}{\bibfnamefont{H.}~\bibnamefont{Okada}},
  \bibinfo{journal}{Phys. Rev. D} \textbf{\bibinfo{volume}{103}},
  \bibinfo{pages}{095011} (\bibinfo{year}{2021}), \eprint{2007.08153}.

\bibitem[{\citenamefont{Ghosh et~al.}(2021)\citenamefont{Ghosh, Guo, Han, and
  Liu}}]{Ghosh:2020ipy}
\bibinfo{author}{\bibfnamefont{T.}~\bibnamefont{Ghosh}},
  \bibinfo{author}{\bibfnamefont{H.-K.} \bibnamefont{Guo}},
  \bibinfo{author}{\bibfnamefont{T.}~\bibnamefont{Han}}, \bibnamefont{and}
  \bibinfo{author}{\bibfnamefont{H.}~\bibnamefont{Liu}},
  \bibinfo{journal}{JHEP} \textbf{\bibinfo{volume}{07}}, \bibinfo{pages}{045}
  (\bibinfo{year}{2021}), \eprint{2012.09758}.

\bibitem[{\citenamefont{Nomura et~al.}(2021)\citenamefont{Nomura, Okada, and
  Yun}}]{Nomura:2020zlm}
\bibinfo{author}{\bibfnamefont{T.}~\bibnamefont{Nomura}},
  \bibinfo{author}{\bibfnamefont{H.}~\bibnamefont{Okada}}, \bibnamefont{and}
  \bibinfo{author}{\bibfnamefont{S.}~\bibnamefont{Yun}},
  \bibinfo{journal}{JHEP} \textbf{\bibinfo{volume}{06}}, \bibinfo{pages}{122}
  (\bibinfo{year}{2021}), \eprint{2012.11377}.

\bibitem[{\citenamefont{Baouche et~al.}(2021)\citenamefont{Baouche, Ahriche,
  Faisel, and Nasri}}]{Baouche:2021wwa}
\bibinfo{author}{\bibfnamefont{N.}~\bibnamefont{Baouche}},
  \bibinfo{author}{\bibfnamefont{A.}~\bibnamefont{Ahriche}},
  \bibinfo{author}{\bibfnamefont{G.}~\bibnamefont{Faisel}}, \bibnamefont{and}
  \bibinfo{author}{\bibfnamefont{S.}~\bibnamefont{Nasri}},
  \bibinfo{journal}{Phys. Rev. D} \textbf{\bibinfo{volume}{104}},
  \bibinfo{pages}{075022} (\bibinfo{year}{2021}), \eprint{2105.14387}.

\bibitem[{\citenamefont{Chowdhury and Saad}(2021)}]{Chowdhury:2021tnm}
\bibinfo{author}{\bibfnamefont{T.~A.} \bibnamefont{Chowdhury}}
  \bibnamefont{and} \bibinfo{author}{\bibfnamefont{S.}~\bibnamefont{Saad}},
  \bibinfo{journal}{JCAP} \textbf{\bibinfo{volume}{10}}, \bibinfo{pages}{014}
  (\bibinfo{year}{2021}), \eprint{2107.11863}.

\bibitem[{\citenamefont{Borah et~al.}(2022)\citenamefont{Borah, Ma, and
  Nanda}}]{Borah:2022phw}
\bibinfo{author}{\bibfnamefont{D.}~\bibnamefont{Borah}},
  \bibinfo{author}{\bibfnamefont{E.}~\bibnamefont{Ma}}, \bibnamefont{and}
  \bibinfo{author}{\bibfnamefont{D.}~\bibnamefont{Nanda}},
  \bibinfo{journal}{Phys. Lett. B} \textbf{\bibinfo{volume}{835}},
  \bibinfo{pages}{137539} (\bibinfo{year}{2022}), \eprint{2204.13205}.

\bibitem[{\citenamefont{Borah et~al.}(2023)\citenamefont{Borah, Ma, and
  Nanda}}]{Borah:2022dbw}
\bibinfo{author}{\bibfnamefont{D.}~\bibnamefont{Borah}},
  \bibinfo{author}{\bibfnamefont{E.}~\bibnamefont{Ma}}, \bibnamefont{and}
  \bibinfo{author}{\bibfnamefont{D.}~\bibnamefont{Nanda}},
  \bibinfo{journal}{Phys. Lett. B} \textbf{\bibinfo{volume}{842}},
  \bibinfo{pages}{137981} (\bibinfo{year}{2023}), \eprint{2212.11847}.

\bibitem[{\citenamefont{Otsuka et~al.}(2022)\citenamefont{Otsuka, Shimomura,
  Tsumura, Uchida, and Yamatsu}}]{Otsuka:2022zdy}
\bibinfo{author}{\bibfnamefont{H.}~\bibnamefont{Otsuka}},
  \bibinfo{author}{\bibfnamefont{T.}~\bibnamefont{Shimomura}},
  \bibinfo{author}{\bibfnamefont{K.}~\bibnamefont{Tsumura}},
  \bibinfo{author}{\bibfnamefont{Y.}~\bibnamefont{Uchida}}, \bibnamefont{and}
  \bibinfo{author}{\bibfnamefont{N.}~\bibnamefont{Yamatsu}},
  \bibinfo{journal}{Phys. Rev. D} \textbf{\bibinfo{volume}{106}},
  \bibinfo{pages}{115033} (\bibinfo{year}{2022}), \eprint{2210.08696}.

\bibitem[{\citenamefont{Frigerio et~al.}(2023)\citenamefont{Frigerio,
  Grimbaum-Yamamoto, and Hambye}}]{Frigerio:2022kyu}
\bibinfo{author}{\bibfnamefont{M.}~\bibnamefont{Frigerio}},
  \bibinfo{author}{\bibfnamefont{N.}~\bibnamefont{Grimbaum-Yamamoto}},
  \bibnamefont{and} \bibinfo{author}{\bibfnamefont{T.}~\bibnamefont{Hambye}},
  \bibinfo{journal}{SciPost Phys.} \textbf{\bibinfo{volume}{15}},
  \bibinfo{pages}{177} (\bibinfo{year}{2023}), \eprint{2212.11918}.

\bibitem[{\citenamefont{Coleppa et~al.}(2024)\citenamefont{Coleppa, Loho, and
  Sarkar}}]{Coleppa:2023vfh}
\bibinfo{author}{\bibfnamefont{B.}~\bibnamefont{Coleppa}},
  \bibinfo{author}{\bibfnamefont{K.}~\bibnamefont{Loho}}, \bibnamefont{and}
  \bibinfo{author}{\bibfnamefont{A.}~\bibnamefont{Sarkar}},
  \bibinfo{journal}{Eur. Phys. J. C} \textbf{\bibinfo{volume}{84}},
  \bibinfo{pages}{144} (\bibinfo{year}{2024}), \eprint{2307.14873}.

\bibitem[{\citenamefont{Chen and Jiang}(2025)}]{Chen:2025ihr}
\bibinfo{author}{\bibfnamefont{S.-L.} \bibnamefont{Chen}} \bibnamefont{and}
  \bibinfo{author}{\bibfnamefont{W.-w.} \bibnamefont{Jiang}}
  (\bibinfo{year}{2025}), \eprint{2512.18568}.

\bibitem[{\citenamefont{Choi et~al.}(2022)\citenamefont{Choi, Lam, and
  Shao}}]{Choi:2022jqy}
\bibinfo{author}{\bibfnamefont{Y.}~\bibnamefont{Choi}},
  \bibinfo{author}{\bibfnamefont{H.~T.} \bibnamefont{Lam}}, \bibnamefont{and}
  \bibinfo{author}{\bibfnamefont{S.-H.} \bibnamefont{Shao}},
  \bibinfo{journal}{Phys. Rev. Lett.} \textbf{\bibinfo{volume}{129}},
  \bibinfo{pages}{161601} (\bibinfo{year}{2022}), \eprint{2205.05086}.

\bibitem[{\citenamefont{Cordova and Ohmori}(2023)}]{Cordova:2022ieu}
\bibinfo{author}{\bibfnamefont{C.}~\bibnamefont{Cordova}} \bibnamefont{and}
  \bibinfo{author}{\bibfnamefont{K.}~\bibnamefont{Ohmori}},
  \bibinfo{journal}{Phys. Rev. X} \textbf{\bibinfo{volume}{13}},
  \bibinfo{pages}{011034} (\bibinfo{year}{2023}), \eprint{2205.06243}.

\bibitem[{\citenamefont{Cordova
  et~al.}(2024{\natexlab{a}})\citenamefont{Cordova, Hong, Koren, and
  Ohmori}}]{Cordova:2022fhg}
\bibinfo{author}{\bibfnamefont{C.}~\bibnamefont{Cordova}},
  \bibinfo{author}{\bibfnamefont{S.}~\bibnamefont{Hong}},
  \bibinfo{author}{\bibfnamefont{S.}~\bibnamefont{Koren}}, \bibnamefont{and}
  \bibinfo{author}{\bibfnamefont{K.}~\bibnamefont{Ohmori}},
  \bibinfo{journal}{Phys. Rev. X} \textbf{\bibinfo{volume}{14}},
  \bibinfo{pages}{031033} (\bibinfo{year}{2024}{\natexlab{a}}),
  \eprint{2211.07639}.

\bibitem[{\citenamefont{Cordova
  et~al.}(2024{\natexlab{b}})\citenamefont{Cordova, Hong, and
  Koren}}]{Cordova:2024ypu}
\bibinfo{author}{\bibfnamefont{C.}~\bibnamefont{Cordova}},
  \bibinfo{author}{\bibfnamefont{S.}~\bibnamefont{Hong}}, \bibnamefont{and}
  \bibinfo{author}{\bibfnamefont{S.}~\bibnamefont{Koren}}
  (\bibinfo{year}{2024}{\natexlab{b}}), \eprint{2402.12453}.

\bibitem[{\citenamefont{Kobayashi and Otsuka}(2024)}]{Kobayashi:2024yqq}
\bibinfo{author}{\bibfnamefont{T.}~\bibnamefont{Kobayashi}} \bibnamefont{and}
  \bibinfo{author}{\bibfnamefont{H.}~\bibnamefont{Otsuka}},
  \bibinfo{journal}{JHEP} \textbf{\bibinfo{volume}{11}}, \bibinfo{pages}{120}
  (\bibinfo{year}{2024}), \eprint{2408.13984}.

\bibitem[{\citenamefont{Kobayashi et~al.}(2024)\citenamefont{Kobayashi, Otsuka,
  and Tanimoto}}]{Kobayashi:2024cvp}
\bibinfo{author}{\bibfnamefont{T.}~\bibnamefont{Kobayashi}},
  \bibinfo{author}{\bibfnamefont{H.}~\bibnamefont{Otsuka}}, \bibnamefont{and}
  \bibinfo{author}{\bibfnamefont{M.}~\bibnamefont{Tanimoto}},
  \bibinfo{journal}{JHEP} \textbf{\bibinfo{volume}{12}}, \bibinfo{pages}{117}
  (\bibinfo{year}{2024}), \eprint{2409.05270}.

\bibitem[{\citenamefont{Kobayashi
  et~al.}(2025{\natexlab{a}})\citenamefont{Kobayashi, Nishioka, Otsuka, and
  Tanimoto}}]{Kobayashi:2025znw}
\bibinfo{author}{\bibfnamefont{T.}~\bibnamefont{Kobayashi}},
  \bibinfo{author}{\bibfnamefont{Y.}~\bibnamefont{Nishioka}},
  \bibinfo{author}{\bibfnamefont{H.}~\bibnamefont{Otsuka}}, \bibnamefont{and}
  \bibinfo{author}{\bibfnamefont{M.}~\bibnamefont{Tanimoto}},
  \bibinfo{journal}{JHEP} \textbf{\bibinfo{volume}{05}}, \bibinfo{pages}{177}
  (\bibinfo{year}{2025}{\natexlab{a}}), \eprint{2503.09966}.

\bibitem[{\citenamefont{Suzuki and Xu}(2025)}]{Suzuki:2025oov}
\bibinfo{author}{\bibfnamefont{M.}~\bibnamefont{Suzuki}} \bibnamefont{and}
  \bibinfo{author}{\bibfnamefont{L.-X.} \bibnamefont{Xu}}
  (\bibinfo{year}{2025}), \eprint{2503.19964}.

\bibitem[{\citenamefont{Liang and Yanagida}(2025)}]{Liang:2025dkm}
\bibinfo{author}{\bibfnamefont{Q.}~\bibnamefont{Liang}} \bibnamefont{and}
  \bibinfo{author}{\bibfnamefont{T.~T.} \bibnamefont{Yanagida}}
  (\bibinfo{year}{2025}), \eprint{2505.05142}.

\bibitem[{\citenamefont{Kobayashi
  et~al.}(2025{\natexlab{b}})\citenamefont{Kobayashi, Otsuka, Tanimoto, and
  Uchida}}]{Kobayashi:2025ldi}
\bibinfo{author}{\bibfnamefont{T.}~\bibnamefont{Kobayashi}},
  \bibinfo{author}{\bibfnamefont{H.}~\bibnamefont{Otsuka}},
  \bibinfo{author}{\bibfnamefont{M.}~\bibnamefont{Tanimoto}}, \bibnamefont{and}
  \bibinfo{author}{\bibfnamefont{H.}~\bibnamefont{Uchida}}
  (\bibinfo{year}{2025}{\natexlab{b}}), \eprint{2505.07262}.

\bibitem[{\citenamefont{Kobayashi
  et~al.}(2025{\natexlab{c}})\citenamefont{Kobayashi, Okada, and
  Otsuka}}]{Kobayashi:2025cwx}
\bibinfo{author}{\bibfnamefont{T.}~\bibnamefont{Kobayashi}},
  \bibinfo{author}{\bibfnamefont{H.}~\bibnamefont{Okada}}, \bibnamefont{and}
  \bibinfo{author}{\bibfnamefont{H.}~\bibnamefont{Otsuka}}
  (\bibinfo{year}{2025}{\natexlab{c}}), \eprint{2505.14878}.

\bibitem[{\citenamefont{Kobayashi
  et~al.}(2025{\natexlab{d}})\citenamefont{Kobayashi, Mita, Otsuka, and
  Sakuma}}]{Kobayashi:2025lar}
\bibinfo{author}{\bibfnamefont{T.}~\bibnamefont{Kobayashi}},
  \bibinfo{author}{\bibfnamefont{H.}~\bibnamefont{Mita}},
  \bibinfo{author}{\bibfnamefont{H.}~\bibnamefont{Otsuka}}, \bibnamefont{and}
  \bibinfo{author}{\bibfnamefont{R.}~\bibnamefont{Sakuma}}
  (\bibinfo{year}{2025}{\natexlab{d}}), \eprint{2506.10241}.

\bibitem[{\citenamefont{Nomura and Okada}(2025{\natexlab{b}})}]{Nomura:2025sod}
\bibinfo{author}{\bibfnamefont{T.}~\bibnamefont{Nomura}} \bibnamefont{and}
  \bibinfo{author}{\bibfnamefont{H.}~\bibnamefont{Okada}}
  (\bibinfo{year}{2025}{\natexlab{b}}), \eprint{2506.16706}.

\bibitem[{\citenamefont{Dong et~al.}(2025)\citenamefont{Dong, Jeric, Kobayashi,
  Nishida, and Otsuka}}]{Dong:2025jra}
\bibinfo{author}{\bibfnamefont{J.}~\bibnamefont{Dong}},
  \bibinfo{author}{\bibfnamefont{T.}~\bibnamefont{Jeric}},
  \bibinfo{author}{\bibfnamefont{T.}~\bibnamefont{Kobayashi}},
  \bibinfo{author}{\bibfnamefont{R.}~\bibnamefont{Nishida}}, \bibnamefont{and}
  \bibinfo{author}{\bibfnamefont{H.}~\bibnamefont{Otsuka}}
  (\bibinfo{year}{2025}), \eprint{2507.02375}.

\bibitem[{\citenamefont{Nomura and Popov}(2025)}]{Nomura:2025yoa}
\bibinfo{author}{\bibfnamefont{T.}~\bibnamefont{Nomura}} \bibnamefont{and}
  \bibinfo{author}{\bibfnamefont{O.}~\bibnamefont{Popov}}
  (\bibinfo{year}{2025}), \eprint{2507.10299}.

\bibitem[{\citenamefont{Chen et~al.}(2025)\citenamefont{Chen, Geng, Okada, and
  Wu}}]{Chen:2025awz}
\bibinfo{author}{\bibfnamefont{J.}~\bibnamefont{Chen}},
  \bibinfo{author}{\bibfnamefont{C.-Q.} \bibnamefont{Geng}},
  \bibinfo{author}{\bibfnamefont{H.}~\bibnamefont{Okada}}, \bibnamefont{and}
  \bibinfo{author}{\bibfnamefont{J.-J.} \bibnamefont{Wu}}
  (\bibinfo{year}{2025}), \eprint{2507.11951}.

\bibitem[{\citenamefont{Okada and Shigekami}(2025)}]{Okada:2025kfm}
\bibinfo{author}{\bibfnamefont{H.}~\bibnamefont{Okada}} \bibnamefont{and}
  \bibinfo{author}{\bibfnamefont{Y.}~\bibnamefont{Shigekami}}
  (\bibinfo{year}{2025}), \eprint{2507.16198}.

\bibitem[{\citenamefont{Kobayashi
  et~al.}(2025{\natexlab{e}})\citenamefont{Kobayashi, Otsuka, and
  Yanagida}}]{Kobayashi:2025thd}
\bibinfo{author}{\bibfnamefont{T.}~\bibnamefont{Kobayashi}},
  \bibinfo{author}{\bibfnamefont{H.}~\bibnamefont{Otsuka}}, \bibnamefont{and}
  \bibinfo{author}{\bibfnamefont{T.~T.} \bibnamefont{Yanagida}}
  (\bibinfo{year}{2025}{\natexlab{e}}), \eprint{2508.12287}.

\bibitem[{\citenamefont{Suzuki et~al.}(2025)\citenamefont{Suzuki, Xu, and
  Zhang}}]{Suzuki:2025bxg}
\bibinfo{author}{\bibfnamefont{M.}~\bibnamefont{Suzuki}},
  \bibinfo{author}{\bibfnamefont{L.-X.} \bibnamefont{Xu}}, \bibnamefont{and}
  \bibinfo{author}{\bibfnamefont{H.~Y.} \bibnamefont{Zhang}}
  (\bibinfo{year}{2025}), \eprint{2508.14970}.

\bibitem[{\citenamefont{Jangid and Okada}(2025)}]{Jangid:2025krp}
\bibinfo{author}{\bibfnamefont{S.}~\bibnamefont{Jangid}} \bibnamefont{and}
  \bibinfo{author}{\bibfnamefont{H.}~\bibnamefont{Okada}}
  (\bibinfo{year}{2025}), \eprint{2508.16174}.

\bibitem[{\citenamefont{Kobayashi
  et~al.}(2025{\natexlab{f}})\citenamefont{Kobayashi, Otsuka, Tanimoto, and
  Yanagida}}]{Kobayashi:2025rpx}
\bibinfo{author}{\bibfnamefont{T.}~\bibnamefont{Kobayashi}},
  \bibinfo{author}{\bibfnamefont{H.}~\bibnamefont{Otsuka}},
  \bibinfo{author}{\bibfnamefont{M.}~\bibnamefont{Tanimoto}}, \bibnamefont{and}
  \bibinfo{author}{\bibfnamefont{T.~T.} \bibnamefont{Yanagida}}
  (\bibinfo{year}{2025}{\natexlab{f}}), \eprint{2510.01680}.

\bibitem[{\citenamefont{Jiang et~al.}(2025)\citenamefont{Jiang, Qu, and
  Ding}}]{Jiang:2025psz}
\bibinfo{author}{\bibfnamefont{Z.}~\bibnamefont{Jiang}},
  \bibinfo{author}{\bibfnamefont{B.-Y.} \bibnamefont{Qu}}, \bibnamefont{and}
  \bibinfo{author}{\bibfnamefont{G.-J.} \bibnamefont{Ding}},
  \bibinfo{journal}{Phys. Rev. D} \textbf{\bibinfo{volume}{112}},
  \bibinfo{pages}{115029} (\bibinfo{year}{2025}), \eprint{2510.07236}.

\bibitem[{\citenamefont{Nomura et~al.}(2025)\citenamefont{Nomura, Okada, and
  Shigekami}}]{Nomura:2025tvz}
\bibinfo{author}{\bibfnamefont{T.}~\bibnamefont{Nomura}},
  \bibinfo{author}{\bibfnamefont{H.}~\bibnamefont{Okada}}, \bibnamefont{and}
  \bibinfo{author}{\bibfnamefont{Y.}~\bibnamefont{Shigekami}}
  (\bibinfo{year}{2025}), \eprint{2510.17156}.

\bibitem[{\citenamefont{Okada and Shoji}(2025)}]{Okada:2025adm}
\bibinfo{author}{\bibfnamefont{H.}~\bibnamefont{Okada}} \bibnamefont{and}
  \bibinfo{author}{\bibfnamefont{Y.}~\bibnamefont{Shoji}}
  (\bibinfo{year}{2025}), \eprint{2512.20891}.

\bibitem[{\citenamefont{Nakai et~al.}(2025)\citenamefont{Nakai, Otsuka,
  Shigekami, and Zhang}}]{Nakai:2025thw}
\bibinfo{author}{\bibfnamefont{Y.}~\bibnamefont{Nakai}},
  \bibinfo{author}{\bibfnamefont{H.}~\bibnamefont{Otsuka}},
  \bibinfo{author}{\bibfnamefont{Y.}~\bibnamefont{Shigekami}},
  \bibnamefont{and} \bibinfo{author}{\bibfnamefont{Z.}~\bibnamefont{Zhang}}
  (\bibinfo{year}{2025}), \eprint{2512.21509}.

\bibitem[{\citenamefont{Catano et~al.}(2012)\citenamefont{Catano, Martinez, and
  Ochoa}}]{Catano:2012kw}
\bibinfo{author}{\bibfnamefont{M.~E.} \bibnamefont{Catano}},
  \bibinfo{author}{\bibfnamefont{R.}~\bibnamefont{Martinez}}, \bibnamefont{and}
  \bibinfo{author}{\bibfnamefont{F.}~\bibnamefont{Ochoa}},
  \bibinfo{journal}{Phys. Rev. D} \textbf{\bibinfo{volume}{86}},
  \bibinfo{pages}{073015} (\bibinfo{year}{2012}), \eprint{1206.1966}.

\bibitem[{\citenamefont{Kajiyama
  et~al.}(2013{\natexlab{b}})\citenamefont{Kajiyama, Okada, and
  Toma}}]{Kajiyama:2012xg}
\bibinfo{author}{\bibfnamefont{Y.}~\bibnamefont{Kajiyama}},
  \bibinfo{author}{\bibfnamefont{H.}~\bibnamefont{Okada}}, \bibnamefont{and}
  \bibinfo{author}{\bibfnamefont{T.}~\bibnamefont{Toma}},
  \bibinfo{journal}{Eur. Phys. J. C} \textbf{\bibinfo{volume}{73}},
  \bibinfo{pages}{2381} (\bibinfo{year}{2013}{\natexlab{b}}),
  \eprint{1210.2305}.

\bibitem[{\citenamefont{Griest and Seckel}(1991)}]{Griest:1990kh}
\bibinfo{author}{\bibfnamefont{K.}~\bibnamefont{Griest}} \bibnamefont{and}
  \bibinfo{author}{\bibfnamefont{D.}~\bibnamefont{Seckel}},
  \bibinfo{journal}{Phys. Rev. D} \textbf{\bibinfo{volume}{43}},
  \bibinfo{pages}{3191} (\bibinfo{year}{1991}).

\bibitem[{\citenamefont{D'Eramo and Thaler}(2010)}]{DEramo:2010keq}
\bibinfo{author}{\bibfnamefont{F.}~\bibnamefont{D'Eramo}} \bibnamefont{and}
  \bibinfo{author}{\bibfnamefont{J.}~\bibnamefont{Thaler}},
  \bibinfo{journal}{JHEP} \textbf{\bibinfo{volume}{06}}, \bibinfo{pages}{109}
  (\bibinfo{year}{2010}), \eprint{1003.5912}.

\bibitem[{\citenamefont{Aghanim et~al.}(2020)}]{Planck:2018vyg}
\bibinfo{author}{\bibfnamefont{N.}~\bibnamefont{Aghanim}} \bibnamefont{et~al.}
  (\bibinfo{collaboration}{Planck}), \bibinfo{journal}{Astron. Astrophys.}
  \textbf{\bibinfo{volume}{641}}, \bibinfo{pages}{A6} (\bibinfo{year}{2020}),
  \bibinfo{note}{[Erratum: Astron.Astrophys. 652, C4 (2021)]},
  \eprint{1807.06209}.

\bibitem[{\citenamefont{Adame et~al.}(2025)}]{DESI:2024hhd}
\bibinfo{author}{\bibfnamefont{A.~G.} \bibnamefont{Adame}} \bibnamefont{et~al.}
  (\bibinfo{collaboration}{DESI}), \bibinfo{journal}{JCAP}
  \textbf{\bibinfo{volume}{07}}, \bibinfo{pages}{028} (\bibinfo{year}{2025}),
  \eprint{2411.12022}.

\bibitem[{\citenamefont{Shao et~al.}(2025)\citenamefont{Shao, Givans, Dunkley,
  Madhavacheril, Qu, Farren, and Sherwin}}]{Shao:2024mag}
\bibinfo{author}{\bibfnamefont{H.}~\bibnamefont{Shao}},
  \bibinfo{author}{\bibfnamefont{J.~J.} \bibnamefont{Givans}},
  \bibinfo{author}{\bibfnamefont{J.}~\bibnamefont{Dunkley}},
  \bibinfo{author}{\bibfnamefont{M.}~\bibnamefont{Madhavacheril}},
  \bibinfo{author}{\bibfnamefont{F.~J.} \bibnamefont{Qu}},
  \bibinfo{author}{\bibfnamefont{G.}~\bibnamefont{Farren}}, \bibnamefont{and}
  \bibinfo{author}{\bibfnamefont{B.}~\bibnamefont{Sherwin}},
  \bibinfo{journal}{Phys. Rev. D} \textbf{\bibinfo{volume}{111}},
  \bibinfo{pages}{083535} (\bibinfo{year}{2025}), \eprint{2409.02295}.

\bibitem[{\citenamefont{Pang et~al.}(2024)\citenamefont{Pang, Zhang, and
  Huang}}]{Pang:2023joc}
\bibinfo{author}{\bibfnamefont{Y.-H.} \bibnamefont{Pang}},
  \bibinfo{author}{\bibfnamefont{X.}~\bibnamefont{Zhang}}, \bibnamefont{and}
  \bibinfo{author}{\bibfnamefont{Q.-G.} \bibnamefont{Huang}},
  \bibinfo{journal}{Chin. Phys. C} \textbf{\bibinfo{volume}{48}},
  \bibinfo{pages}{065102} (\bibinfo{year}{2024}), \eprint{2312.07188}.

\bibitem[{\citenamefont{Maki et~al.}(1962)\citenamefont{Maki, Nakagawa, and
  Sakata}}]{Maki:1962mu}
\bibinfo{author}{\bibfnamefont{Z.}~\bibnamefont{Maki}},
  \bibinfo{author}{\bibfnamefont{M.}~\bibnamefont{Nakagawa}}, \bibnamefont{and}
  \bibinfo{author}{\bibfnamefont{S.}~\bibnamefont{Sakata}},
  \bibinfo{journal}{Prog. Theor. Phys.} \textbf{\bibinfo{volume}{28}},
  \bibinfo{pages}{870} (\bibinfo{year}{1962}).

\bibitem[{\citenamefont{Abe et~al.}(2024)}]{KamLAND-Zen:2024eml}
\bibinfo{author}{\bibfnamefont{S.}~\bibnamefont{Abe}} \bibnamefont{et~al.}
  (\bibinfo{collaboration}{KamLAND-Zen}) (\bibinfo{year}{2024}),
  \eprint{2406.11438}.

\bibitem[{\citenamefont{Abgrall et~al.}(2021)}]{LEGEND:2021bnm}
\bibinfo{author}{\bibfnamefont{N.}~\bibnamefont{Abgrall}} \bibnamefont{et~al.}
  (\bibinfo{collaboration}{LEGEND}) (\bibinfo{year}{2021}),
  \eprint{2107.11462}.

\bibitem[{\citenamefont{Adhikari et~al.}(2022)}]{nEXO:2021ujk}
\bibinfo{author}{\bibfnamefont{G.}~\bibnamefont{Adhikari}} \bibnamefont{et~al.}
  (\bibinfo{collaboration}{nEXO}), \bibinfo{journal}{J. Phys. G}
  \textbf{\bibinfo{volume}{49}}, \bibinfo{pages}{015104}
  (\bibinfo{year}{2022}), \eprint{2106.16243}.

\bibitem[{\citenamefont{Aker et~al.}(2025)}]{KATRIN:2024cdt}
\bibinfo{author}{\bibfnamefont{M.}~\bibnamefont{Aker}} \bibnamefont{et~al.}
  (\bibinfo{collaboration}{KATRIN}), \bibinfo{journal}{Science}
  \textbf{\bibinfo{volume}{388}}, \bibinfo{pages}{adq9592}
  (\bibinfo{year}{2025}), \eprint{2406.13516}.

\bibitem[{\citenamefont{Afanaciev et~al.}(2024)}]{MEGII:2023ltw}
\bibinfo{author}{\bibfnamefont{K.}~\bibnamefont{Afanaciev}}
  \bibnamefont{et~al.} (\bibinfo{collaboration}{MEG II}),
  \bibinfo{journal}{Eur. Phys. J. C} \textbf{\bibinfo{volume}{84}},
  \bibinfo{pages}{216} (\bibinfo{year}{2024}), \bibinfo{note}{[Erratum:
  Eur.Phys.J.C 84, 1042 (2024)]}, \eprint{2310.12614}.

\bibitem[{\citenamefont{Aubert et~al.}(2010)}]{BaBar:2009hkt}
\bibinfo{author}{\bibfnamefont{B.}~\bibnamefont{Aubert}} \bibnamefont{et~al.}
  (\bibinfo{collaboration}{BaBar}), \bibinfo{journal}{Phys. Rev. Lett.}
  \textbf{\bibinfo{volume}{104}}, \bibinfo{pages}{021802}
  (\bibinfo{year}{2010}), \eprint{0908.2381}.

\bibitem[{\citenamefont{Abdesselam et~al.}(2021)}]{Belle:2021ysv}
\bibinfo{author}{\bibfnamefont{A.}~\bibnamefont{Abdesselam}}
  \bibnamefont{et~al.} (\bibinfo{collaboration}{Belle}),
  \bibinfo{journal}{JHEP} \textbf{\bibinfo{volume}{10}}, \bibinfo{pages}{19}
  (\bibinfo{year}{2021}), \eprint{2103.12994}.

\bibitem[{\citenamefont{Fan}(2022)}]{Fan:2022oyb}
\bibinfo{author}{\bibfnamefont{X.}~\bibnamefont{Fan}}, Ph.D. thesis,
  \bibinfo{school}{Harvard U.} (\bibinfo{year}{2022}).

\bibitem[{\citenamefont{Fan et~al.}(2023)\citenamefont{Fan, Myers, Sukra, and
  Gabrielse}}]{Fan:2022eto}
\bibinfo{author}{\bibfnamefont{X.}~\bibnamefont{Fan}},
  \bibinfo{author}{\bibfnamefont{T.~G.} \bibnamefont{Myers}},
  \bibinfo{author}{\bibfnamefont{B.~A.~D.} \bibnamefont{Sukra}},
  \bibnamefont{and}
  \bibinfo{author}{\bibfnamefont{G.}~\bibnamefont{Gabrielse}},
  \bibinfo{journal}{Phys. Rev. Lett.} \textbf{\bibinfo{volume}{130}},
  \bibinfo{pages}{071801} (\bibinfo{year}{2023}), \eprint{2209.13084}.

\bibitem[{\citenamefont{Navas et~al.}(2024)}]{ParticleDataGroup:2024cfk}
\bibinfo{author}{\bibfnamefont{S.}~\bibnamefont{Navas}} \bibnamefont{et~al.}
  (\bibinfo{collaboration}{Particle Data Group}), \bibinfo{journal}{Phys. Rev.
  D} \textbf{\bibinfo{volume}{110}}, \bibinfo{pages}{030001}
  (\bibinfo{year}{2024}).

\bibitem[{\citenamefont{Morel et~al.}(2020)\citenamefont{Morel, Yao, Clad{\'e},
  and Guellati-Kh{\'e}lifa}}]{Morel:2020dww}
\bibinfo{author}{\bibfnamefont{L.}~\bibnamefont{Morel}},
  \bibinfo{author}{\bibfnamefont{Z.}~\bibnamefont{Yao}},
  \bibinfo{author}{\bibfnamefont{P.}~\bibnamefont{Clad{\'e}}},
  \bibnamefont{and}
  \bibinfo{author}{\bibfnamefont{S.}~\bibnamefont{Guellati-Kh{\'e}lifa}},
  \bibinfo{journal}{Nature} \textbf{\bibinfo{volume}{588}}, \bibinfo{pages}{61}
  (\bibinfo{year}{2020}).

\bibitem[{\citenamefont{Kong and Matchev}(2006)}]{Kong:2005hn}
\bibinfo{author}{\bibfnamefont{K.}~\bibnamefont{Kong}} \bibnamefont{and}
  \bibinfo{author}{\bibfnamefont{K.~T.} \bibnamefont{Matchev}},
  \bibinfo{journal}{JHEP} \textbf{\bibinfo{volume}{01}}, \bibinfo{pages}{038}
  (\bibinfo{year}{2006}), \eprint{hep-ph/0509119}.

\bibitem[{\citenamefont{Barbieri et~al.}(2006)\citenamefont{Barbieri, Hall, and
  Rychkov}}]{Barbieri:2006dq}
\bibinfo{author}{\bibfnamefont{R.}~\bibnamefont{Barbieri}},
  \bibinfo{author}{\bibfnamefont{L.~J.} \bibnamefont{Hall}}, \bibnamefont{and}
  \bibinfo{author}{\bibfnamefont{V.~S.} \bibnamefont{Rychkov}},
  \bibinfo{journal}{Phys. Rev. D} \textbf{\bibinfo{volume}{74}},
  \bibinfo{pages}{015007} (\bibinfo{year}{2006}), \eprint{hep-ph/0603188}.

\bibitem[{\citenamefont{Esteban et~al.}(2024)\citenamefont{Esteban,
  Gonzalez-Garcia, Maltoni, Martinez-Soler, Pinheiro, and
  Schwetz}}]{Esteban:2024eli}
\bibinfo{author}{\bibfnamefont{I.}~\bibnamefont{Esteban}},
  \bibinfo{author}{\bibfnamefont{M.~C.} \bibnamefont{Gonzalez-Garcia}},
  \bibinfo{author}{\bibfnamefont{M.}~\bibnamefont{Maltoni}},
  \bibinfo{author}{\bibfnamefont{I.}~\bibnamefont{Martinez-Soler}},
  \bibinfo{author}{\bibfnamefont{J.~a.~P.} \bibnamefont{Pinheiro}},
  \bibnamefont{and} \bibinfo{author}{\bibfnamefont{T.}~\bibnamefont{Schwetz}},
  \bibinfo{journal}{JHEP} \textbf{\bibinfo{volume}{12}}, \bibinfo{pages}{216}
  (\bibinfo{year}{2024}), \eprint{2410.05380}.

\bibitem[{\citenamefont{Bodrov}(2024)}]{Bodrov:2024wrw}
\bibinfo{author}{\bibfnamefont{D.}~\bibnamefont{Bodrov}},
  \bibinfo{journal}{Int. J. Mod. Phys. A} \textbf{\bibinfo{volume}{39}},
  \bibinfo{pages}{2442006} (\bibinfo{year}{2024}), \eprint{2405.16512}.

\bibitem[{\citenamefont{Funakoshi et~al.}(2025)\citenamefont{Funakoshi,
  Kobayashi, and Otsuka}}]{Funakoshi:2024uvy}
\bibinfo{author}{\bibfnamefont{S.}~\bibnamefont{Funakoshi}},
  \bibinfo{author}{\bibfnamefont{T.}~\bibnamefont{Kobayashi}},
  \bibnamefont{and} \bibinfo{author}{\bibfnamefont{H.}~\bibnamefont{Otsuka}},
  \bibinfo{journal}{JHEP} \textbf{\bibinfo{volume}{04}}, \bibinfo{pages}{183}
  (\bibinfo{year}{2025}), \eprint{2412.12524}.

\end{thebibliography}

\end{document}